\documentclass[a4paper, 11pt]{article}

\usepackage{subcaption}
\usepackage{booktabs}
\usepackage{longtable}
\usepackage{array}
\usepackage{multirow}
\usepackage{caption}
\usepackage{adjustbox}
\usepackage{pdflscape}
\usepackage{siunitx}

\usepackage{ xcolor}
\usepackage{geometry}
\usepackage{float}

\usepackage{amsmath, amssymb, amsthm, amsfonts}
\usepackage{graphicx, color}
\usepackage{booktabs}
\usepackage{setspace}
\usepackage{geometry}
\usepackage{hyperref}
\theoremstyle{remark}
\newtheorem{theorem}{Theorem}

\newtheorem{remark}{Remark}

\title{PCA-Guided Quantile Sampling: Preserving Data Structure in Large-Scale Subsampling}
\author{Foo Hui-Mean and Yuan-chin Ivan Chang\\Institute of Statistical Science\\ Academia Sinica, Taipei, Taiwan 11529}
\date{\small\today}

\begin{document}

\maketitle

\begin{abstract}
We introduce Principal Component Analysis-guided Quantile Sampling (PCA-QS), a novel sampling framework designed to preserve both the statistical and geometric structure of large-scale datasets. Unlike conventional PCA, which reduces dimensionality at the cost of interpretability, PCA-QS retains the original feature space while using leading principal components solely to guide a quantile-based stratification scheme. This principled design ensures that sampling remains representative without distorting the underlying data semantics. We establish rigorous theoretical guarantees, deriving convergence rates for empirical quantiles, Kullback–Leibler divergence, and Wasserstein distance, thus quantifying the distributional fidelity of PCA-QS samples. Practical guidelines for selecting the number of principal components, quantile bins, and sampling rates are provided based on these results. Extensive empirical studies on both synthetic and real-world datasets show that PCA-QS consistently outperforms simple random sampling, yielding better structure preservation and improved downstream model performance. Together, these contributions position PCA-QS as a scalable, interpretable, and theoretically grounded solution for efficient data summarization in modern machine learning workflows.\\


\end{abstract}

\section{Introduction}

The exponential growth of modern datasets poses a fundamental challenge: how to reduce data volume while preserving both statistical fidelity and structural integrity. Conventional sampling strategies, although effective for data reduction, often compromise representativeness, distort key distributions, and degrade the performance of downstream models. To address these limitations, we propose Principal Component Analysis-guided Quantile Sampling (PCA-QS), a principled framework for subsampling that preserves the intrinsic structure of complex datasets.

In contrast to conventional PCA, which projects data into a lower-dimensional subspace—often sacrificing interpretability—PCA-QS retains the full original feature space. Instead, it leverages the leading principal components purely as structural guides to define a quantile-based stratification scheme. Sampling is then conducted within these quantile strata, ensuring that the selected subset faithfully captures major variance patterns, distributional characteristics, and the underlying geometric structure of the data.

The proposed method is supported by rigorous theoretical analysis. By building on established results for empirical quantiles, Kullback–Leibler (KL) divergence, and Wasserstein distance, we demonstrate that PCA-QS exhibits well-defined convergence behavior: as the sample size grows, the distribution of the sampled subset converges uniformly to that of the full dataset. These guarantees provide a robust theoretical foundation for the method’s representativeness and structural preservation.

PCA-QS makes several notable contributions:
First, it introduces a novel sampling paradigm that combines the interpretability of the original feature space with the structural insight provided by principal components. Second, it offers a theoretical framework for parameter selection, including the number of principal components and quantile bins, grounded in explicit convergence rates. Third, it demonstrates empirically—across synthetic and real-world datasets—that PCA-QS consistently outperforms standard simple random sampling (SRS) in maintaining structural fidelity and improving the performance of downstream machine learning models. Finally, it provides a scalable, interpretable, and theoretically justified solution for efficient data summarization in modern machine learning workflows.

The remainder of this paper is organized as follows: Section~\ref{sec:method} reviews related work and details the PCA-QS methodology. Section~\ref{sec:experiments} presents empirical evaluations on synthetic and real-world datasets. Section~\ref{sec:conclusion} discusses broader implications and concludes the study.

\section{Method}
\label{sec:method}

Principal Component Analysis (PCA) is a classical technique for dimensionality reduction that projects high-dimensional data onto a lower-dimensional subspace that retains the maximal variance. Given a dataset \(\mathbf{X} \in \mathbb{R}^{n \times p}\), the PCA transformation is typically obtained via the Singular Value Decomposition (SVD):
\(
\mathbf{X} = \mathbf{U} \mathbf{D} \mathbf{V}^\top,
\)
where \(\mathbf{U} \in \mathbb{R}^{n \times p}\) contains the left singular vectors, \(\mathbf{V} \in \mathbb{R}^{p \times p}\) the right singular vectors, and \(\mathbf{D} \in \mathbb{R}^{p \times p}\) is a diagonal matrix of singular values. The projection of the dataset onto the top \(k\) principal components yields the reduced representation:
\(
\mathbf{Z} = \mathbf{U}_k \mathbf{D}_k,
\)
where \(\mathbf{U}_k\) and \(\mathbf{D}_k\) contain the first \(k\) columns and leading singular values, respectively. While PCA is effective in capturing dominant variance directions, it does not inherently reduce the number of data points. To address this, the PCA-guided Quantile Sampling (PCA-QS) framework augments PCA with a stratified sampling mechanism to achieve both data summarization and structural fidelity.

\paragraph{Property of Quantiles:}

Let \( X_1, X_2, \dots, X_n \) be an i.i.d.\ sample from a distribution with cumulative distribution function (CDF) \( F(x) \). The empirical CDF is defined as \( F_n(x) = \frac{1}{n} \sum_{i=1}^{n} \mathbf{1}(X_i \leq x) \). For a quantile level \( p \in (0,1) \), the true quantile is \( q_p = F^{-1}(p) \), and the empirical quantile is \( q_p^n = F_n^{-1}(p) \).

By the Glivenko--Cantelli theorem \cite{vanderVaartWellner1996}, the empirical CDF converges uniformly to the true CDF as \( n \to \infty \), i.e., \( \sup_x |F_n(x) - F(x)| = O(n^{-1/2}) \). Consequently, the quantile estimation error satisfies \( q_p^n - q_p = O(n^{-1/2}) \).

\subsection{PCA-Guided Quantile Sampling Method}
\paragraph{Sampling in Principal Component Space:}
PCA-guided Quantile Sampling (PCA-QS) stratifies data along principal component directions using quantile partitions to facilitate representative sampling. For the case of a single principal component (\(\text{num.\ of PCs} = 1\)) and \(g\) quantile groups, the data is partitioned into intervals as follows:
\begin{equation}
[\text{min}, Q_1), [Q_1, Q_2), \ldots, [Q_{g-1}, \text{max}],
\end{equation}
ensuring that each group contains an approximately equal proportion of the data.

\paragraph{Quantile Assignment for Multiple Components}
In the general case with multiple principal components (\(\text{num.\ of PCs} > 1\)), quantile stratification extends across a multi-dimensional subspace. For each principal component \(z_j\), the data is divided into quantile bins:
\[
\text{Quantile Groups for Component } j: [\text{Min}, Q_1), [Q_1, Q_2), \ldots, [Q_{g-1}, \text{Max}].
\]
Composite quantile groups are then formed by combining the bin assignments across the selected components. These composite groups are uniquely encoded to facilitate balanced sampling in the reduced PCA space.

Sampling within each composite quantile group is typically random to promote unbiased representation. The number of retained samples per group is defined as:
\[
\text{Sample Size per Group} = \min \left( \left\lceil \text{retention rate} \times N_g \right\rceil, N_g \right),
\]
where \(N_g\) is the size of the group, and the retention rate controls the overall proportion of retained data. This process preserves both distributional balance and structural integrity. The random selection step may also be replaced by adaptive sampling strategies tailored to downstream tasks.

Before analyzing the theoretical properties of the PCA-QS procedure, we briefly review the relevant convergence characteristics of PCA projections and quantile estimators. A complete procedural description is provided in Appendix~\ref{app:pcaqs_sampling}.

\subsection{Large Sample Properties of PCA-QS}

Let the population covariance matrix be \( \boldsymbol{\Sigma} = \mathbb{E}[(X - \mu)(X - \mu)^T] \), and denote the sample covariance matrix as \( \hat{\boldsymbol{\Sigma}} = \frac{1}{n-1} \sum_{i=1}^{n} (X_i - \bar{X})(X_i - \bar{X})^T \). If \( X \sim \mathcal{N}(\boldsymbol{\mu}, \boldsymbol{\Sigma}) \), then \( n \hat{\boldsymbol{\Sigma}} \sim \mathcal{W}(\boldsymbol{\Sigma}, n-1) \), where \( \mathcal{W}(\cdot, \cdot) \) denotes the Wishart distribution. The estimated eigenvectors \( \hat{\boldsymbol{V}} \) of \( \hat{\boldsymbol{\Sigma}} \) converge to the true eigenvectors \( \boldsymbol{V} \) at the rate \( \|\hat{\boldsymbol{V}} - \boldsymbol{V}\|_F = O(n^{-1/2}) \).

Now consider the case where only the first principal component is used. Let \( Y_i = \boldsymbol{v}_1^T X_i \) for \( i = 1, \dots, n \), and suppose the range is partitioned into \( m \) quantile bins. Then the number of observations falling into each bin satisfies \( n_j = p_j n + O(n^{1/2}) \), where \( p_j \) is the mass of the true distribution in that bin, yielding a relative error of \( (n_j - p_j n)/(p_j n) = O(n^{-1/2}) \).

When using the top \( k \) components, let \( \boldsymbol{Y}_i = (\boldsymbol{v}_1^T X_i, \dots, \boldsymbol{v}_k^T X_i) \) be the projected scores for sample \( i \). The empirical quantile function over these projections, denoted \( q_p^n(\boldsymbol{Y}, k) \), satisfies the multivariate central limit theorem: \( \sqrt{n}(q_p^n(\boldsymbol{Y}, k) - q_p(\boldsymbol{Y}, k)) \xrightarrow{d} \mathcal{N}(\mathbf{0}, \boldsymbol{\Sigma}_q(k)) \), where \( \boldsymbol{\Sigma}_q(k) \) is the asymptotic variance. This leads to the following uniform convergence result:

\begin{theorem}[Uniform Convergence of Multiple Quantiles]
\label{UnifQS}
For any quantile function \( q_p^n(\boldsymbol{Y}, k) \), we have \( \sup_p |q_p^n(\boldsymbol{Y}, k) - q_p(\boldsymbol{Y}, k)| = O(n^{-1/2}) \). This implies that quantiles estimated via PCA-QS converge uniformly across the top \( k \) principal components.
\end{theorem}

\begin{remark}
This result ensures that PCA-QS preserves statistical representativeness as the sample size increases. See Appendix~\ref{app:intuition} for an intuitive explanation and visual illustration.
\end{remark}

Furthermore, PCA-QS achieves rapid convergence in distributional fidelity. The Kullback–Leibler (KL) divergence between the sampled and full distributions decays at a rate of \( O(n^{-1}) \), while the Wasserstein distance decays at \( O(n^{-1/d}) \), reflecting preservation of geometric structure even in high dimensions.

\begin{remark}
These results imply that as more samples are drawn via PCA-QS, the retained subset becomes increasingly similar to the original data, both statistically and geometrically. See Appendix~\ref{app:intuition} for conceptual analogies.
\end{remark}

\begin{proof}[Proof of Theorem \ref{UnifQS}]
Let \( F_{\boldsymbol{Y}, k} \) be the joint CDF of the top \( k \) principal components, and \( F_{\boldsymbol{Y}, k}^n(\boldsymbol{y}) = \frac{1}{n} \sum_{i=1}^{n} \mathbf{1}(\boldsymbol{Y}_i \leq \boldsymbol{y}) \) be its empirical estimate. By the Dvoretzky–Kiefer–Wolfowitz (DKW) inequality \cite{Massart1990}, the uniform error satisfies \( \sup_{\boldsymbol{y}} |F_{\boldsymbol{Y}, k}^n(\boldsymbol{y}) - F_{\boldsymbol{Y}, k}(\boldsymbol{y})| = O_p(n^{-1/2}) \).

Using the quantile inversion property, we have:
\[
F_{\boldsymbol{Y}, k}(q_p^n) - F_{\boldsymbol{Y}, k}(q_p) \approx F_{\boldsymbol{Y}, k}^n(q_p^n) - p = O(n^{-1/2}).
\]
Since \( F_{\boldsymbol{Y}, k} \) is differentiable with bounded density, the inverse function theorem implies that \( |q_p^n - q_p| = O(n^{-1/2}) \) for fixed \( p \). Uniform convergence over all \( p \in (0, 1) \) then follows by Kolmogorov’s uniform law of large numbers, completing the proof.
\end{proof}

\subsection{Distributional Approximation in PCA-QS}

Beyond quantile-level accuracy, PCA-QS also achieves strong distributional fidelity. In this section, we examine how the divergence between the empirical and true distributions behaves as the sample size increases.

\subsubsection{KL Divergence Decay in PCA-QS}

The \emph{Kullback–Leibler (KL) divergence} quantifies how much the empirical distribution \( P_n \), derived via PCA-QS, deviates from the true distribution \( P \). For continuous densities \( f_n \) and \( f \), it is defined as:
\[
D_{\mathrm{KL}}(f_n \,\|\, f) = \int f_n(\boldsymbol{y}) \log \frac{f_n(\boldsymbol{y})}{f(\boldsymbol{y})} \, d\boldsymbol{y}.
\]

To analyze \( D_{\mathrm{KL}}(P_n \| P) \), let \( \boldsymbol{Y} = (\boldsymbol{v}_1^T X, \dots, \boldsymbol{v}_k^T X) \) denote the first \( k \) principal components, projecting the original data \( X \) into a reduced subspace. The marginal density of \( \boldsymbol{Y} \), denoted \( f_{\boldsymbol{Y}} \), is given by \( f_{\boldsymbol{Y}}(\boldsymbol{y}) = \int f_X(\boldsymbol{x}) \, d\boldsymbol{x}_{\perp} \), where \( \boldsymbol{x}_{\perp} \) integrates out the orthogonal complement. Since PCA preserves second-order structure, we assume \( f_{\boldsymbol{Y}} \) is approximately Gaussian.

Let \( f_n \) be the kernel density estimator (KDE) of \( f_{\boldsymbol{Y}} \) using the PCA-QS samples. From KDE theory \cite{Silverman1986}, we have \( \sup_{\boldsymbol{y}} | f_n(\boldsymbol{y}) - f_{\boldsymbol{Y}}(\boldsymbol{y}) | = O_p(n^{-1/2}) \). Furthermore, the expected \( L_2 \)-distance between \( f_n \) and \( f \) satisfies \( \mathbb{E}[\| f_n - f \|^2] = O(n^{-1}) \).

Applying \emph{Pinsker’s inequality} \cite{CoverThomas2006}, we obtain:
\[
D_{\mathrm{KL}}(P_n \| P) \leq \frac{1}{2} \mathbb{E} \left[ \| f_n - f \|^2 \right] = O(n^{-1}).
\]

This result demonstrates that PCA-QS achieves rapid convergence in KL divergence as the sample size grows. Notably, the rate \( O(n^{-1}) \) is faster than the quantile convergence rate of \( O(n^{-1/2}) \), underscoring the strength of PCA-QS in approximating full data distributions under a probabilistic lens.

\subsubsection{Wasserstein Distance Decay in PCA-QS}

While KL divergence quantifies the discrepancy between probability densities, the \emph{Wasserstein distance} captures geometric differences between distributions. This makes it especially relevant for evaluating how well the spatial structure of data is preserved under sampling.

For probability measures \( P \) and \( P_n \) on \( \mathbb{R}^d \), the \( p \)-Wasserstein distance is defined as:
\[
W_p(P_n, P) = \left( \inf_{\gamma \in \Gamma(P_n, P)} \mathbb{E}_{\gamma} \left[ \| \boldsymbol{y} - \boldsymbol{y}' \|^p \right] \right)^{1/p},
\]
where \( \Gamma(P_n, P) \) is the set of all couplings of \( P_n \) and \( P \). For the commonly used case \( p = 2 \), this simplifies to:
\[
W_2(P_n, P) = \left( \inf_{\gamma \in \Gamma(P_n, P)} \mathbb{E}_{\gamma} \left[ \| \boldsymbol{y} - \boldsymbol{y}' \|^2 \right] \right)^{1/2}.
\]

Let \( \boldsymbol{Y} = (\boldsymbol{v}_1^T X, \dots, \boldsymbol{v}_k^T X) \) be the projection of data \( X \) onto its first \( k \) principal components, with marginal density \( f_{\boldsymbol{Y}} \) defined by:
\[
f_{\boldsymbol{Y}}(\boldsymbol{y}) = \int f_X(\boldsymbol{x}) \, d\boldsymbol{x}_{\perp}.
\]
Since PCA preserves the second-order structure of \( X \), PCA-QS samples approximate the distribution of \( \boldsymbol{Y} \) well.

From empirical measure theory and concentration inequalities,   
\(
W_2(P_n, P) = O(n^{-1/d}),
\)
where \( d \) is the dimension of the space over which the approximation is measured. This rate reflects the well-known curse of dimensionality, where convergence slows in high dimensions.  (\cite{Dudley1999,BoucheronLugosiMassart2013} \cite{Talagrand1996,  vanderVaartWellner1996}.)

However, PCA-QS mitigates this effect by reducing the effective dimension from \( d \) to \( k \ll d \). Thus, even though the Wasserstein convergence rate is slower than KL divergence (\( O(n^{-1}) \)), it provides complementary guarantees: KL decay ensures fidelity in density approximation, while Wasserstein decay reflects preservation of geometric structure.

\begin{remark}
The dual decay rates—\( O(n^{-1}) \) for KL divergence and \( O(n^{-1/d}) \) for Wasserstein distance—jointly establish that PCA-QS provides both distributional and spatial fidelity. By operating in a reduced-dimensional space, PCA-QS accelerates convergence and maintains generalization robustness in large-scale settings.
\end{remark}

\begin{remark}[Comparison of KL Divergence and Wasserstein Distance]
KL divergence measures differences in probability densities and decays at a fixed rate of \( O(n^{-1}) \), ensuring PCA-QS accurately preserves the distributional structure. In contrast, Wasserstein distance quantifies geometric misalignment and decays more slowly at \( O(n^{-1/d}) \), where \( d \) is the dimensionality. However, by projecting data onto a lower-dimensional subspace, PCA-QS effectively reduces \( d \), accelerating Wasserstein convergence. Together, these measures highlight that PCA-QS preserves both probabilistic and geometric fidelity.
\end{remark}

\subsection{Selection of the Number of Principal Components in PCA-QS}

Selecting the number of principal components (\( k \)) in PCA-based Quantile Sampling (PCA-QS) is essential for balancing computational cost and fidelity to the original data distribution. The value of \( k \) directly influences convergence behavior, including KL divergence decay, Wasserstein stability, and quantile estimation accuracy.

\subsubsection{Criteria for Selecting \( k \)}

PCA transforms data into an orthogonal basis where each component captures a portion of total variance. Let \( \lambda_1, \dots, \lambda_d \) be the eigenvalues of the covariance matrix \( \boldsymbol{\Sigma} \). The cumulative variance explained by the first \( k \) components is:
\[
\rho_k = \frac{\sum_{j=1}^{k} \lambda_j}{\sum_{j=1}^{d} \lambda_j}.
\]
A common heuristic is to choose the smallest \( k \) such that \( \rho_k \geq 0.90 \) or \( 0.95 \), ensuring that most variability is retained.

From a distributional perspective, KL divergence decays at rate \( D_{\mathrm{KL}}(P_n^{(k)} \| P) = O(k^{-1}) \), implying that increasing \( k \) enhances density approximation, but with diminishing returns. In contrast, Wasserstein distance decays more slowly as \( W_2(P_n^{(k)}, P) = O(n^{-1/k}) \), where \( k \) acts as the effective dimension. Thus, PCA helps accelerate convergence by reducing \( k \ll d \).

Another useful criterion is the spectral gap between eigenvalues, defined as \( \lambda_k - \lambda_{k+1} \). A large spectral gap indicates a natural truncation point, beyond which additional components add minimal variance. Selecting \( k \) at the largest spectral gap ensures efficient compression without substantial loss.

\subsubsection{Cross-Validation Based on Sampling Performance}

An alternative is to empirically validate PCA-QS across different \( k \) by assessing the mean squared error (MSE) in quantile estimation:
\[
\text{MSE}(q_p^n(\boldsymbol{Y}, k)) = \mathbb{E}\left[(q_p^n(\boldsymbol{Y}, k) - q_p(\boldsymbol{Y}))^2\right].
\]
The optimal \( k \) minimizes this error while keeping the dimension low.

\paragraph{Theoretical Justification}
Choosing an appropriate \( k \) ensures that PCA-QS retains critical distributional features. Moderate \( k \) values enable fast KL divergence decay for statistical fidelity and reduce effective dimensionality, improving Wasserstein convergence. Spectral gap analysis guards against overfitting to noise by identifying informative components.

\paragraph{Practical Justification}
In applied settings, computational efficiency often constrains the choice of \( k \). Monitoring when KL divergence stabilizes offers a natural stopping rule. Likewise, Wasserstein distance helps assess spatial consistency in the sampled data. Finally, ensuring that the selected \( k \) captures a meaningful share of variance improves interpretability and downstream analysis.

In sum, PCA-QS requires careful selection of \( k \) to balance variance retention, distributional convergence, and efficiency. A principled choice of \( k \) ensures that PCA-QS remains robust and scalable for large-scale data sampling.

\subsection{Balancing Principal Components and Quantiles}

In PCA-based Quantile Sampling (PCA-QS), both the number of principal components (\( k \)) and the number of quantile bins (\( m \)) play a crucial role in determining sampling performance. This section explores their theoretical relationship and offers practical guidance for their selection.

\subsubsection{Theoretical Relationship Between \( k \) and \( m \)}

PCA projects data onto a lower-dimensional subspace defined by the top \( k \) components, while quantile stratification partitions the projected data into \( m \) bins per dimension. The total approximation error in PCA-QS can be decomposed into three sources: sampling error (\( O(n^{-1/2}) \)), projection error (\( O(k^{-1}) \)), and quantile discretization error (\( O(m^{-1}) \)). To minimize the combined error while preserving efficiency, one should balance the projection and discretization terms, i.e., \( O(k^{-1}) \approx O(m^{-1}) \).

\subsubsection{Trade-off Between \( k \) and \( m \)}

Increasing \( k \) reduces projection error and better captures data variability, but increases computational cost and necessitates finer stratification (larger \( m \)) to preserve detail. Conversely, increasing \( m \) refines stratification but requires more data and becomes ineffective if the projection is too coarse (i.e., \( k \) too small). Hence, optimal performance depends on jointly tuning these parameters.

\paragraph{Guidelines for Choosing \( k \) and \( m \)}  
In practice, selection depends on the trade-off between computational efficiency and distributional accuracy. When efficiency is prioritized, a smaller \( k \) may be used, with \( m \) chosen such that \( O(k^{-1}) \approx O(m^{-1}) \). For higher fidelity, \( k \) should be increased first to reduce projection loss, followed by increasing \( m \) to achieve finer stratification. When the sample size \( n \) is small, overly large \( m \) should be avoided, as sampling error may dominate.

\paragraph{Conclusion Remark}  
The optimal selection of \( k \) and \( m \) ensures that dimensionality reduction and stratification remain aligned. The guideline \( O(k^{-1}) \approx O(m^{-1}) \) offers a practical rule for balancing structural integrity and efficiency in PCA-QS.

\subsection{Effects of the Number of Quantiles in PCA-QS}

The number of quantiles (\( m \)) controls the resolution of stratified sampling in PCA-QS. Increasing \( m \) refines stratification and improves sampling granularity but also increases computational overhead and risks overfitting, especially with limited data. The quantile approximation error decays as \( O(m^{-1}) \), yet higher values of \( m \) require more samples to ensure empirical stability and meaningful partitions.

In high-dimensional settings, if \( m \ll k \), stratification may fail to capture local density variations, degrading sampling quality. Conversely, overly large \( m \) with small \( n \) results in unstable quantiles due to sparse bins.

\begin{table}[h]
    \centering
    \resizebox{\textwidth}{!}{
    \begin{tabular}{l l l}
    \toprule
    Factor & Rate or Condition & Effect on PCA-QS \\
    \midrule
    Quantile Approximation Error \( \epsilon_q(k, m) \) & \( O(m^{-1}) \) & Finer stratification, better distributional representation \\
    Empirical Quantile Stability \( q_p^n \) & \( O(n^{-1/2}) \) & Larger \( m \) needs more data to avoid noise \\
    Computational Cost & \( O(m \log m) \) & Cost of sorting and partitioning increases with \( m \) \\
    Quantile Distortion (High Dimensions) & \( O(k/m) \) & Too few bins (\( m \ll k \)) can miss local structure \\
    Sampling Bias Reduction & \( O(m^{-1}) \) & More bins reduce bias but may increase variance \\
    \bottomrule
    \end{tabular}
    }
    \caption{Influence of Quantile Count \( m \) on PCA-QS Performance Metrics}
\end{table}

Choosing \( m \) should balance stratification resolution with sample size \( n \) and dimension \( k \). When \( n \) is small, large \( m \) may lead to unstable quantiles. In high-dimensional contexts, ensuring that \( k/m \) remains bounded avoids excessive quantile distortion. Cross-validation or model-specific performance metrics can help identify an optimal \( m \).

These findings emphasize that PCA-QS performance depends critically on both sample size and quantile selection. Larger samples improve quantile estimation and reduce bias, while KL divergence and Wasserstein distance offer validation metrics for structural fidelity. Section~\ref{sec:selection_example} provides a practical illustration of how to use these theoretical insights to choose optimal PCA-QS parameters.

\begin{remark}
Instead of relying on random subsampling, PCA-QS partitions data along principal components into quantile-defined groups, ensuring balanced and structure-preserving representation. The formal quantile assignment procedure is detailed in Appendix~\ref{app:pcaqs_sampling}.
\end{remark}

\subsection{Effect of Sample Size in PCA-QS}

This section analyzes how the sample size (\( n \)) affects the accuracy, stability, and convergence behavior of PCA-based Quantile Sampling (PCA-QS). Theoretical convergence rates and their implications are summarized in Table~\ref{tab:sample_size}.

Quantile estimates converge at a rate of \( O(n^{-1/2}) \), which improves estimation precision slowly. The stability of PCA components also improves at the same rate, yielding more consistent projections. For multi-dimensional stratification using \( k \) principal components, uniform quantile convergence remains at \( O(n^{-1/2}) \), ensuring robust bin assignment as \( n \) increases.

In contrast, KL divergence decays faster at \( O(n^{-1}) \), indicating rapid improvement in distributional approximation. Wasserstein distance decays more slowly at \( O(n^{-1/d}) \), where \( d \) is the data dimension, but this convergence is accelerated by PCA, which reduces the effective dimension.

\begin{table}[h]
    \centering
    \resizebox{\textwidth}{!}{
    \begin{tabular}{l l l}
    \toprule
    Factor & Convergence Rate & Impact on PCA-QS \\
    \midrule
    Empirical Quantiles \( q_p^n \) & \( O(n^{-1/2}) \) & Accuracy improves gradually with \( n \) \\
    PCA Components \( \hat{\boldsymbol{V}} \) & \( O(n^{-1/2}) \) & Stability of projections increases \\
    Multiple PCs: \( \hat{\lambda}_j, \hat{\boldsymbol{v}}_j \) & \( O(n^{-1/2}) \) & Multi-dimensional quantile robustness improves \\
    Quantile Asymptotic Normality & \( \mathcal{N}(\mathbf{0}, \boldsymbol{\Sigma}_q(k)) \) & Distributional shape stabilizes \\
    Uniform Quantile Convergence & \( O(n^{-1/2}) \) & Consistency across \( k \) components \\
    KL Divergence \( D_{\text{KL}} \) & \( O(n^{-1}) \) & Rapid improvement in distributional match \\
    Wasserstein Distance \( W_2(P_n, P) \) & \( O(n^{-1/d}) \) & Slower convergence unless PCA is used \\
    \bottomrule
    \end{tabular}
    }
    \caption{Impact of Sample Size on PCA-QS Convergence and Sampling Fidelity}
    \label{tab:sample_size}
\end{table}

In practice, PCA-QS becomes more stable and accurate as \( n \) increases. If the primary goal is distributional fidelity, KL divergence provides a sensitive metric due to its faster decay rate. For applications where geometric alignment is critical, monitoring the Wasserstein distance is more appropriate. The sample size also mediates trade-offs between computational cost and statistical precision, making it a key factor in designing scalable and reliable PCA-QS pipelines.

\subsection{Practical Guidelines for Applying PCA-QS}

Building on the theoretical convergence results and parameter trade-offs discussed in prior sections, this subsection outlines practical considerations for applying PCA-based Quantile Sampling (PCA-QS) effectively. A well-structured application ensures a balance between computational efficiency and statistical accuracy.

\paragraph{Preprocessing and Data Normalization}

Before applying PCA-QS, the dataset should be standardized — centering each feature to zero mean and scaling to unit variance — to ensure that PCA functions correctly. Handle missing values with appropriate imputation techniques, and examine outliers that could distort the principal components. Proper preprocessing ensures stability in both PCA transformation and quantile stratification.

\paragraph{Choosing the Number of Principal Components (\( k \))}

Selecting \( k \), the number of retained principal components, is key to preserving structural integrity with minimal complexity. A common approach is to select \( k \) such that the cumulative variance explained \( \rho_k \) exceeds 90\% or 95\%. Alternatively, the spectral gap heuristic — selecting \( k \) where the eigenvalue drop \( \lambda_k - \lambda_{k+1} \) is largest — provides a data-driven cutoff. In high dimensions, cross-validation may help assess PCA-QS stability across different \( k \) values, especially in terms of KL divergence and quantile estimation accuracy.

\paragraph{Selecting the Number of Quantiles (\( m \)) for Stratification}

The number of quantile bins \( m \) controls the granularity of stratified sampling. Choosing \( m \) too small results in poor resolution, while overly large \( m \) introduces variance and computational overhead. To balance projection and stratification errors, set \( m \) such that \( O(m^{-1}) \approx O(k^{-1}) \), aligning the granularity of sampling with the retained PCA detail. Larger values of \( k \) necessitate increased \( m \) for fidelity, while smaller datasets call for conservative \( m \) values to avoid instability.

\begin{figure}[h]
    \centering
    \includegraphics[width=0.65\textwidth]{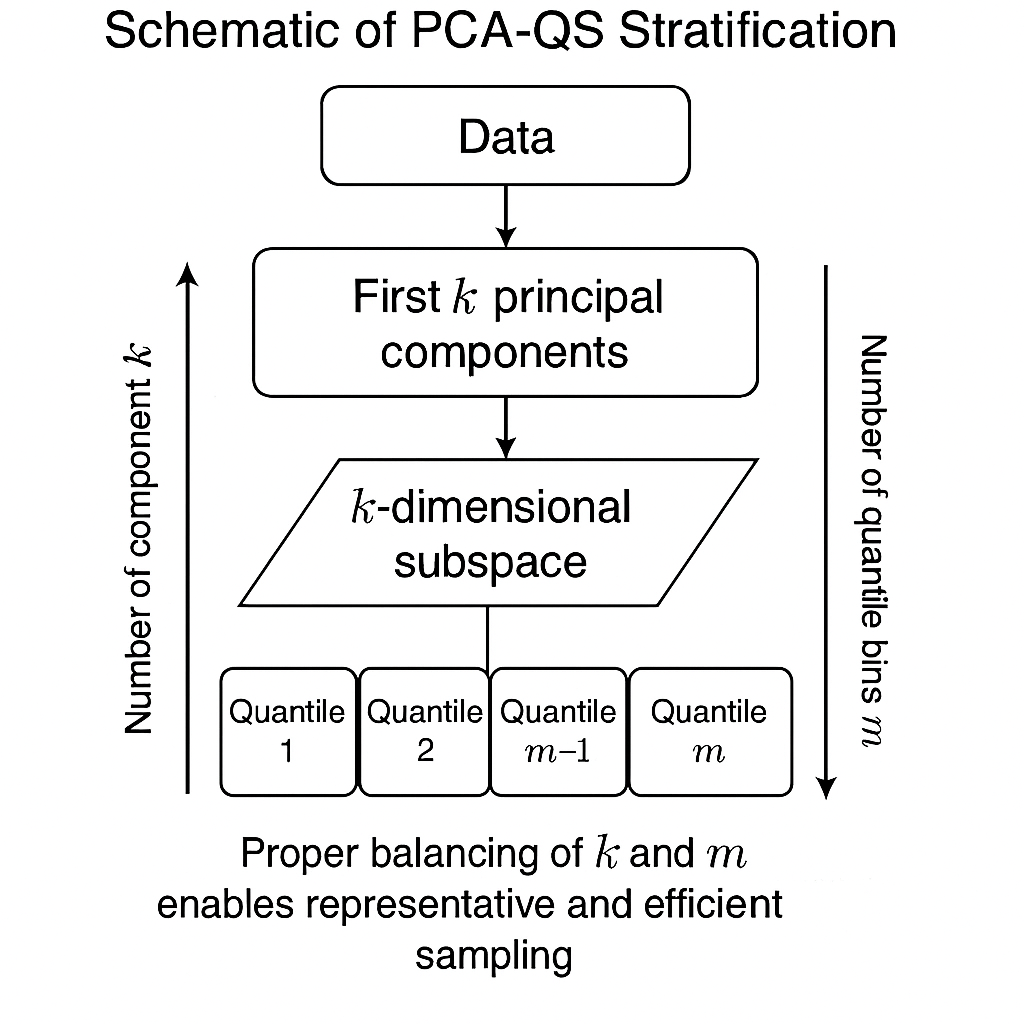}
    \caption{Schematic of PCA-QS stratification. The first \( k \) principal components define a subspace in which data are stratified into quantile-based groups. The resolution depends on the number of components and quantile bins. Proper balancing of \( k \) and \( m \) enables representative and efficient sampling.}
    \label{fig:pcaqs_stratification}
\end{figure}

\paragraph{Ensuring Distributional Fidelity}

PCA-QS is designed to maintain both statistical and geometric properties of the dataset. KL divergence decay at \( O(n^{-1}) \) verifies that the sampled distribution retains key probabilistic features. Meanwhile, Wasserstein distance at \( O(n^{-1/d}) \) reflects spatial fidelity and benefits from PCA’s dimensionality reduction. Evaluating both metrics provides a comprehensive view of distributional accuracy.

\paragraph{Computational Considerations}

Scalability of PCA-QS is critical for large datasets. Use randomized PCA for approximate decomposition in high dimensions, or incremental PCA in streaming settings. Efficient quantile computation (e.g., Greenwald-Khanna summaries) helps manage stratification overhead. Together, these strategies make PCA-QS feasible in real-world pipelines.

\paragraph{Validating PCA-QS Sampling Performance}

To confirm performance, compare the empirical quantiles of PCA-QS samples with those of the full dataset. Assess metrics such as quantile error, variance retention, and the behavior of KL divergence and Wasserstein distance under sampling. Cross-validation over partitions of the dataset improves robustness by ensuring consistency across subsamples.

PCA-QS provides a flexible and computationally efficient approach to data reduction while preserving distributional structure. Success hinges on selecting \( k \) to capture sufficient variance and choosing \( m \) to ensure accurate stratification. Validation via multiple distributional metrics ensures that PCA-QS can be confidently deployed for large-scale data analysis.

{
\subsubsection{Example: Optimal Parameter Selection Based on Convergence Rates}
\label{sec:selection_example}

To illustrate how theoretical convergence rates guide parameter selection in PCA-based Quantile Sampling (PCA-QS), we consider a hypothetical dataset with \( n = 100{,}000 \) samples and \( d = 50 \) features. The goal is to choose optimal values for the number of principal components \( k \), number of quantile bins \( m \), and the sampling retention rate \( \delta \) to achieve efficient, structure-preserving subsampling.

\paragraph{Number of Principal Components \( k \)}
The number of components \( k \) should capture sufficient variance while limiting computational cost. Let \( \rho_k = \sum_{j=1}^k \lambda_j / \sum_{j=1}^d \lambda_j \), where \( \lambda_j \) are eigenvalues of the covariance matrix. We select the smallest \( k \) satisfying both:
\[
\rho_k \geq V \quad \text{and} \quad k \leq \frac{d}{4}.
\]
For \( V = 85\% \) and \( d = 50 \), empirical results suggest \( k \in \{12, \ldots, 15\} \).

\paragraph{Choosing the Number of Quantiles \( m \)}
The quantile count \( m \) should balance projection error \( O(k^{-1}) \) and quantization error \( O(m^{-1}) \). A natural guideline is:
\[
m \approx \max(k, 15),
\]
which for \( k = 12\text{–}15 \), yields \( m \in \{15, \ldots, 20\} \).

\paragraph{Selecting the Deduction Rate \( \delta \)}
The deduction rate controls sampling intensity. KL divergence decays rapidly as \( O(n^{-1}) \), allowing for aggressive reduction, whereas Wasserstein distance \( O(n^{-1/d}) \) requires more cautious downsampling to preserve geometry. For large \( n \), empirical evidence supports:
\[
\delta = 0.05 \quad \text{(default)}, \quad \delta = 0.10 \quad \text{(for higher fidelity)}.
\]

\paragraph{Computational Considerations}
The full cost includes PCA computation \( O(ndk) \) (or \( O(nd) \) for randomized PCA), quantile sorting \( O(n \log m) \), and sampling \( O(n) \). For large-scale applications, Incremental PCA or Randomized PCA, and efficient quantile estimation (e.g., Greenwald–Khanna) are recommended.

\paragraph{Summary of Recommended Parameters}
Using the above criteria, the recommended parameters for the given dataset are:
\[
k = \min \left\{ \arg\min_k (\rho_k \geq V), \frac{d}{4} \right\}, \quad
m = \max(k, 15), \quad
\delta \in \{0.05, 0.10\}.
\]
For \( d = 50 \) and \( V = 85\% \), we obtain \( k \in \{12, \ldots, 15\} \), \( m \in \{15, \ldots, 20\} \), and \( \delta = 0.05 \) (or \( 0.10 \) for more precision).

\paragraph{PCA-QS Parameter Selection as a Trade-Off Optimization}

Parameter tuning in PCA-QS can be viewed as a constrained optimization that balances distributional fidelity and computational efficiency. Projection error decays at rate \( O(k^{-1}) \), while quantile approximation error decays as \( O(m^{-1}) \), suggesting that \( m \) should scale roughly with \( k \). KL divergence decays faster at \( O(n^{-1}) \), allowing for aggressive data reduction via small \( \delta \), whereas Wasserstein distance decays more slowly at \( O(n^{-1/d}) \), benefiting from dimension reduction through PCA.

In practice, heuristics such as setting \( m \approx k \) and choosing \( \delta \in [0.05, 0.10] \) perform well across many settings. Nonetheless, in high-dimensional or small-sample scenarios, data-specific tuning may yield further improvements. This underscores the value of adaptive strategies for optimizing PCA-QS parameters in applied settings.

For a detailed derivation of the quantile assignment and optimization of parameter settings, please refer to the Supplementary Material.
}

\section{Numerical Study}
\label{sec:experiments}

\subsection{Synthesized Data}

\paragraph{Properties of Retentive Data Sets}

This section evaluates how well \textit{PCA-based Quantile Sampling (PCA-QS)} and \textit{Simple Random Sampling (SRS)} preserve the statistical structure of high-dimensional data. Specifically, we examine their ability to retain covariance structure and distributional properties relative to the full dataset.

We simulate a synthetic dataset with $100{,}000$ samples and 50 features using a block-wise correlated covariance matrix, optionally enhanced with a Gaussian Mixture Model (GMM) to test robustness under multimodal structures.

PCA-QS is applied with a retention rate of $\delta = 0.05$, using quantile binning over several principal components. SRS selects samples uniformly without regard to structure. We compare both sampling strategies using geometric and probabilistic distance metrics. Technical details of the PCA-QS stratification procedure are provided in the Supplementary material.

\paragraph{Quantitative Comparison of Covariance and Distributional Distances}

We compare the performance of \textit{PCA-based Quantile Sampling (PCA-QS)} and \textit{Simple Random Sampling (SRS)} across different numbers of retained principal components: 3, 5, 10, and a dynamic choice based on preserving 70\% of cumulative variance. Each sampling method produces a reduced dataset, which is evaluated against the original data using similarity metrics such as covariance fidelity, KL divergence, and Wasserstein distance. Results are averaged over 1,000 replications for each configuration to ensure statistical stability.

\subsection*{Synthetic Data Generation}

The synthetic data are constructed for binary classification using a structured Gaussian mixture model. Each instance consists of a feature vector \( x \in \mathbb{R}^{30} \) and a class label \( y \in \{0, 1\} \), with a class imbalance of 90:10. The 30-dimensional feature space includes 20 informative variables that are statistically dependent on the label, 5 redundant variables formed as linear combinations of the informative ones, 2 repeated copies, and 3 pure noise variables drawn independently from standard normal distributions.

Class-conditional informative features are sampled as \( x_{y=0}^{(\text{inf})} \sim \mathcal{N}(\mu_0, I) \) and \( x_{y=1}^{(\text{inf})} \sim \mathcal{N}(\mu_1, I) \), where \( \|\mu_1 - \mu_0\| = 1.5 \cdot \sqrt{20} \approx 6.7 \). This structure ensures a moderate separation between the classes while introducing dependencies and noise that resemble real-world data. The full dataset contains one million samples and is standardized to zero mean and unit variance prior to sampling.

This experimental design provides a controlled setting for assessing how well SRS and PCA-QS retain the statistical and geometric structure of the original data under various sampling configurations.

\paragraph{Quantitative Comparison of Covariance and Distributional Distances}

We compare the performance of \textit{PCA-based Quantile Sampling (PCA-QS)} and \textit{Simple Random Sampling (SRS)} across different numbers of retained principal components: 3, 5, 10, and a dynamic choice based on preserving 70\% of cumulative variance. Each sampling method produces a reduced dataset, which is evaluated against the original data using similarity metrics such as covariance fidelity, KL divergence, and Wasserstein distance. Results are averaged over 1,000 replications for each configuration to ensure statistical stability.

\paragraph{Evaluation}

For PCA-QS, the principal component projections are computed using \textbf{Truncated SVD}, a scalable PCA variant well-suited for high-dimensional or sparse data. We consider both fixed numbers of principal components (3, 5, and 10) and a dynamic configuration where the number of components is chosen to explain at least 70\% of total variance. Each retained component is divided into $Q = 10$ quantile bins, and stratified sampling is performed across these bins to preserve the structure of the projected space.

To assess how well the sampled subsets approximate the full dataset, we employ a set of statistical and geometric distance metrics. These include:  
(i) the {\it Jensen-Shannon (JS) divergence}, defined as $\text{JS}(P \| Q) = [\text{KL}(P \| M) + \text{KL}(Q \| M)]/2$, where $M = (P + Q)/2$;  
(ii) the {\it Kullback-Leibler (KL) divergence}, $\text{KL}(P \| Q) = \sum_i P(i) \log \frac{P(i)}{Q(i)}$;  
(iii) the {\it Energy Distance}, given by $D_E(X, Y) = 2\mathbb{E}\|X - Y\| - \mathbb{E}\|X - X'\| - \mathbb{E}\|Y - Y'\|$;  
(iv) the {\it Maximum Mean Discrepancy (MMD)}, where $\text{MMD}^2 = \mathbb{E}[k(x,x')] + \mathbb{E}[k(y,y')] - 2\mathbb{E}[k(x,y)]$ with an RBF kernel $k(x,y) = \exp(-\gamma \|x - y\|^2)$ with $\gamma=1.0$;  
(v) the {\it Mahalanobis Distance}, computed as $D_M(x, \mu, \Sigma) = \sqrt{(x - \mu)^T \Sigma^{-1} (x - \mu)}$; and  
(vi) a {\it Pairwise Distance Difference} metric, defined as the average absolute difference between all pairwise distances in the original and sampled datasets:  
\[
\frac{1}{n(n-1)} \sum_{i \neq j} \left| D^{\text{orig}}_{ij} - D^{\text{proj}}_{ij} \right|,
\]
where $D^{\text{orig}}_{ij}$ and $D^{\text{proj}}_{ij}$ denote the pairwise distances between samples $i$ and $j$ in the full and sampled datasets, respectively.
These complementary metrics offer a comprehensive evaluation of how well SRS and PCA-QS preserve both distributional and geometric properties of the original data.

\begin{figure}[htbp]
    \centering

    \begin{subfigure}[b]{0.45\textwidth}
        \includegraphics[width=\linewidth]{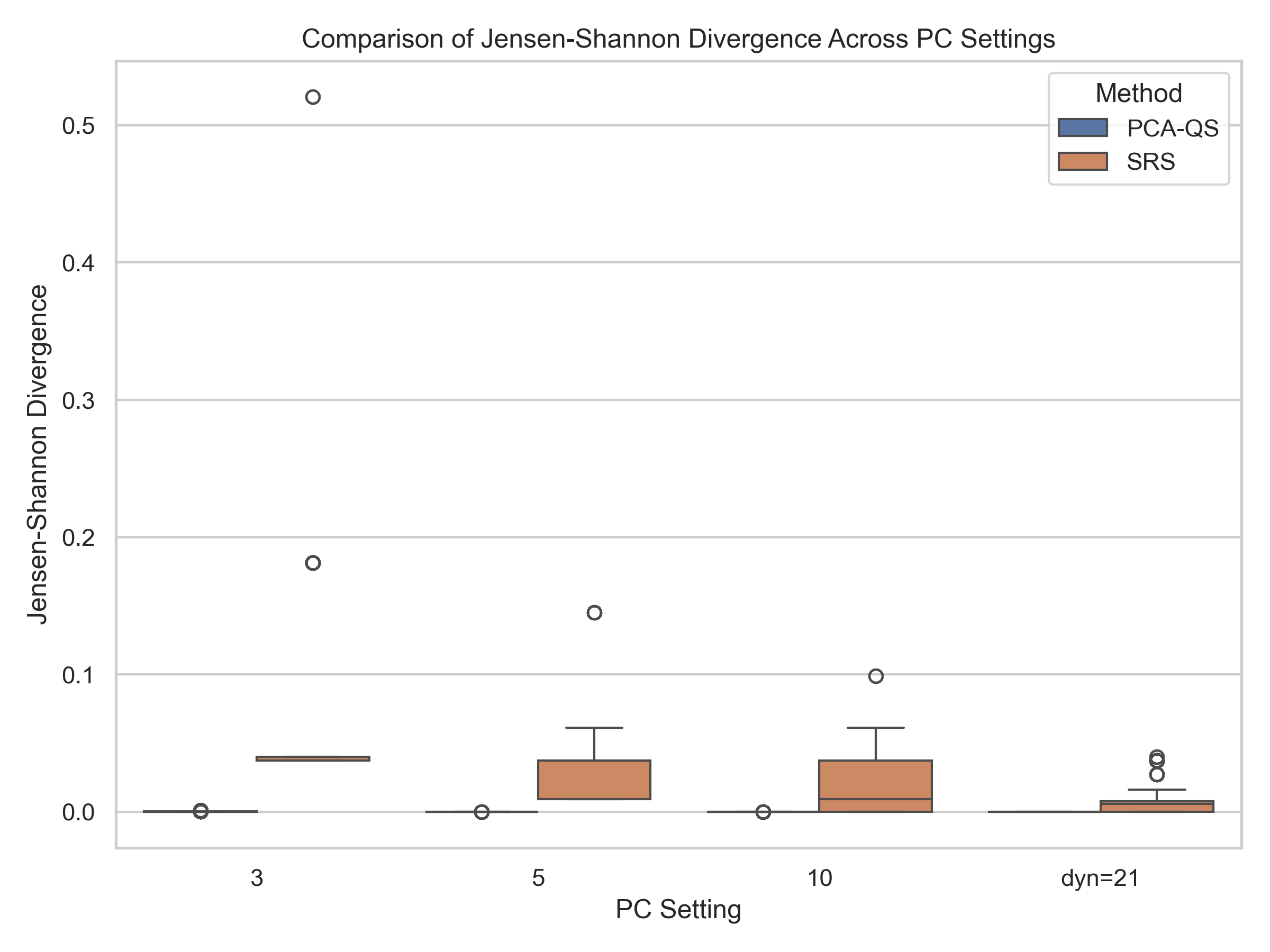}
        \caption{Jensen-Shannon Divergence}\label{JS}
    \end{subfigure}
    \hfill
    \begin{subfigure}[b]{0.45\textwidth}
        \includegraphics[width=\linewidth]{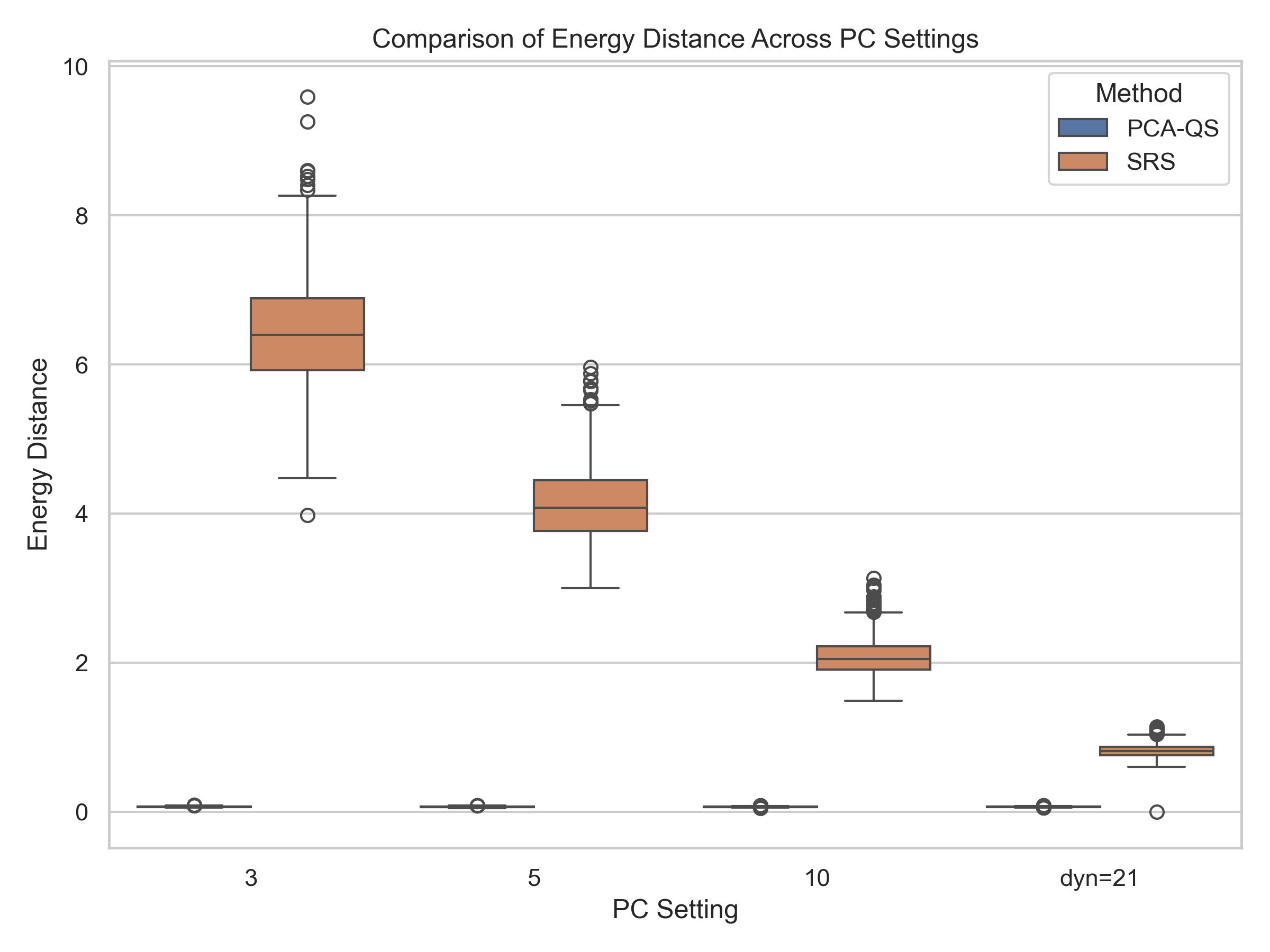}
        \caption{Energy Distance}\label{ED}
    \end{subfigure}

    \vspace{0.5cm}

    \begin{subfigure}[b]{0.45\textwidth}
        \includegraphics[width=\linewidth]{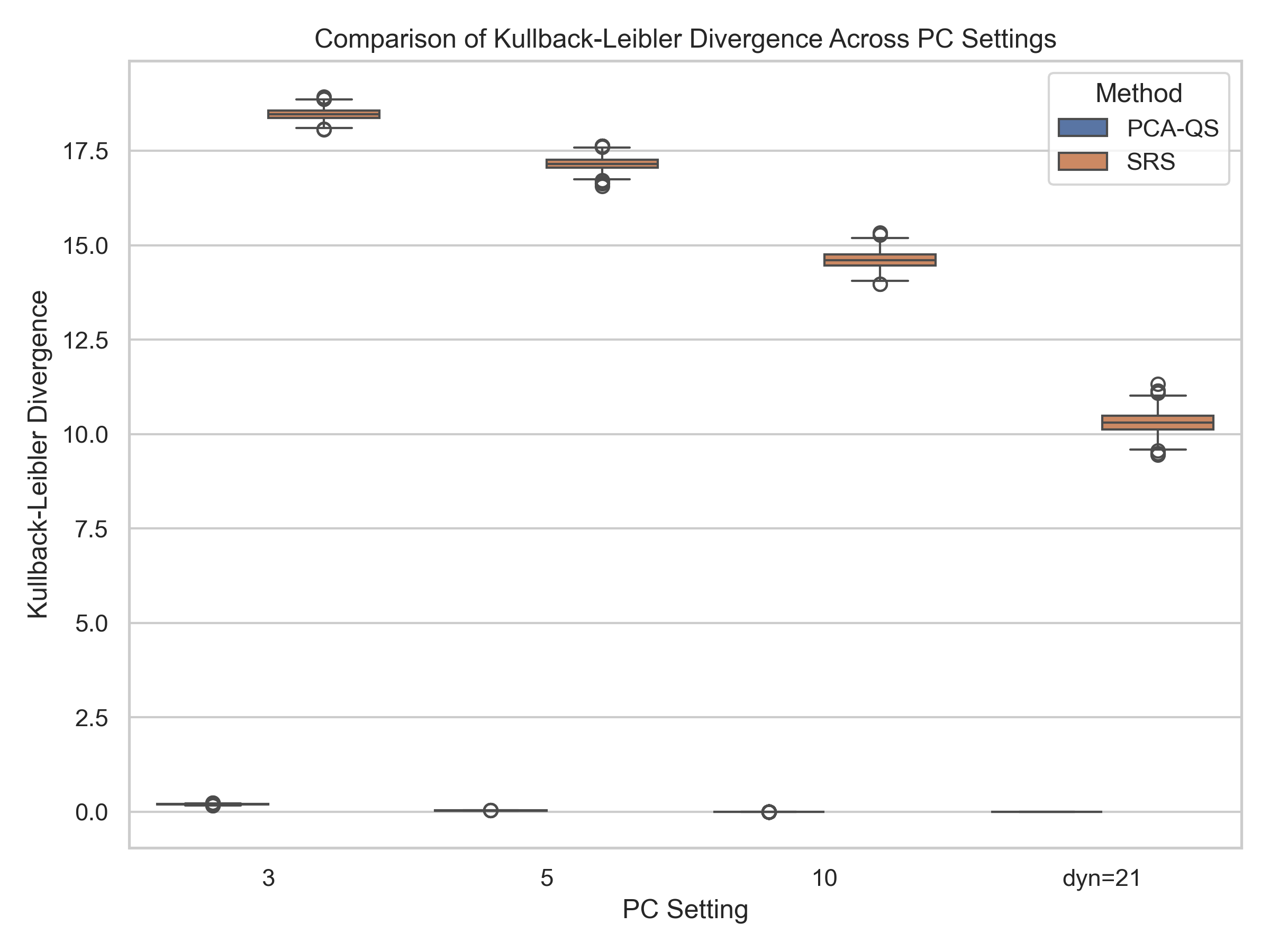}
        \caption{Kullback-Leibler Divergence}\label{KL}
    \end{subfigure}
    \hfill
    \begin{subfigure}[b]{0.45\textwidth}
        \includegraphics[width=\linewidth]{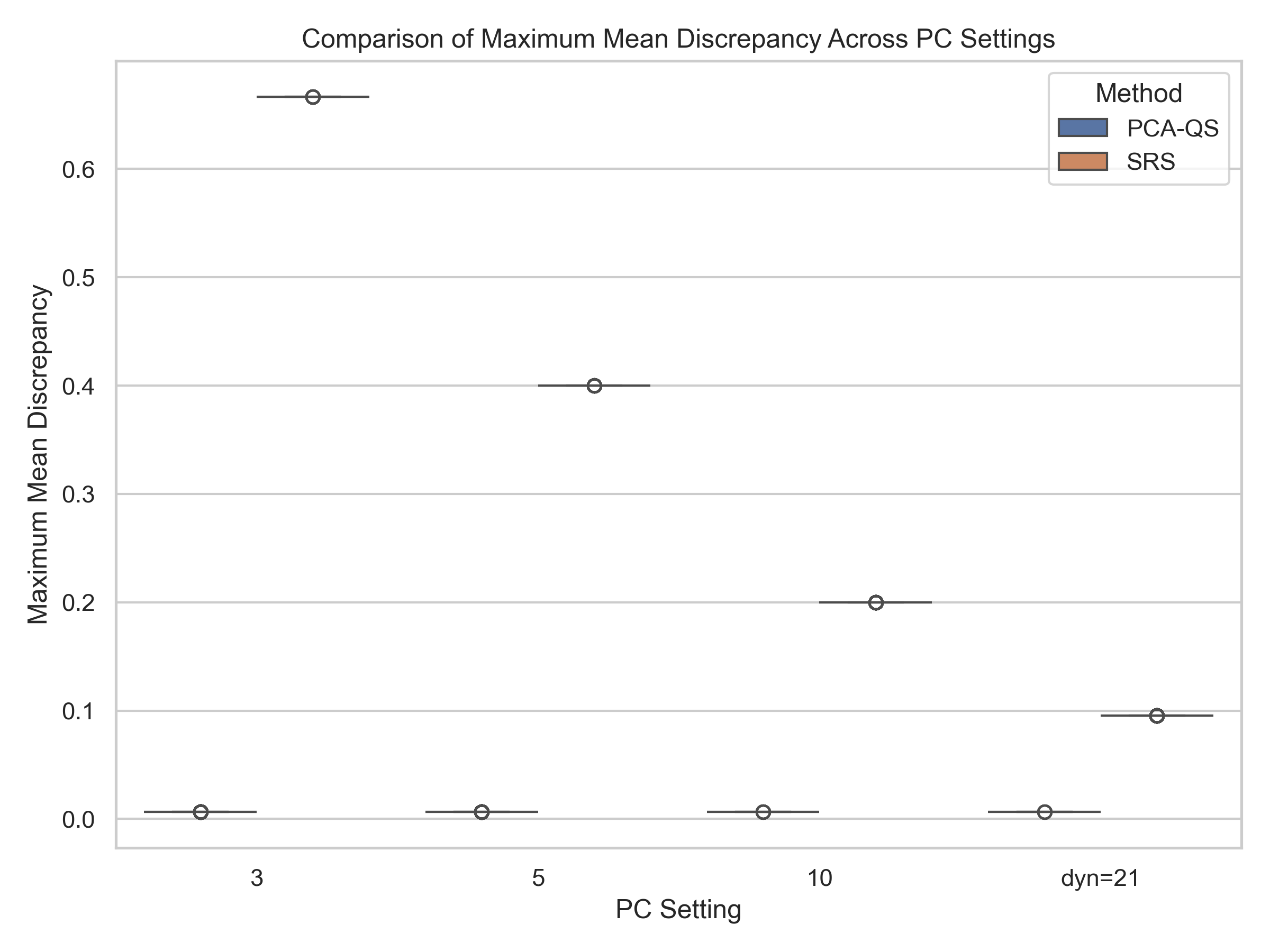}
        \caption{Maximum Mean Discrepancy}\label{MMD}
    \end{subfigure}

    \vspace{0.5cm}

    \begin{subfigure}[b]{0.45\textwidth}
        \includegraphics[width=\linewidth]{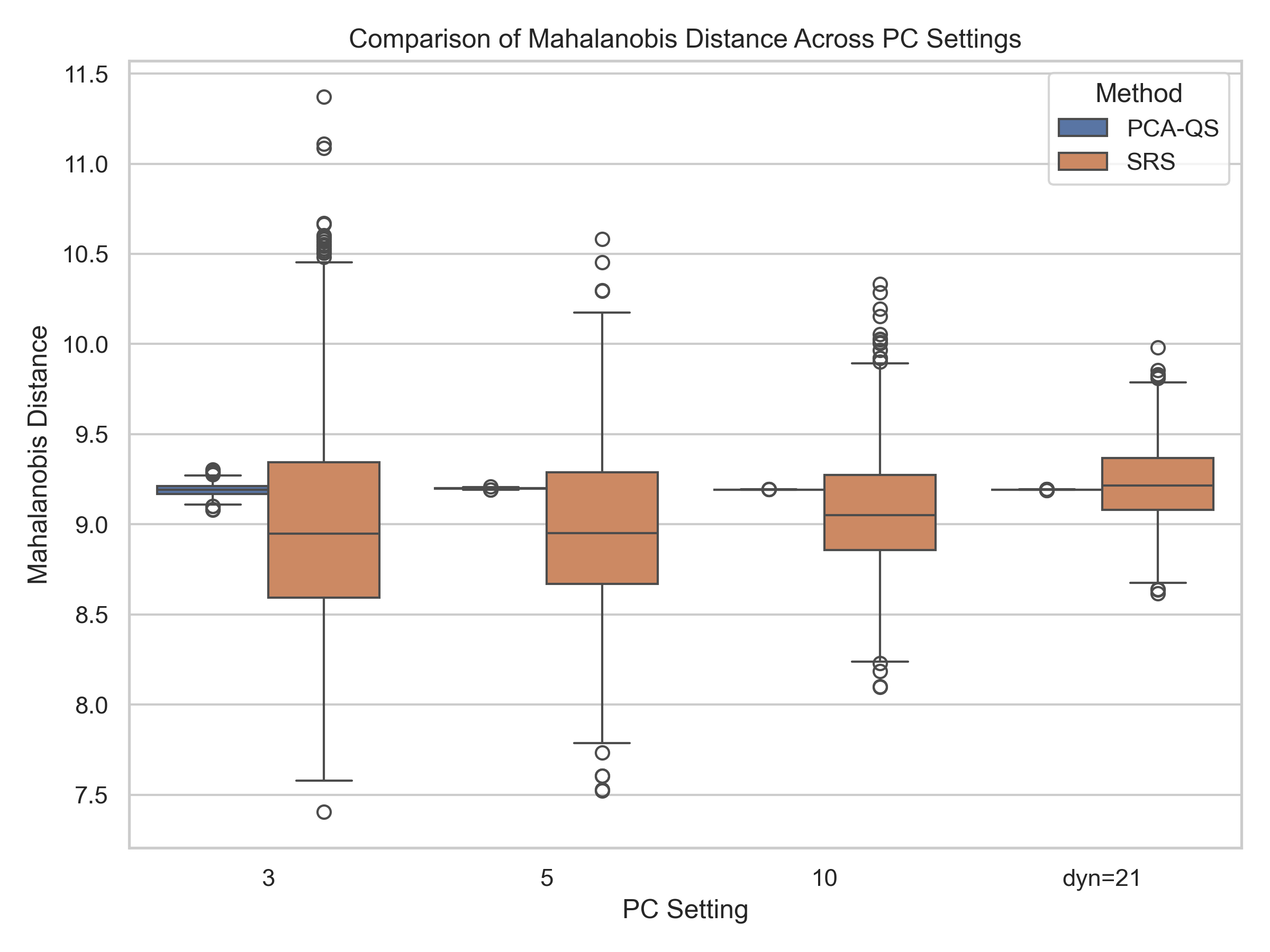}
        \caption{Mahalanobis Distance}\label{MD}
    \end{subfigure}
    \hfill
    \begin{subfigure}[b]{0.45\textwidth}
        \includegraphics[width=\linewidth]{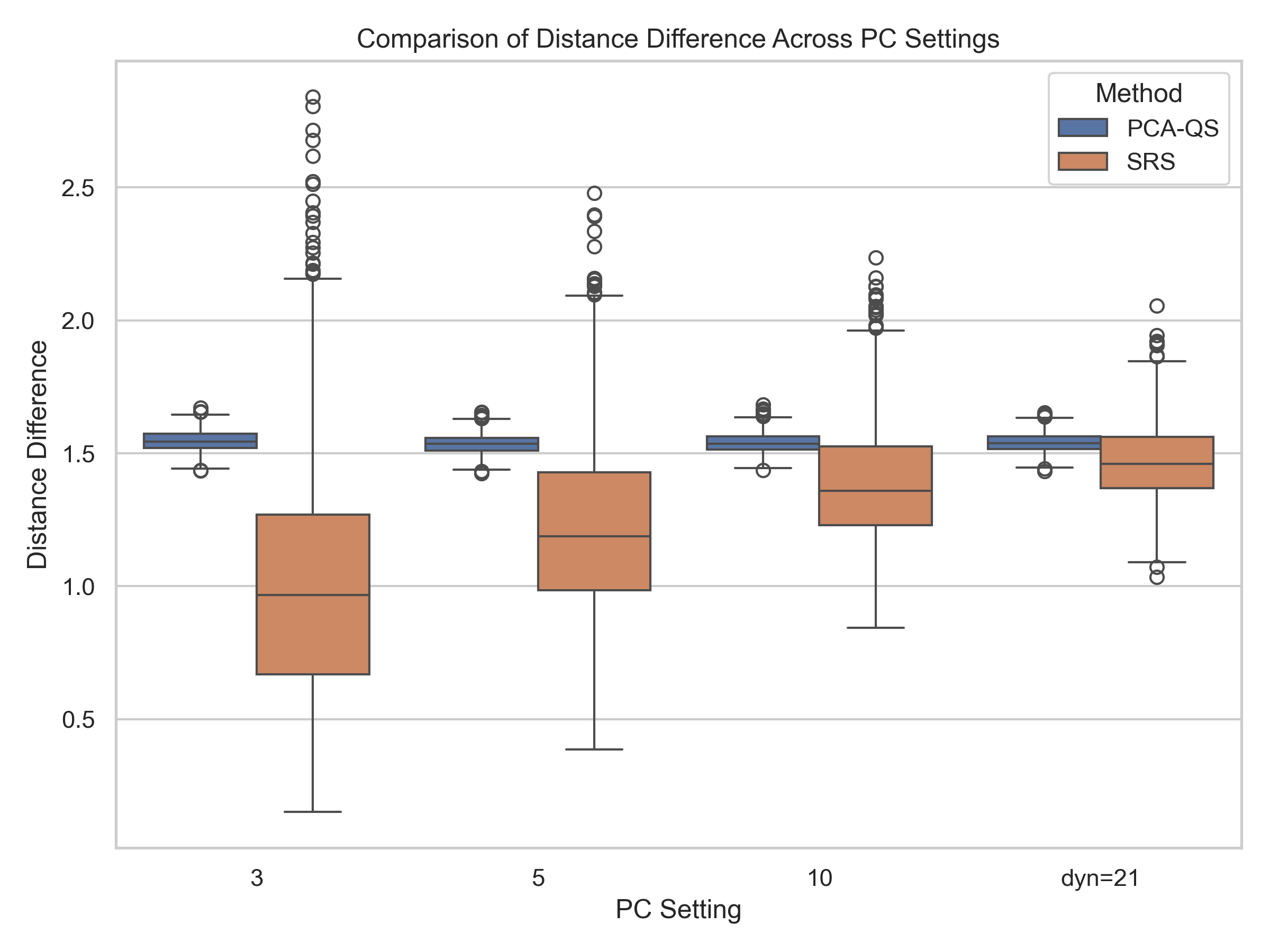}
        \caption{Distance Difference}\label{DD}
    \end{subfigure} 
    \caption{Comparison of PCA-QS and SRS across PC Settings using various statistical similarity metrics.}
    \label{fig:metric_comparison}
\end{figure}

Figure \ref{fig:metric_comparison} presents box-plots of each metrics for PCA-QS and SRS.
Each box-plot provides a visual distribution of metric values over 1000 runs, stratified by the setting of PC's.
Figures demonstrate clear, consistent trends:
For every metric, PCA-QS has narrower  IQRs (inter-quartile range) and lower medians, visually reinforcing its statistical robustness.
SRS results are more dispersed, particularly at low PC settings, confirming its sensitivity and stochastic inconsistency. 

Figure \ref{JS} shows the Jensen-Shannon Divergence (JS Divergence), which
measures the similarity between label distributions in the retained datasets.
PCA-QS showed near-zero values across all PC settings, indicating high preservation of label distribution.
SRS had significantly higher JS divergence, reflecting inconsistency and loss in label similarity.
Figure \ref{ED} presents box-plots for energy distance -- 
a metric based on the statistical distribution of distances.
PCA-QS had consistently lower values, while SRS showed large fluctuations and higher energy distances, indicating less structural preservation.
Figure \ref{KL} is Kullback-Leibler divergence (KL Divergence), which is a commonly used
 probabilistic measure of divergence between distributions.
KL divergence for PCA-QS remained close to zero, suggesting minimal distortion.
SRS values were significantly large, implying substantial information loss.
Maximum mean discrepancy (MMD) quantifies distributional similarity in terms of means of feature mappings.
In Figure \ref{MMD}, we see that
PCA-QS yielded extremely low MMD values, demonstrating almost no change in underlying data distributions.
SRS was again consistently higher and fixed, suggesting less robustness in selection.
 We use Mahalanobis distance to
evaluate the distance between distributions considering correlations.
Figure \ref{MD} shows that PCA-QS had more consistent Mahalanobis distances, whereas SRS exhibited more variability.
Figure \ref{DD} is distance difference is used to
assess average pairwise distance differences between original and retained datasets.
PCA-QS again performed better, maintaining closer distance profiles to the original data.
SRS showed greater variance, especially for low PC settings like 3 and 5.

Although PCA-QS outperforms SRS in distributional and divergence-based metrics (JS, KL, MMD, Energy), it incurs slightly higher average distance distortion. This suggests a trade-off: PCA-QS better preserves overall structure and class distributions, while SRS may retain local distances more faithfully.

\begin{table}[h!]
\caption{Comparison of PCA-QS and SRS across PC Settings using various statistical similarity metrics.}
\begin{tabular}{llrrrr}
\hline
 PCs   & Metric                      &  PCA-QS          & SRS              &      p-value \\
\hline
 3            & Jensen-Shannon    & 0.0001 (0.0002) & 0.0449 (0.0376)  & 1.02e-193 \\
              & Distance Difference         & 1.5462 (0.0382) & 1.0152 (0.4601)  & 1.68e-185 \\
             & Energy Distance             & 0.0684 (0.0057) & 6.4367 (0.7005)  & 0            \\
              & Kullback-Leibler  & 0.1972 (0.0111) & 18.4693 (0.1427) & 0            \\
              & Maximum Mean     & 0.0067 (0.0000) & 0.6667 (0.0000)  & 0            \\
              & Mahalanobis Distance        & 9.1905 (0.0317) & 8.9954 (0.5860)  & 1.11e-24  \\
\midrule
 5            & Jensen-Shannon   & 0.0000 (0.0000) & 0.0308 (0.0203)  & 2.46e-261 \\
              & Distance Difference         & 1.5349 (0.0370) & 1.2217 (0.3462)  & 9.93e-132 \\
              & Energy Distance             & 0.0660 (0.0056) & 4.1291 (0.5101)  & 0            \\
              & Kullback-Leibler  & 0.0332 (0.0040) & 17.1552 (0.1638) & 0            \\
              & Maximum Mean     & 0.0067 (0.0000) & 0.4000 (0.0000)  & 0            \\
              & Mahalanobis Distance        & 9.1997 (0.0029) & 8.9671 (0.4555)  & 2.76e-52  \\  
\midrule
 10           & Jensen-Shannon    & 0.0000 (0.0000) & 0.0187 (0.0186)  & 1.16e-154  \\
            & Distance Difference         & 1.5395 (0.0357) & 1.3877 (0.2359)  & 1.66e-75  \\
            & Energy Distance             & 0.0656 (0.0049) & 2.0787 (0.2465)  & 0            \\
             & Kullback-Leibler e & 0.0000 (0.0000) & 14.6066 (0.2142) & 0            \\
             & Maximum Mean     & 0.0067 (0.0000) & 0.2000 (0.0000)  & 0            \\
             & Mahalanobis Distance        & 9.1922 (0.0006) & 9.0769 (0.3227)  & 6.15e-28  \\
\midrule
 dyn=21       & Jensen-Shannon    & 0.0000 (0.0000) & 0.0074 (0.0108)  & 3.13e-84  \\
        & Distance Difference         & 1.5406 (0.0367) & 1.4679 (0.1441)  & 3.76e-49  \\
        & Energy Distance             & 0.0685 (0.0045) & 0.8208 (0.0873)  & 0            \\
        & Kullback-Leibler  & 0.0000 (0.0000) & 10.3052 (0.2723) & 0            \\
        & Maximum Mean     & 0.0067 (0.0000) & 0.0952 (0.0000)  & 0            \\
        & Mahalanobis Distance        & 9.1917 (0.0005) & 9.2239 (0.2181)  & 3.42e-06  \\
\hline
\end{tabular}
\end{table}

\begin{remark}
PCA-QS actually shows higher Distance Difference values than SRS in all PC settings.
This contradicts the general trend observed in other metrics, where QS was consistently better.
It is because while PCA-QS preserves global structure and distributions (as shown by low JS, KL, MMD), it may distort pairwise distances slightly more than SRS.
This could stem from the dimensional compression in PCA, which sacrifices some fine-grained inter-point relationships to retain global variance.
SRS retains actual points, hence it may better preserve exact distances — even though it loses higher-level structure.
\end{remark}

\subsubsection{Logistic Regression with Synthesized Data}

This experiment evaluates the effectiveness of PCA-based Quantile Sampling (PCA-QS) versus Simple Random Sampling (SRS) in preserving predictive structure for supervised learning. We use logistic regression on a synthetic dataset designed to exhibit complex feature interactions and imbalanced class distributions.

\paragraph{Data Generation}
We simulate a binary classification task using a two-component Gaussian Mixture Model (GMM). The input matrix \( X \in \mathbb{R}^{n \times d} \) contains \( n = 100{,}000 \) samples and \( d = 50 \) features, with class labels \( y \in \{0, 1\}^n \) drawn from a Bernoulli distribution with class priors \( P(y = 0) = 0.9 \), \( P(y = 1) = 0.1 \). The class-conditional distributions are:
\[
X \mid y = 0 \sim \mathcal{N}(\mathbf{0}, \mathbf{I}), \quad
X \mid y = 1 \sim \mathcal{N}(0.5 \cdot \mathbf{1}, 1.2 \cdot \mathbf{I}),
\]
where \( \mathbf{I} \in \mathbb{R}^{50 \times 50} \) is the identity matrix.

To emulate nonlinear and interaction effects, we augment each feature vector with:
(1) All pairwise interaction terms \( x_i x_j \), for \( i < j \), yielding \( \binom{50}{2} = 1225 \) features, and
(2) Sine-transformed features \( \sin(x_i) \), adding 50 features.
This results in a final dimensionality of \( d_{\text{final}} = 1325 \).

\paragraph{Sampling Strategy}
From the full dataset, a subset of 5000 samples is drawn using either PCA-QS or SRS. PCA-QS involves projecting data onto the top principal components that retain at least 70\% of total variance, followed by stratified sampling across 10 quantile bins formed from the PCA scores. To reduce sampling noise, nearest-neighbor smoothing with \( k = 5 \) is optionally applied during bin assignment. A test set of 1000 independent samples is reserved for model evaluation.
All experiments use fixed random seeds for reproducibility.

\paragraph{Sampling and Classification Strategy:}
For SRS, training samples are drawn uniformly at random from the full dataset. For  PCA-QS, the procedure involves three steps: (1) applying PCA to the normalized data matrix, (2) discretizing the top principal components into quantile bins, and (3) performing stratified sampling across these composite quantile groups to ensure proportional coverage in the reduced PCA space.

To assess predictive performance, we apply five widely used classifiers to the retentive subsets obtained via PCA-QS and SRS:
\texttt{logistic regression}, \texttt{k-nearest neighbors (kNN)}, \texttt{random forest (RF)} , \texttt{support vector machine (SVM)}, and \texttt{Extreme Gradient Boosting (XGBoost)}. Each model is trained on a sample of 5,000 observations and evaluated on a held-out test set of 1,000 observations.
Classification performance is measured using multiple evaluation metrics: accuracy, area under the ROC curve (AUC), F1 score, true positive rate (TPR), false positive rate (FPR), true negative rate (TNR), and false negative rate (FNR). These metrics offer complementary perspectives on classification effectiveness under balanced and imbalanced class conditions.

\begin{figure}[th!]
\centering
\includegraphics[width=0.95\textwidth]{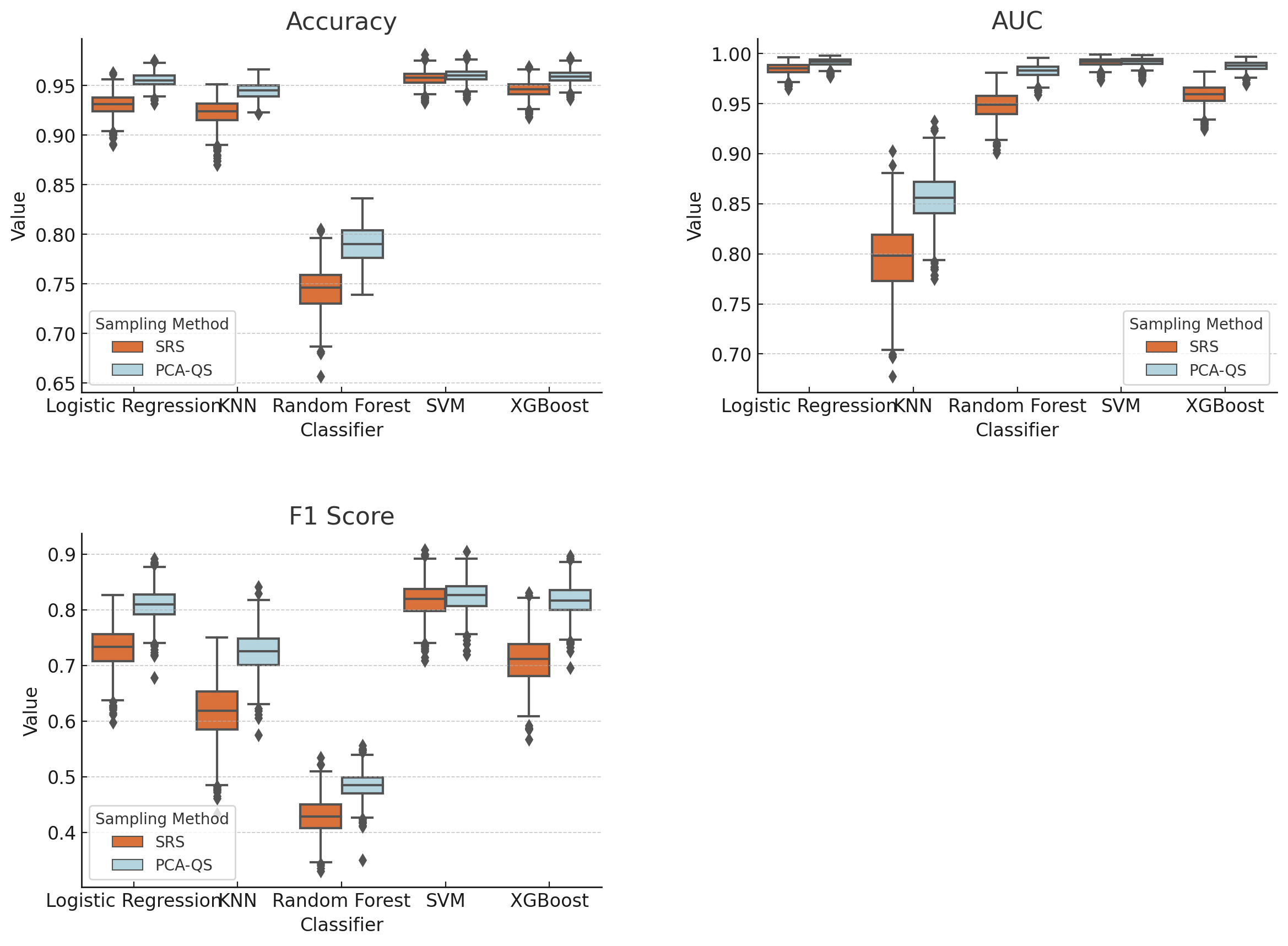}
\caption{Visual comparison of Accuracy, AUC, and F1 Score across classifiers (Logistic Regression, KNN, Random Forest, SVM, XGBoost). PCA-QS typically yields higher or more stable results, particularly for AUC and F1 metrics.}\label{fig: ACC_AUC_F1}
\end{figure}

\begin{figure}[h!]
\centering
\includegraphics[width=0.95\textwidth]{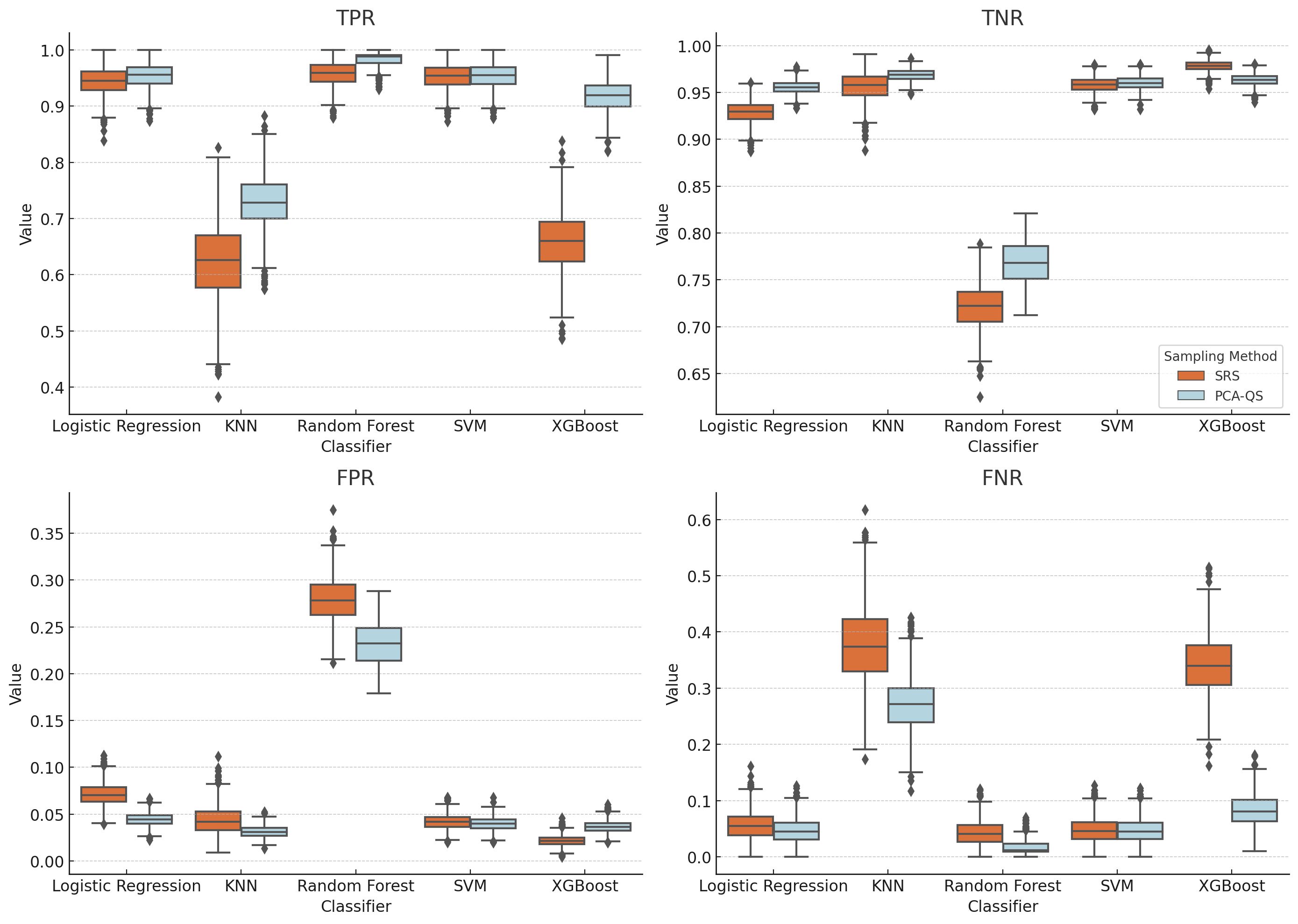}
\caption{Comparison of TPR, TNR, FPR, and FNR distributions. PCA-QS improves sensitivity and specificity across classifiers, reducing false positive and false negative rates.}\label{fig:tpr_tne-fpr_fnr}
\end{figure}

The following box-plots provide some visual Comparison.
Figures \ref{fig: ACC_AUC_F1} and \ref{fig:tpr_tne-fpr_fnr} show the box-plots of commonly used classification metrics -- accuracy, AUC, F1m true positive rate (TPR), true negative rate (TNR), false positive rate (FPR) and false negative rate (FNR).  From these box-plots, we can see that the methods with PCA-QS data set consistently outperform the methods with SRS samples.

\paragraph{Summary of Results}

This study compares PCA-based Quantile Sampling (PCA-QS) and Simple Random Sampling (SRS) across five classifiers: logistic regression, KNN, random forest, SVM, and XGBoost, using synthetic datasets with correlated feature structures. Classification performance was evaluated using accuracy, AUC, F1-score, and confusion-matrix-based metrics, including true positive rate (TPR), true negative rate (TNR), false positive rate (FPR), and false negative rate (FNR).

Results show that PCA-QS consistently outperforms SRS across most classifiers and metrics. It delivers higher AUC and F1 scores, especially for KNN, random forest, and XGBoost. Additionally, PCA-QS achieves better TPR and TNR, and lower FPR and FNR. Standard deviations across replications are also smaller, highlighting its stability. These improvements are statistically significant, with most p-values below 0.001.

The effectiveness of PCA-QS comes from its stratified sampling in the PCA-transformed space. Unlike the uniform randomness of SRS, PCA-QS ensures representation across the principal components, capturing structural variation more reliably. This helps improve both generalization and representativeness, especially in high-dimensional or heterogeneous data scenarios.

In summary, PCA-QS is a robust and efficient subsampling technique that preserves essential data characteristics and enhances classification performance, making it a strong alternative to traditional random sampling.  We also include a python code for logistic regression with synthesized data in Supplementary material 

\begin{remark}
The choice of PCA-QS parameters should reflect the goal of the application. If computational efficiency is the main concern, a higher deduction rate of 2--5\%, a smaller number of principal components (around 6 to 10), and 10 to 15 quantile groups are often sufficient. For statistical representativeness, more variance should be retained---typically 85\% to 95\%---by selecting 12 to 15 components and 15 to 20 quantile bins. When geometric preservation is important, such as in clustering or anomaly detection, 12 to 20 components and 20 to 30 quantiles can better maintain local structure. If interpretability is the goal, fewer components (around 6 to 10) and a slightly lower variance threshold (80\% to 85\%) keep the reduced data more understandable. In large-scale or real-time applications, a moderate setup---such as 8 to 12 components, 10 to 15 quantiles, and a 2--5\% deduction rate---offers a balanced trade-off between efficiency and accuracy.

Overall, PCA-QS is adaptable and should be configured based on the analysis objectives. Its flexibility allows it to be tailored to different domains while maintaining both computational and statistical integrity.
\end{remark}

\subsection{Recommended Settings and Consistency}

Our empirical results align closely with the theoretical convergence guarantees of PCA-QS. Across diverse datasets and tasks, PCA-QS consistently achieves lower distributional distances (KL divergence, Energy distance, Mahalanobis distance, and MMD) than simple random sampling (SRS). The advantage is more pronounced in higher-dimensional settings, demonstrating its effectiveness in mitigating the curse of dimensionality. Empirical tests confirm that selecting 12--15 principal components captures sufficient variance and achieves strong performance, consistent with theoretical recommendations. Larger sample sizes reduce divergence at the expected rate (e.g., KL decay of $O(n^{-1})$), and using 15--20 quantile groups provides a practical balance between local structure preservation and computational cost. A deduction rate of 5--10\% reliably maintains representativeness while significantly reducing data size.

\textbf{Recommended parameter choices} based on both theory and empirical validation are as follows:
(1) Principal components ($k$): 12--15,
(2) Quantile groups ($m$): 15--20,
(3) Deduction rate ($\delta$): 5--10\%, and
(4) Minimum sample size ($n$): at least 1000.

These settings provide a practical default for practitioners seeking a good trade-off between computational efficiency and statistical fidelity. Detailed derivations and extended results are available in the supplementary appendix.

\subsection{Real Data: Thematic Comparison:Structure Preserving of Retentive Data Sets}

\subsubsection{Imbalanced Datasets}
\paragraph {CreditCard}
The UCI Credit Card dataset is a well-known dataset for binary classification tasks related to credit risk assessment. It contains data on credit card clients in Taiwan and is primarily used to predict whether a client will default on their payment in the next month. The dataset includes 30,000 instances, each representing an individual client, with 23 features capturing demographic information (such as age, gender, and education), credit history (such as past payment behavior), and billing information over six months. The target variable is binary, indicating whether the client defaulted on the next month’s payment (1) or not (0). All features are numerical, though some represent categorical concepts encoded as integers. This dataset is widely used in financial analytics, fraud detection, and machine learning research focused on imbalanced classification problems and model interpretability in credit scoring systems.

\begin{figure}[htbp]
\centering

\begin{subfigure}[b]{0.49\textwidth}
    \includegraphics[width=\textwidth]{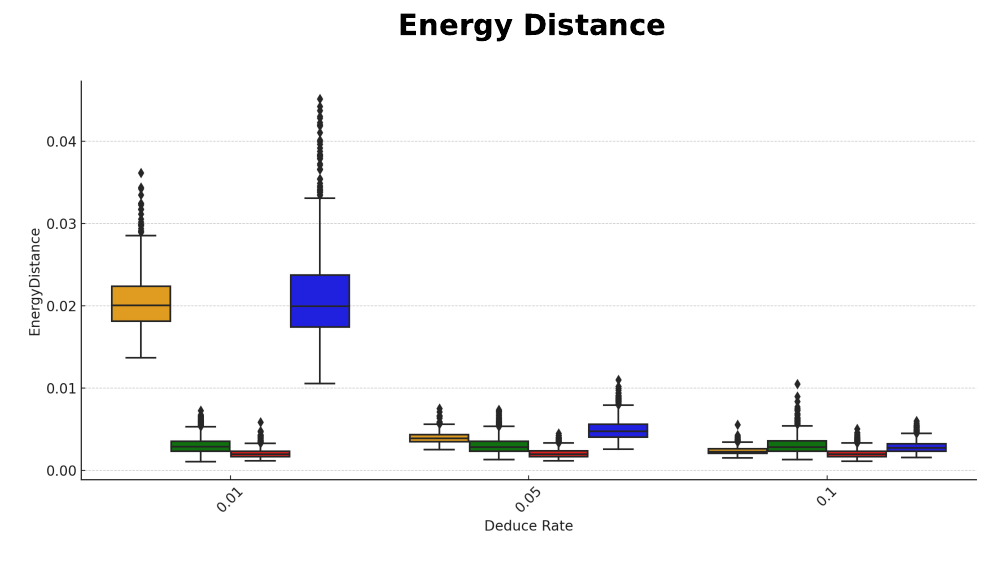}
    \caption{Energy Distance}
\end{subfigure}
\hfill
\begin{subfigure}[b]{0.49\textwidth}
    \includegraphics[width=\textwidth]{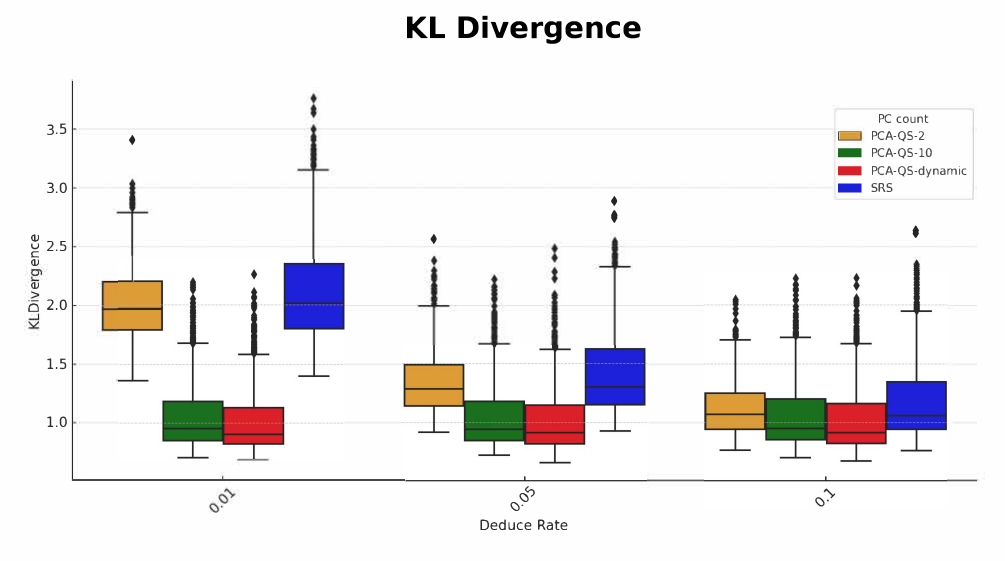}
    \caption{KL Divergence}
\end{subfigure}

\vspace{0.5cm}

\begin{subfigure}[b]{0.49\textwidth}
    \includegraphics[width=\textwidth]{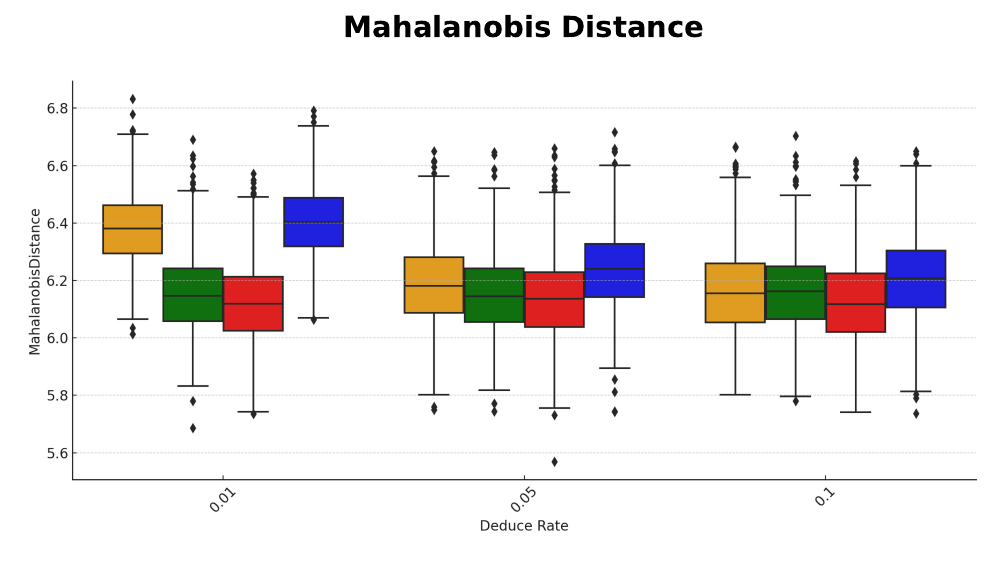}
    \caption{Mahalanobis Distance}
\end{subfigure}
\hfill
\begin{subfigure}[b]{0.49\textwidth}
    \includegraphics[width=\textwidth]{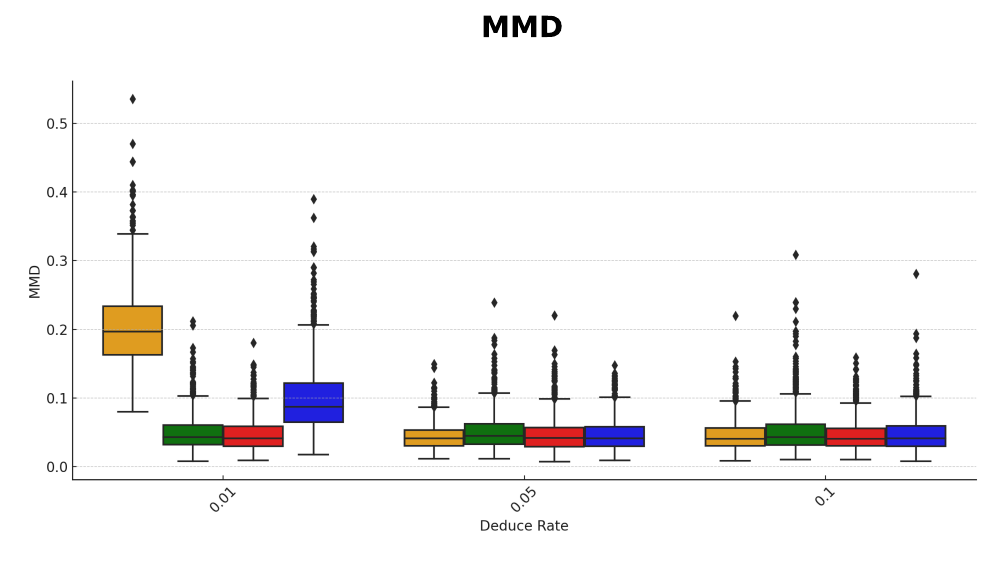}
    \caption{MMD}
\end{subfigure}

\caption{UCI Credit Card Dataset}
\end{figure}

\paragraph{Magic Gamma Telescope}
The MAGIC Gamma Telescope dataset from the UCI Machine Learning Repository is designed for binary classification tasks related to high-energy astrophysics. It simulates data from a ground-based atmospheric Cherenkov gamma telescope, aiming to distinguish between gamma rays (signal) and hadronic cosmic rays (background noise) based on the characteristics of particle showers in the atmosphere. Each instance in the dataset represents an event and includes 10 numerical features derived from the image parameters of the recorded showers, such as length, width, size, and conciseness. The target variable is categorical, indicating whether the event is a gamma ray or a hadron. The dataset contains 19,020 instances, with all features being continuous and real-valued. It is widely used for testing classification algorithms, especially in the context of imbalanced datasets, as the two classes are not evenly distributed. This dataset provides a practical scenario for machine learning applications in physics and signal detection.

\begin{figure}[htbp]
\centering

\begin{subfigure}[b]{0.49\textwidth}
    \includegraphics[width=\textwidth]{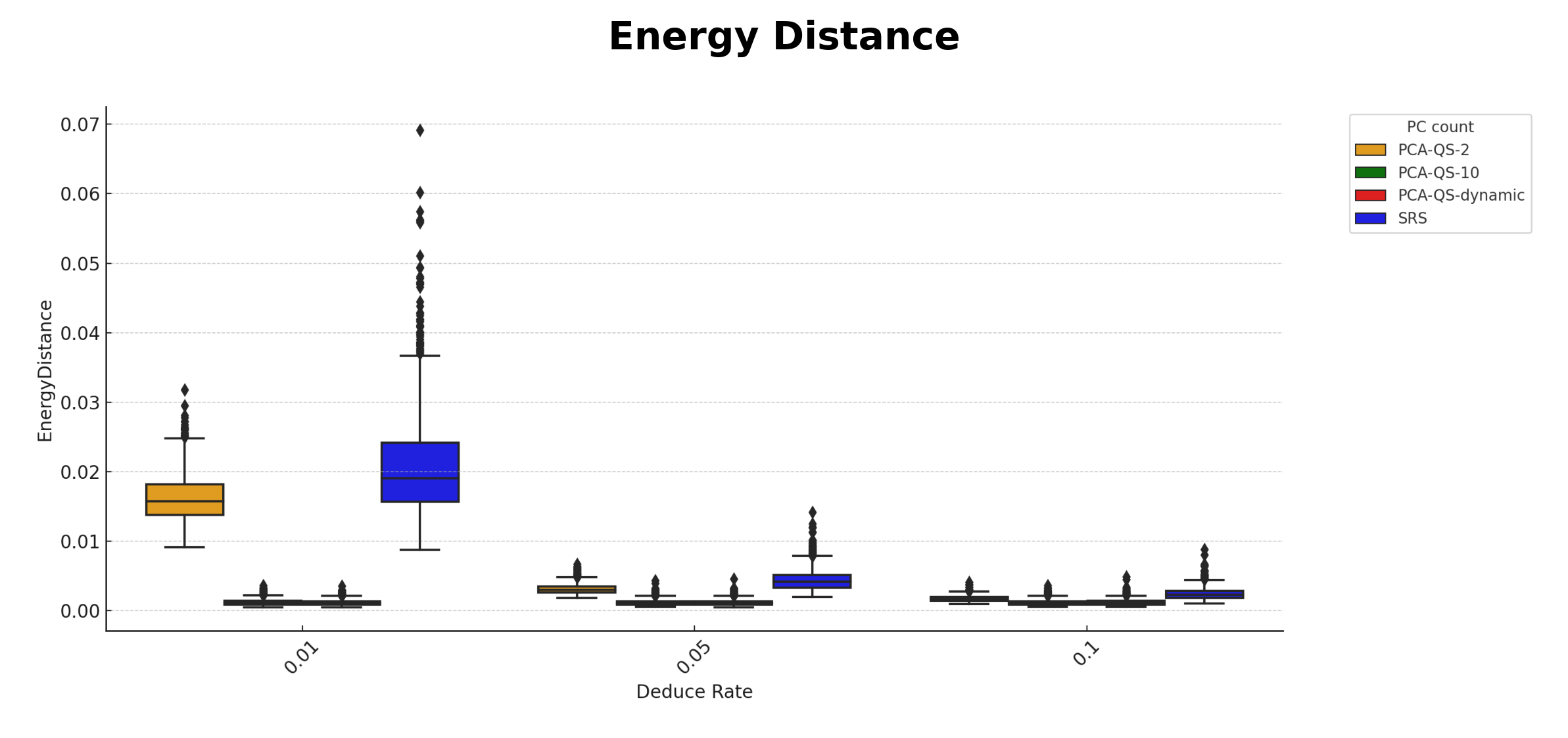}
\end{subfigure}
\hfill
\begin{subfigure}[b]{0.49\textwidth}
    \includegraphics[width=\textwidth]{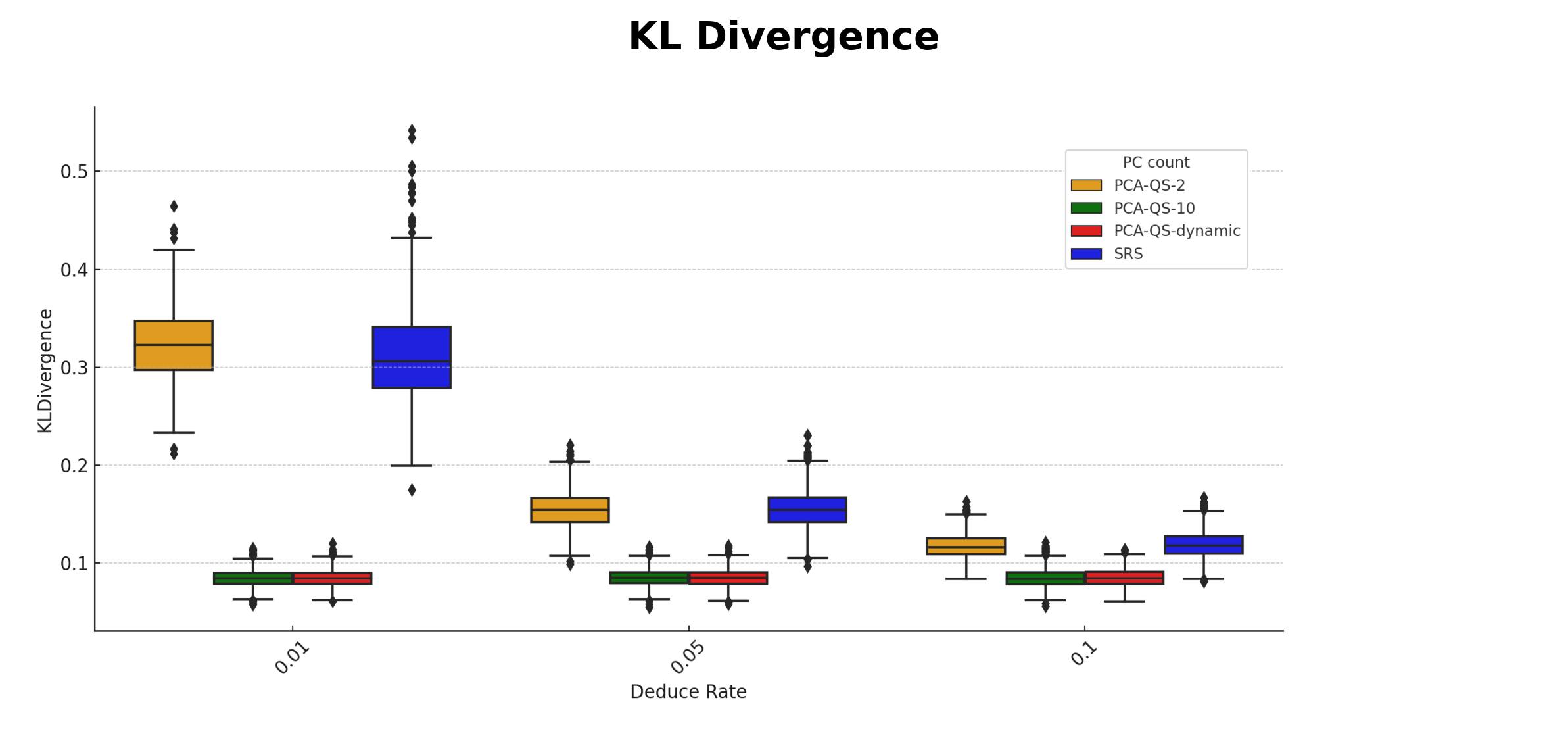}
\end{subfigure}


\begin{subfigure}[b]{0.49\textwidth}
    \includegraphics[width=\textwidth]{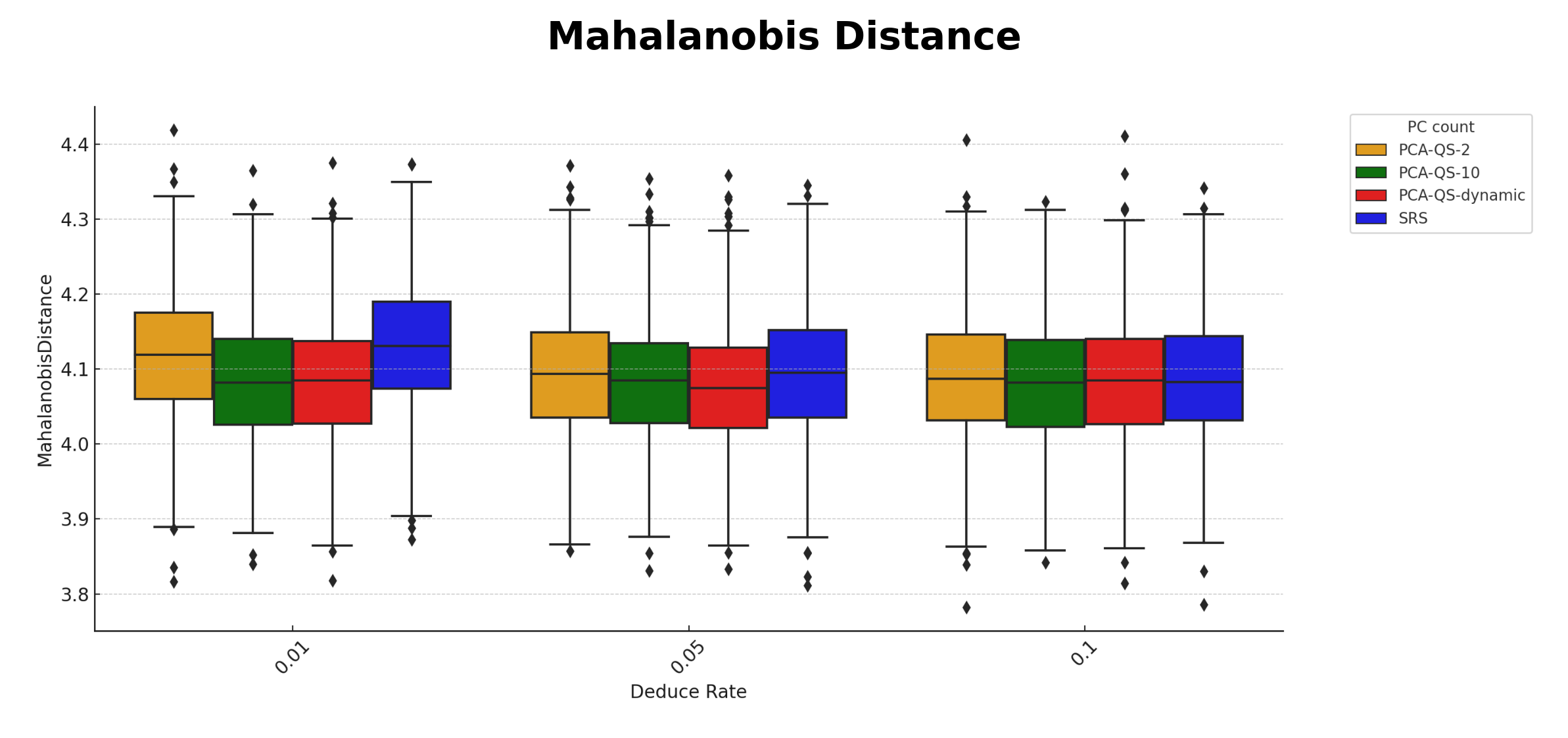}
\end{subfigure}
\hfill
\begin{subfigure}[b]{0.49\textwidth}
    \includegraphics[width=\textwidth]{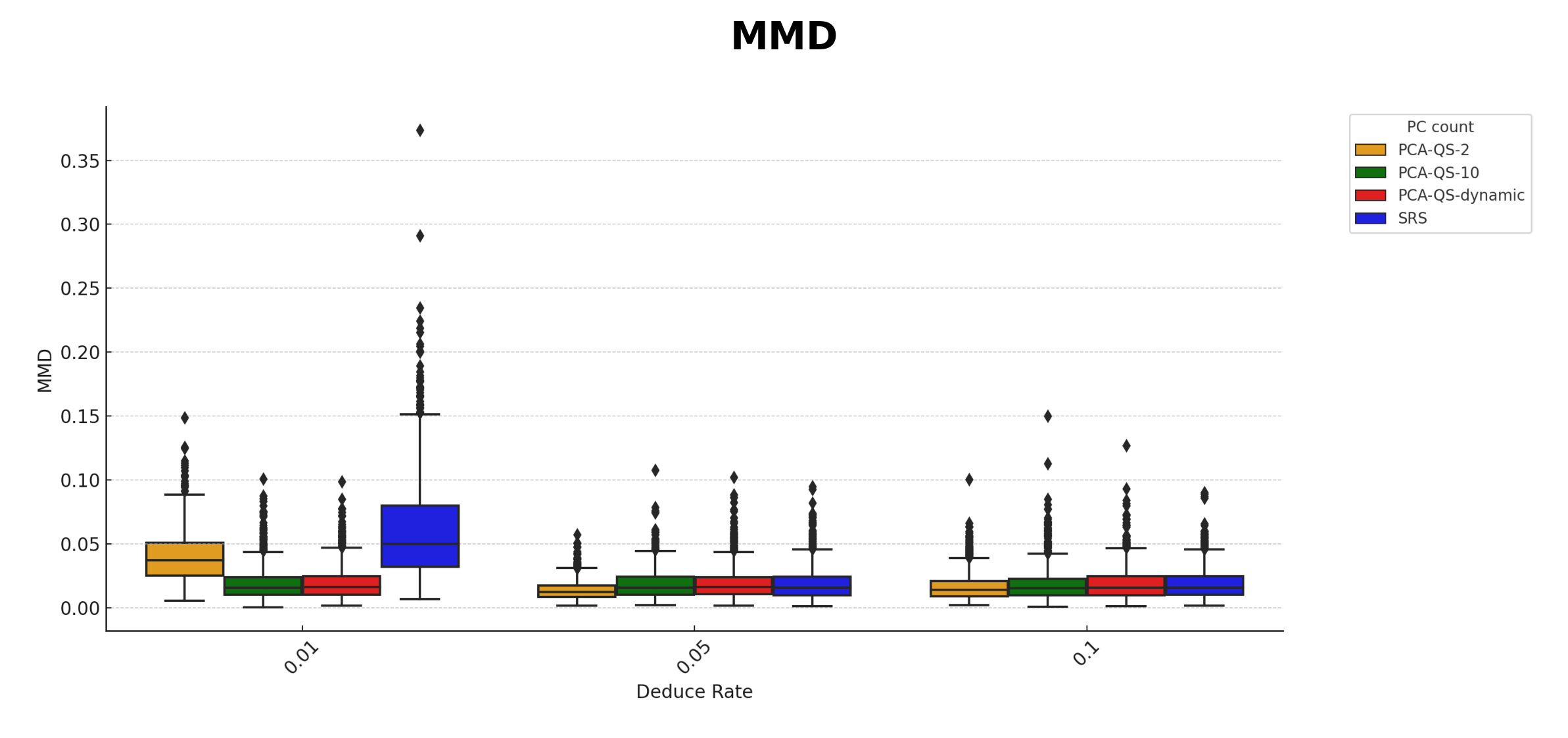}
\end{subfigure}

\caption{Distance Metrics for Two Sampling Methods with Magic Gamma Telescope Data}
\end{figure}

\subsubsection{Time-Series and Signal Data}
\paragraph{Epileptic Seizure Recognition}
The Epileptic Seizure Recognition dataset from the UCI Machine Learning Repository is designed for multiclass classification tasks related to the detection of epileptic seizures using EEG (electroencephalogram) signals. Each instance in the dataset represents a 1-second segment of EEG data, recorded using a single electrode from a patient. The data is preprocessed into 178 time-ordered measurements per segment, along with one target label. The target variable includes five classes, where class 1 represents seizure activity and classes 2 through 5 represent different non-seizure states recorded during various normal activities. The dataset contains 11,500 instances, each with 178 numerical features and one categorical label. This dataset is widely used in biomedical signal processing and machine learning applications focused on neurological disorder detection, offering a clean, high-dimensional, and balanced example for time-series classification problems.

\begin{figure}[htbp]
\centering

\begin{subfigure}[b]{0.49\textwidth}
    \includegraphics[width=\textwidth]{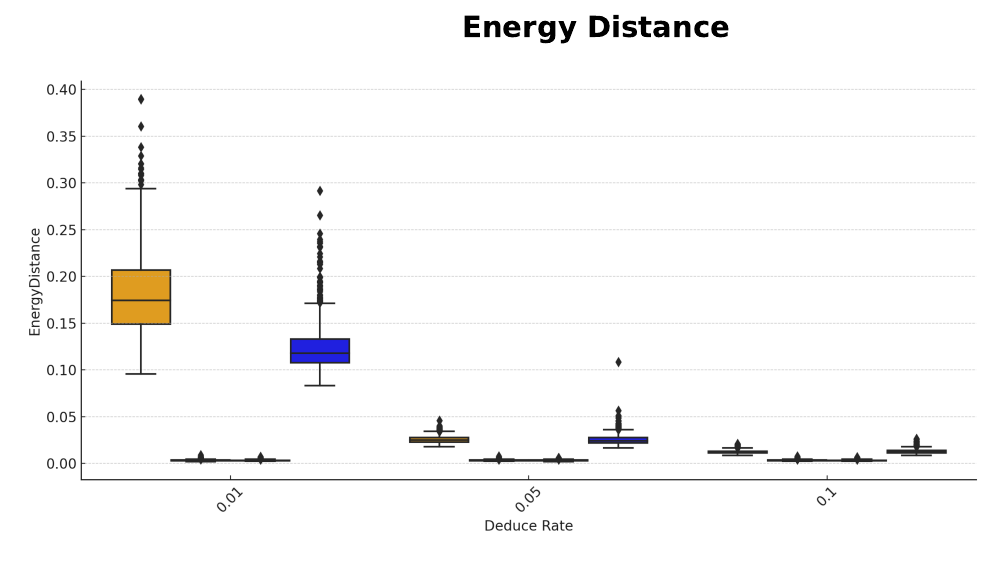}
    \caption{Energy Distance}
\end{subfigure}
\hfill
\begin{subfigure}[b]{0.49\textwidth}
    \includegraphics[width=\textwidth]{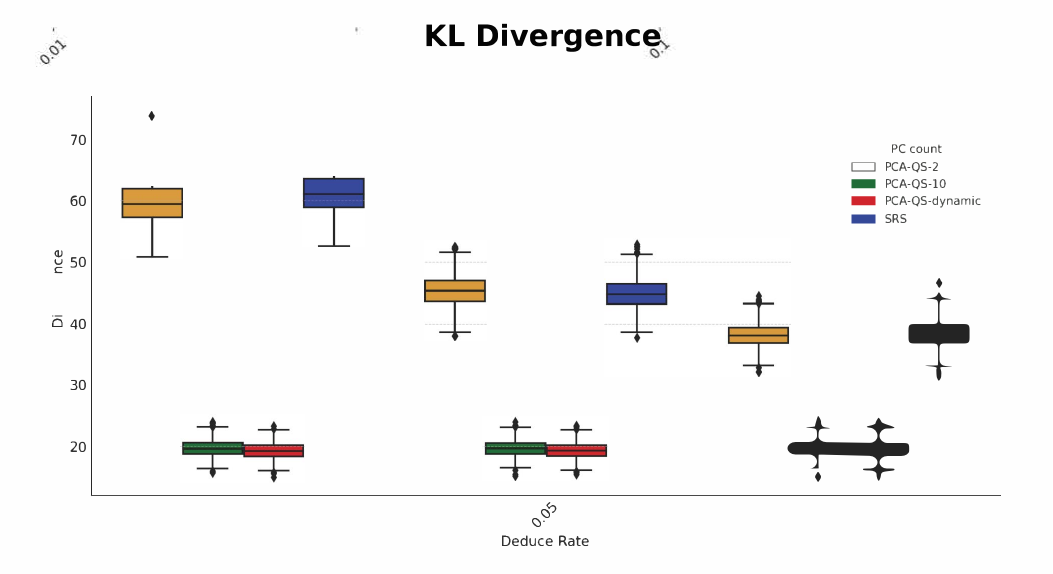}
    \caption{KL Divergence}
\end{subfigure}

\vspace{0.5cm}

\begin{subfigure}[b]{0.49\textwidth}
    \includegraphics[width=\textwidth]{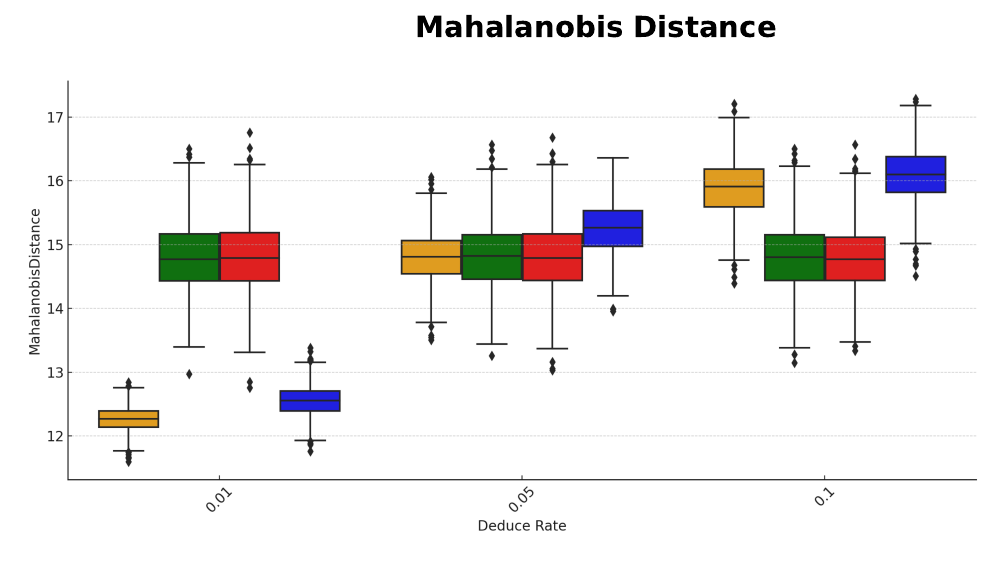}
    \caption{Mahalanobis Distance}
\end{subfigure}
\hfill
\begin{subfigure}[b]{0.49\textwidth}
    \includegraphics[width=\textwidth]{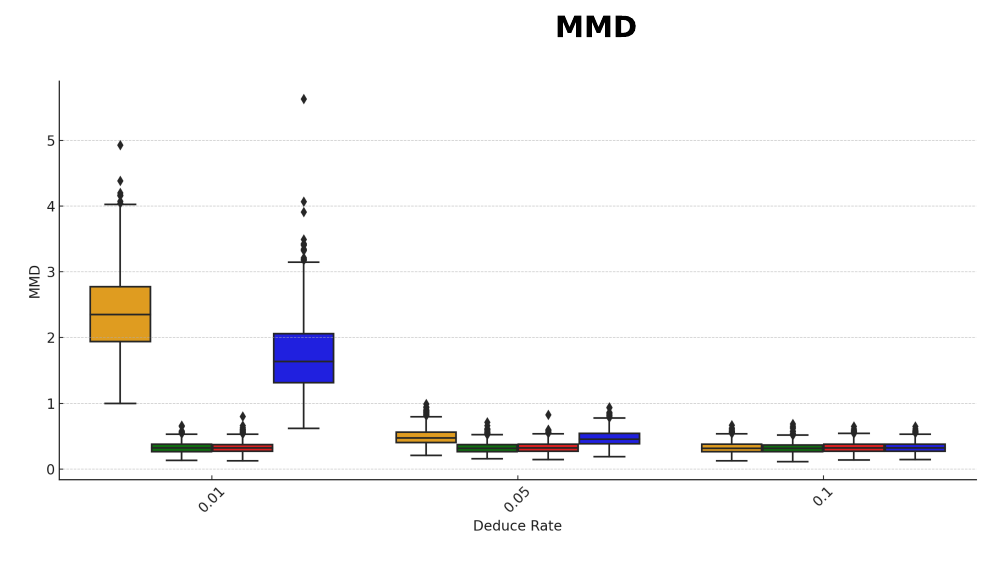}
    \caption{MMD}
\end{subfigure}

\caption{Epileptic Seizure Dataset}
\end{figure}

\paragraph{EEG Eye State}

The EEG Eye State dataset from the UCI Machine Learning Repository is used for binary classification tasks involving the detection of a person's eye state—whether open or closed—based on EEG (electroencephalogram) signals. The data was collected from one participant using an Emotiv EEG headset with 14 sensors while performing normal daily activities. Each instance includes 14 continuous numerical features corresponding to the EEG readings from the sensors, along with a binary target variable: 0 for eyes open and 1 for eyes closed. The dataset contains 14,980 instances in total. It is commonly used to test real-time brain-computer interface applications, signal processing models, and lightweight classification algorithms suitable for wearable or embedded systems.

\begin{figure}[htbp]
\centering

\begin{subfigure}[b]{0.49\textwidth}
    \includegraphics[width=\textwidth]{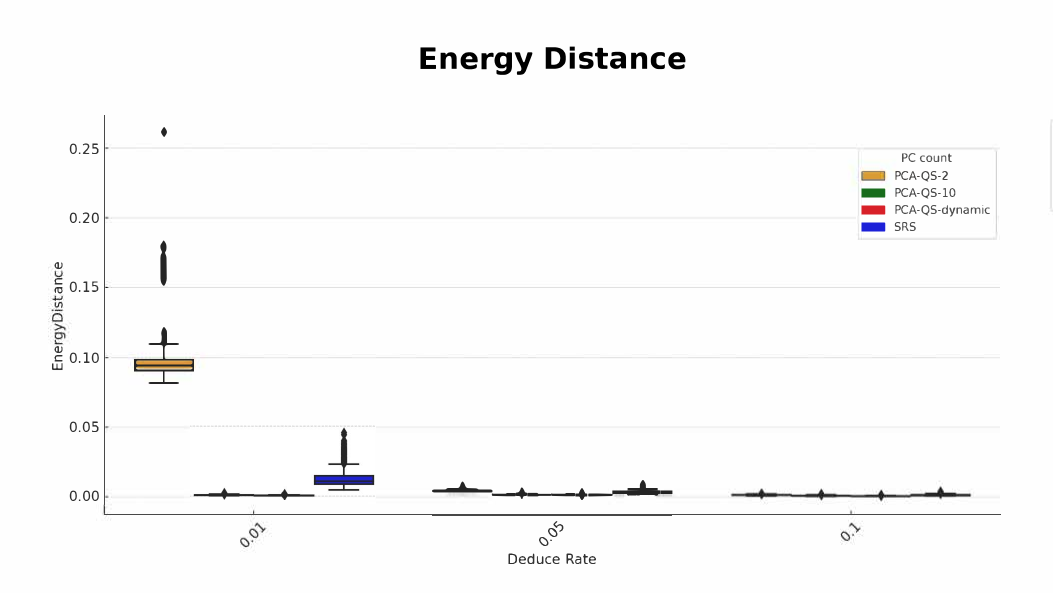}
    \caption{Energy Distance}
\end{subfigure}
\hfill
\begin{subfigure}[b]{0.49\textwidth}
    \includegraphics[width=\textwidth]{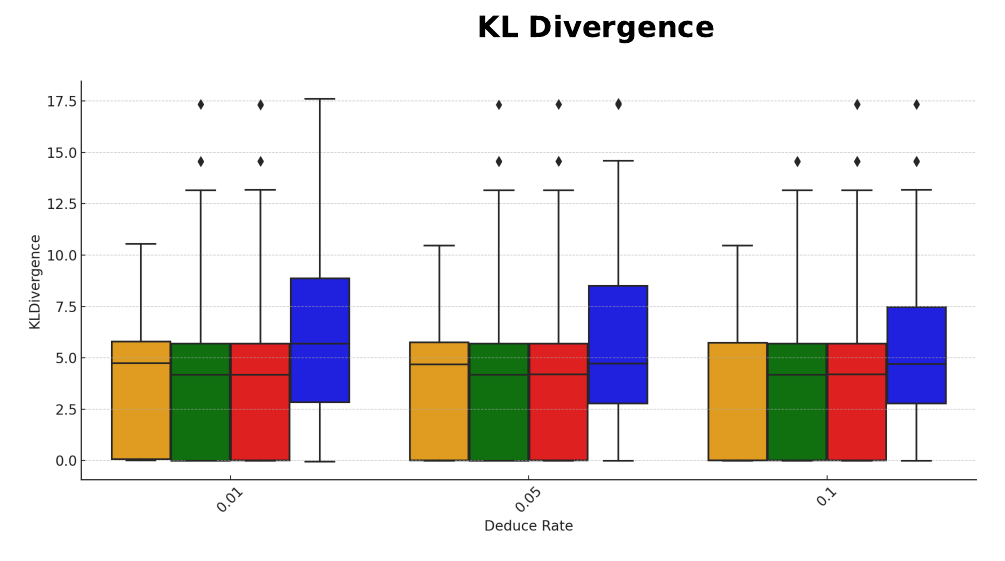}
    \caption{KL Divergence}
\end{subfigure}

\vspace{0.5cm}

\begin{subfigure}[b]{0.49\textwidth}
    \includegraphics[width=\textwidth]{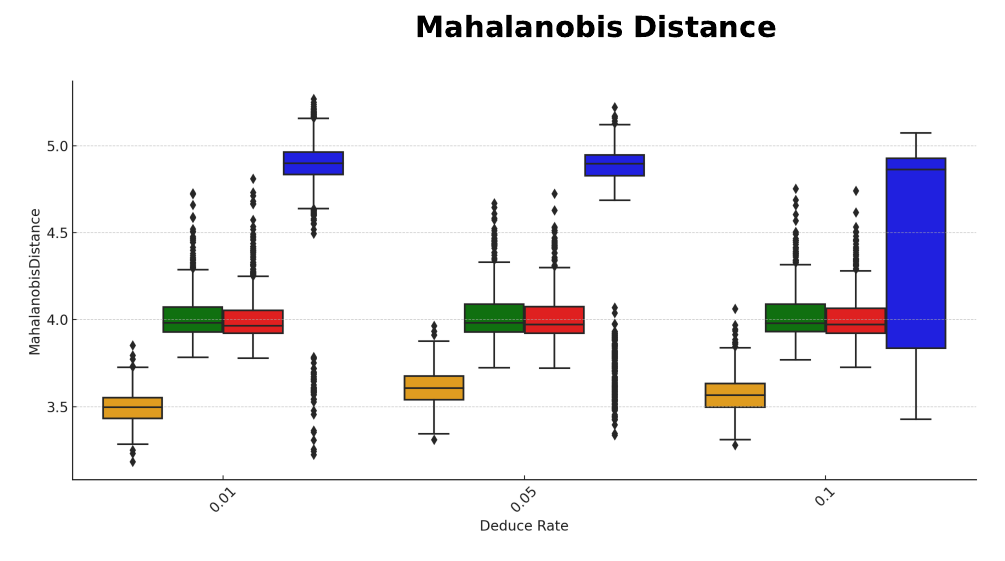}
    \caption{Mahalanobis Distance}
\end{subfigure}
\hfill
\begin{subfigure}[b]{0.49\textwidth}
    \includegraphics[width=\textwidth]{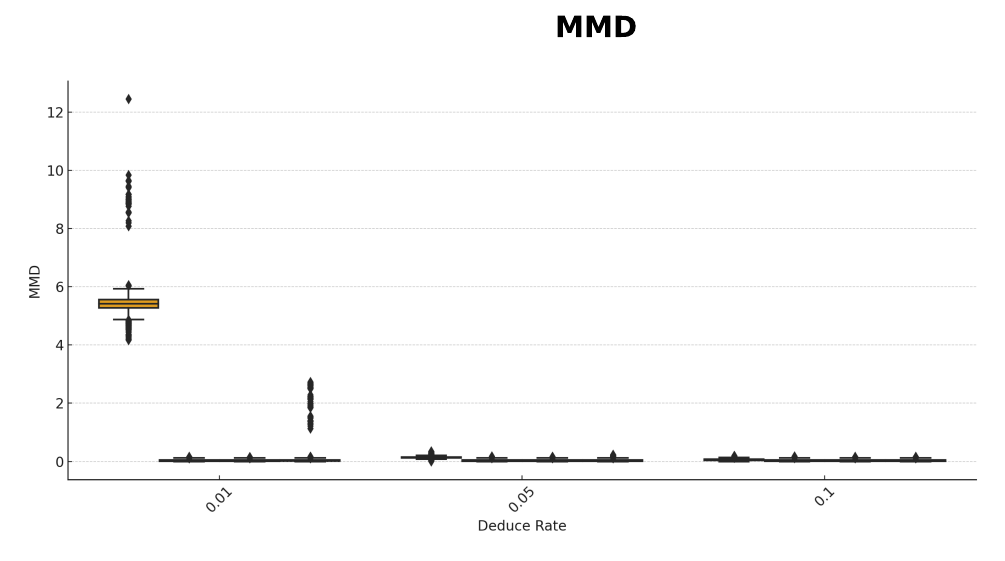}
    \caption{MMD}
\end{subfigure}

\caption{EEG Eye State Dataset}
\end{figure}


\subsubsection{Large-Scale Tabular Data}
\paragraph{Higgs Data set}
The HIGGS dataset from the UCI Machine Learning Repository is a large-scale benchmark dataset created for binary classification tasks in high-energy particle physics. It simulates the process of distinguishing between signal events that originate from the decay of a Higgs boson and background events that do not. Each instance represents a collision event and includes 28 numerical features: 21 low-level kinematic properties measured directly by the particle detectors, and 7 high-level features derived from those measurements through complex physics calculations. The target variable is binary, indicating whether the event corresponds to a Higgs boson (signal) or not (background). The dataset contains 11 million instances in total, making it suitable for testing the scalability and performance of machine learning models on very large datasets. It has become a popular benchmark for deep learning and large-scale classification research, especially in scientific computing contexts where interpretability and computational efficiency are critical.

\begin{figure}[htbp]
\centering

\begin{subfigure}[b]{0.49\textwidth}
    \includegraphics[width=\textwidth]{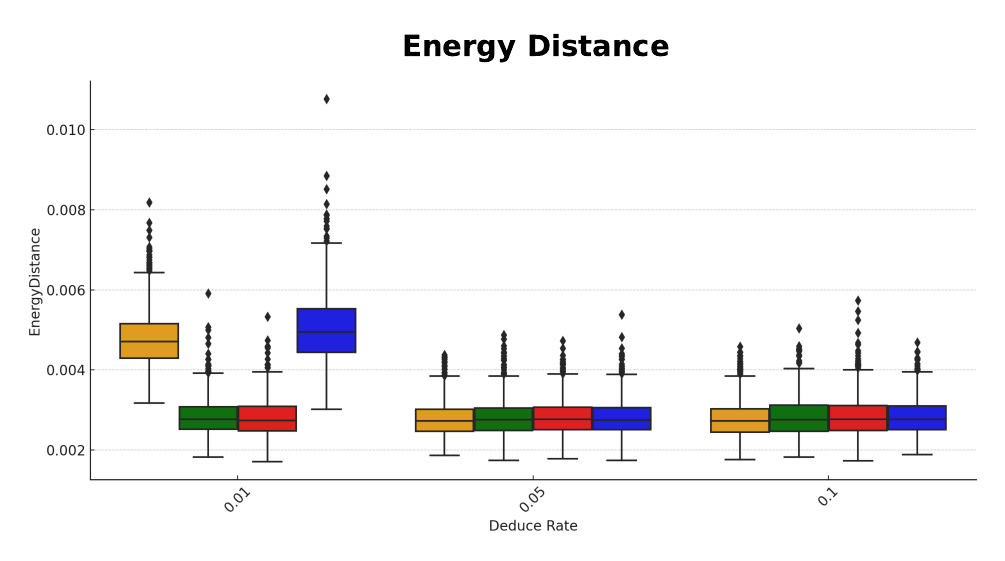}
\end{subfigure}
\hfill
\begin{subfigure}[b]{0.49\textwidth}
    \includegraphics[width=\textwidth]{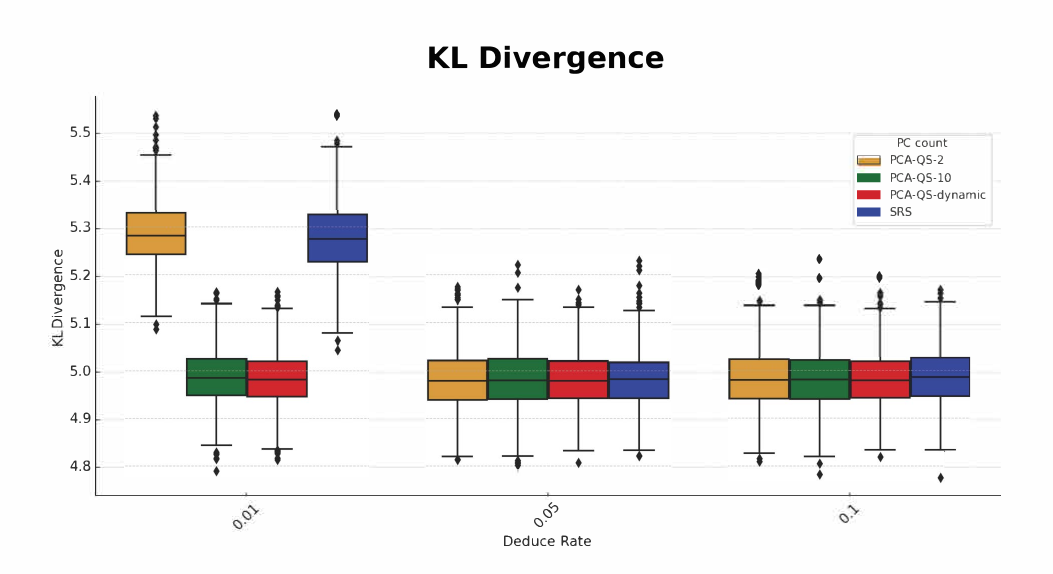}
\end{subfigure}


\begin{subfigure}[b]{0.49\textwidth}
    \includegraphics[width=\textwidth]{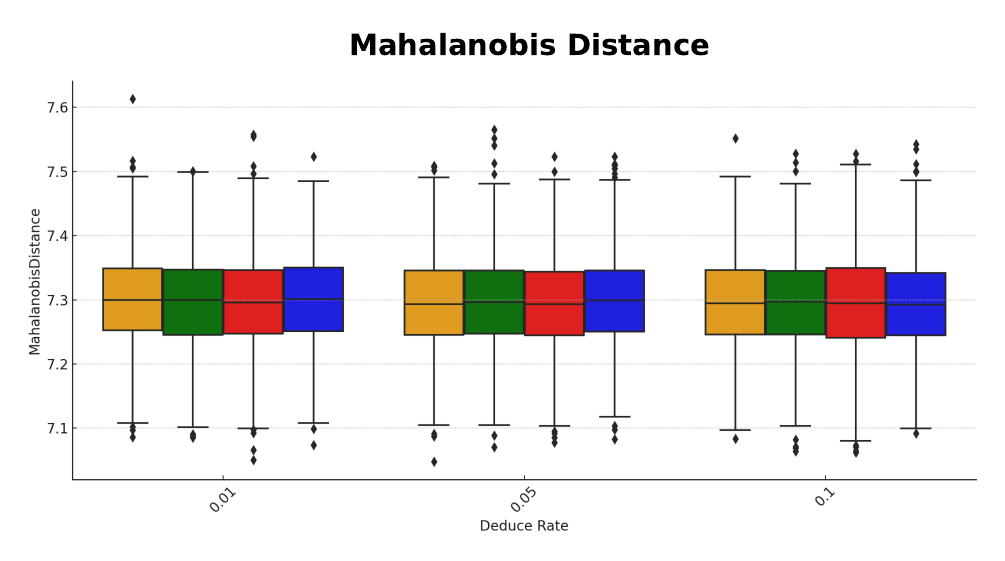}
\end{subfigure}
\hfill
\begin{subfigure}[b]{0.49\textwidth}
    \includegraphics[width=\textwidth]{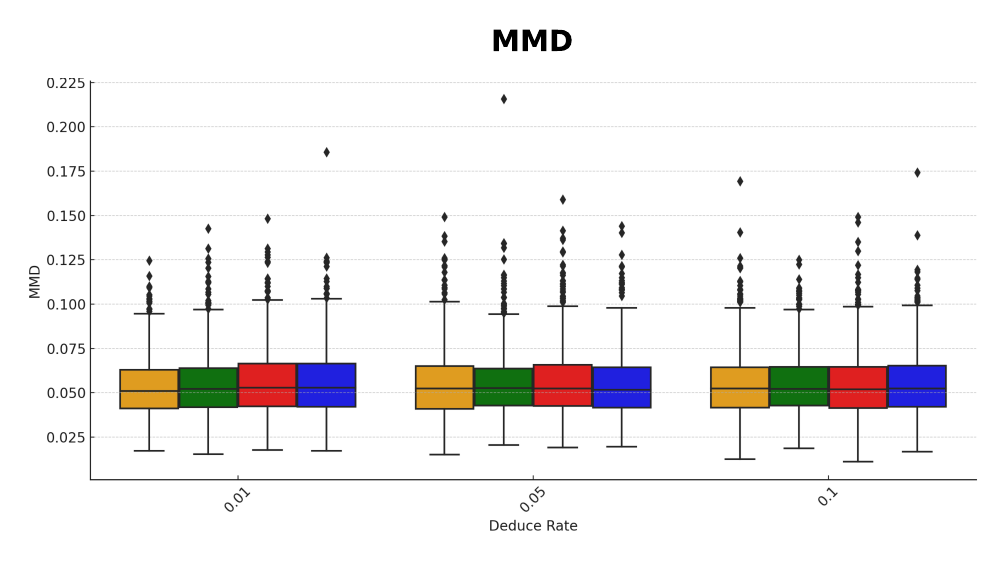}
\end{subfigure}

\caption{Distance Metrics for Two Sampling Methods with Higgs Data}
\end{figure}

\paragraph{Year Prediction MSD}
The YearPredictionMSD dataset from the UCI Machine Learning Repository is a large-scale benchmark dataset used for regression tasks, specifically aimed at predicting the release year of songs based on their audio characteristics. It is derived from the Million Song Dataset and contains 515,345 instances, each representing a song. Each instance includes 90 numerical audio features extracted from signal processing techniques, such as timbre averages and covariances. The first column in each row is the target variable, which is the year the song was released, while the remaining columns are the input features. All data are continuous and numerical. The dataset is split into a training set with 463,715 samples and a test set with 51,630 samples. This dataset is frequently used to benchmark regression models and study high-dimensional, real-world data in the context of music analysis and machine learning. It is available at the UCI Machine Learning Repository.

\begin{figure}[htbp]
\centering

\begin{subfigure}[b]{0.49\textwidth}
    \includegraphics[width=\textwidth]{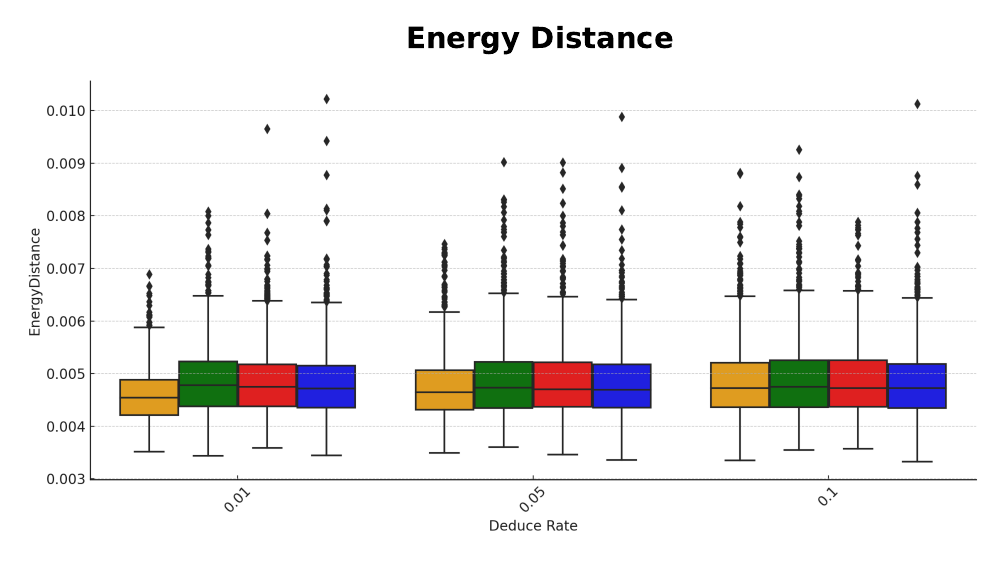}
    \caption{Energy Distance}
\end{subfigure}
\hfill
\begin{subfigure}[b]{0.49\textwidth}
    \includegraphics[width=\textwidth]{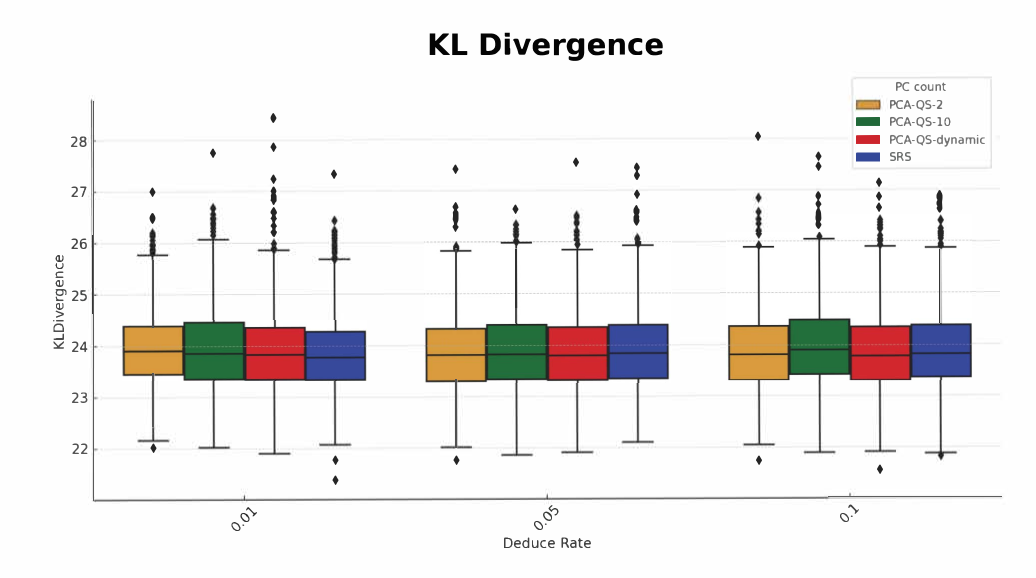}
    \caption{KL Divergence}
\end{subfigure}

\vspace{0.5cm}

\begin{subfigure}[b]{0.49\textwidth}
    \includegraphics[width=\textwidth]{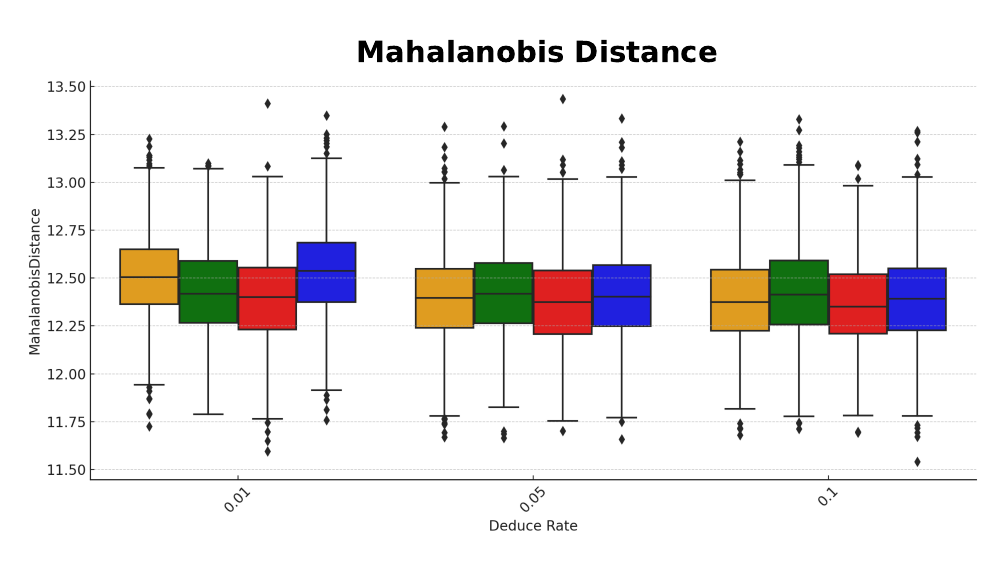}
    \caption{Mahalanobis Distance}
\end{subfigure}
\hfill
\begin{subfigure}[b]{0.49\textwidth}
    \includegraphics[width=\textwidth]{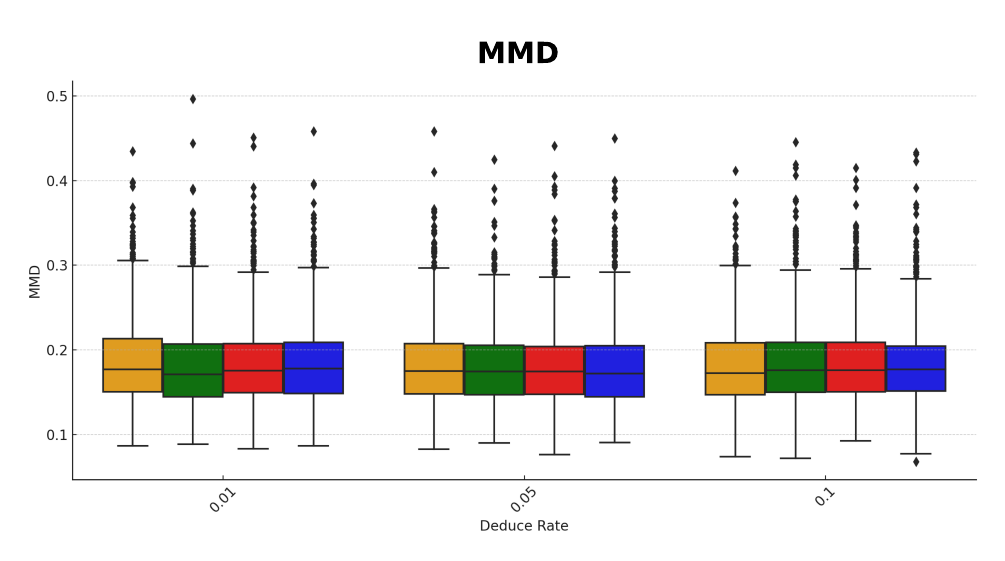}
    \caption{MMD}
\end{subfigure}

\caption{Distance Metrics of Two Sampling Methods with Year Prediction MSD Dataset}
\end{figure}


\subsubsection{Feature Complexity (Text/NLP)}
\paragraph{Online News}
The Online News Popularity dataset from the UCI Machine Learning Repository is a benchmark dataset used to predict the popularity of news articles published on the Mashable website. Each instance in the dataset represents a single news article and includes a set of numerical features that describe the article’s content, metadata, and social engagement metrics. These features cover aspects such as the number of words, presence of images or videos, natural language processing (NLP) indicators, and publication timing. The target variable is the number of times the article was shared on social media, making this a regression problem in its original form, though it is often used for binary classification by thresholding the share count. The dataset contains 39,644 instances and 61 features, all of which are numerical. It provides a rich testbed for predictive modeling, feature selection, and text-driven analytics in the field of digital media and social engagement prediction.

\begin{figure}[htbp]
\centering

\begin{subfigure}[b]{0.49\textwidth}
    \includegraphics[width=\textwidth]{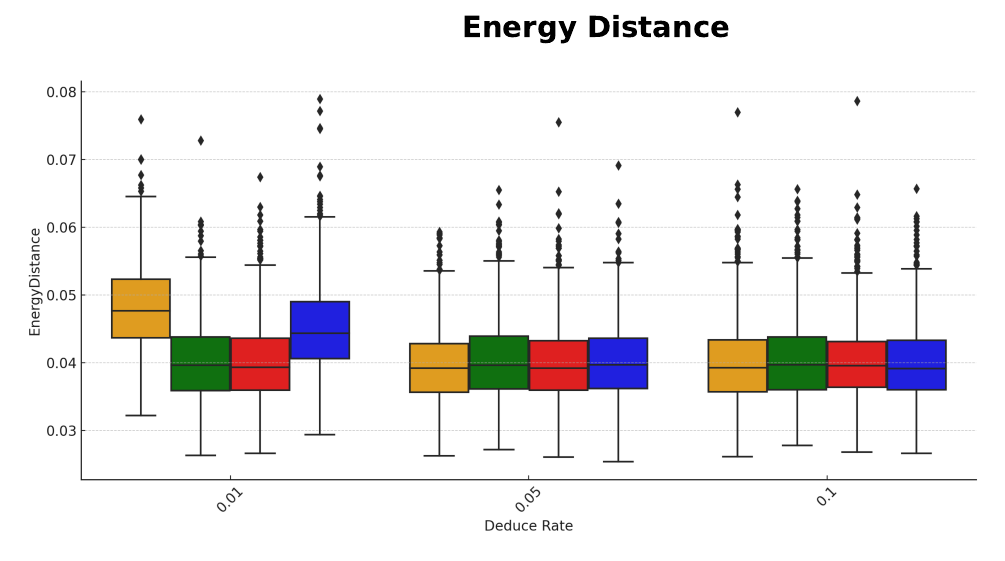}
    \caption{Energy Distance}
\end{subfigure}
\hfill
\begin{subfigure}[b]{0.49\textwidth}
    \includegraphics[width=\textwidth]{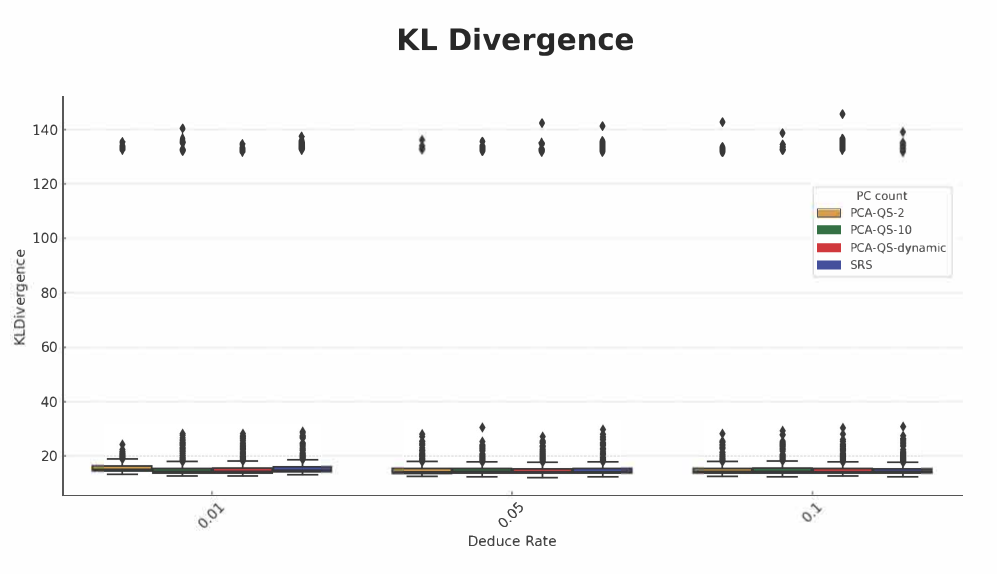}
    \caption{KL Divergence}
\end{subfigure}

\vspace{0.5cm}

\begin{subfigure}[b]{0.49\textwidth}
    \includegraphics[width=\textwidth]{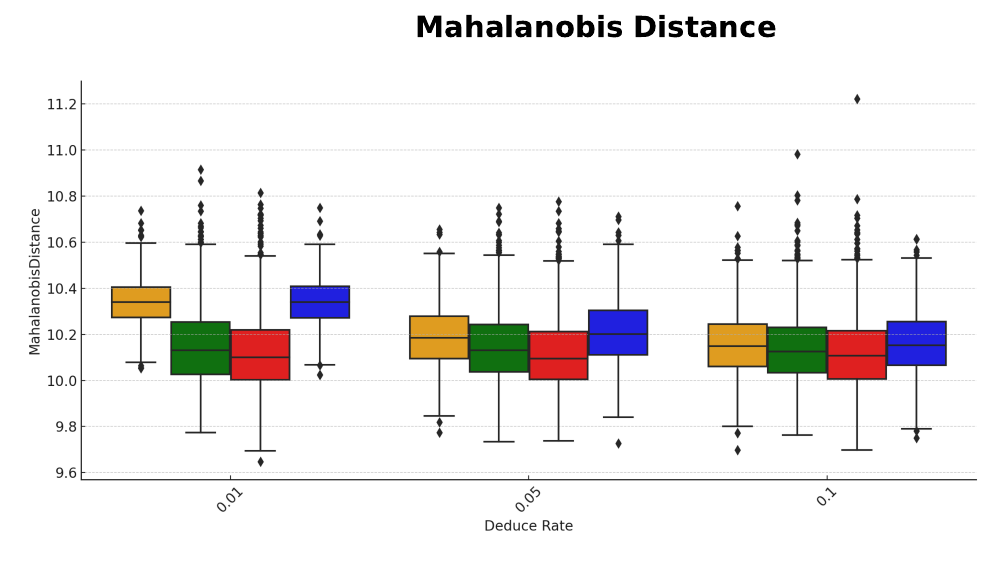}
    \caption{Mahalanobis Distance}
\end{subfigure}
\hfill
\begin{subfigure}[b]{0.49\textwidth}
    \includegraphics[width=\textwidth]{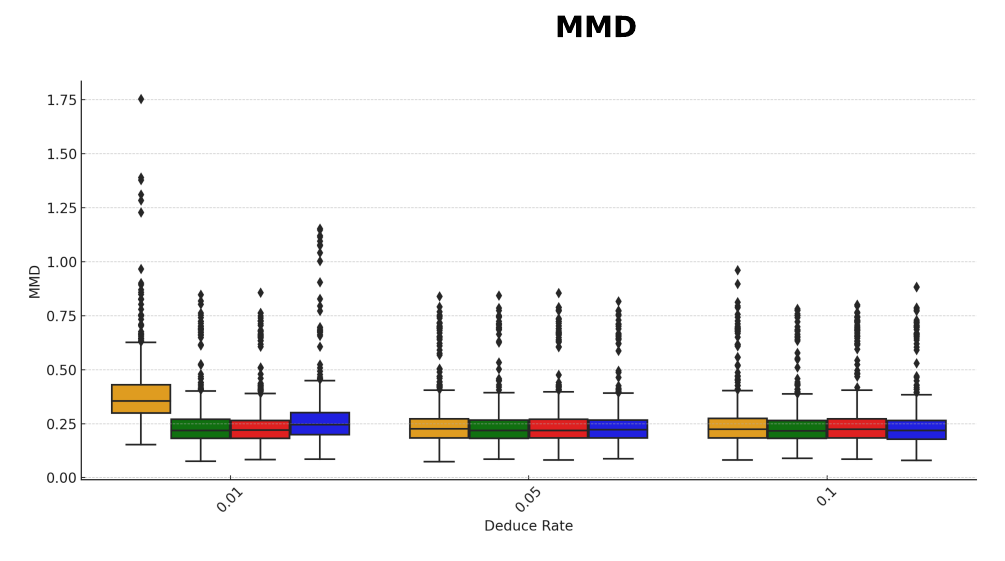}
    \caption{MMD}
\end{subfigure}

\caption{Distance Metrics of Two Sampling Methods with Online News Dataset}
\end{figure}

\subsubsection*{Summary of Real Data Structure-Preserving Comparisons}

Across all real-world datasets grouped by class imbalance, time-series nature, scale, and feature complexity, the results consistently show that PCA-QS retains the structural and distributional properties of the original datasets more faithfully than Simple Random Sampling (SRS). Specifically, PCA-QS achieves lower Kullback-Leibler divergence, Energy distance, Mahalanobis distance, and Maximum Mean Discrepancy, indicating better alignment in both global and local feature spaces.

This advantage is particularly notable for imbalanced and high-dimensional settings (e.g., CreditCard, MagicGamma), as well as for signal-rich time-series data (EEG, Epileptic Seizure) where local temporal structures are important. Even for massive tabular data (Higgs, YearPredictionMSD) and complex NLP-related features (Online News), PCA-QS demonstrates stable and reliable structure preservation. These findings validate PCA-QS as a robust sampling strategy for downstream analyses where retaining the underlying data geometry is critical.

{
\subsection{Logistic Regression and Other Popular Classification Models}

\paragraph{Credit Default Dataset}

This classification experiment revisits the \textbf{Credit Default Dataset} already introduced in the matrix comparison section. Recall that this dataset provides client credit information used to predict whether a payment default will occur. 

For model evaluation, we compare \textbf{PCA-QS} (with three principal component configurations: \texttt{fixed\_5}, \texttt{fixed\_10}, and \texttt{dynamic\_0.7}) against \textbf{SRS} (Simple Random Sampling). The goal is to quantify how well each sampling method preserves predictive structure for supervised learning.

From these experiment, we observe that:
- \textit{AUC Performance}: PCA-QS achieves consistently higher and more stable AUCs than SRS. Dynamic PC and fixed-10 configurations yield the best discrimination.
- \textit{Precision vs. Recall}: SRS occasionally peaks in precision but at the cost of poorer and unstable recall. PCA-QS shows superior balance, crucial for reliable default detection.
- \textit{F1 Score Stability}: PCA-QS, particularly fixed-10, maintains lower variance, indicating robust trade-offs between precision and recall.
- \textit{False Rates and Reliability}: PCA-QS minimizes both false positives and negatives more consistently than SRS, reducing risk in credit scoring systems.

\begin{table}[h!]
\centering
\caption{Mean (std) of AUC, F1, Recall, and Precision over 1000 runs for each method/configuration on Credit Default.}
\label{tab:performance_precision_creditdefault}
\tiny
\begin{tabular}{llllll}
\toprule
Method &   PC Config & AUC & F1 & Recall & Precision \\
\midrule
PCA-QS & Fixed 5 & 0.9122 (0.0205) & 0.5854 (0.0249) & 0.8297 (0.0269) & 0.4530 (0.0285) \\
 & Fixed 10 & 0.9691 (0.0050) & 0.5974 (0.0114) & 0.9143 (0.0069) & 0.4438 (0.0124) \\
 & Dynamic =7 & 0.9807 (0.0008) & 0.6048 (0.0050) & 0.9256 (0.0026) & 0.4492 (0.0055) \\
SRS & - & 0.9055 (0.0221) & 0.6112 (0.0321) & 0.8113 (0.0289) & 0.4918 (0.0401) \\
\bottomrule
\end{tabular}
\end{table}

\begin{table}[h!]
\centering
\caption{TPR, TNR, FPR, FNR, and Threshold over 1000 runs for each method on Credit Default.}
\label{tab:proportional_creditdefault}
\tiny
\begin{tabular}{lllllll}
\toprule
Method & PC Config & TPR & TNR & FPR & FNR & Threshold \\
\midrule
PCA-QS & Fixed 5 & 0.8297 (0.0269) & 0.9758 (0.0029) & 0.0242 (0.0029) & 0.1703 (0.0269) & 0.0148 (0.0008) \\
 & Fixed 10 & 0.9143 (0.0069) & 0.9725 (0.0014) & 0.0275 (0.0014) & 0.0857 (0.0069) & 0.0167 (0.0005) \\
 & Dynamic =7 & 0.9256 (0.0026) & 0.9728 (0.0006) & 0.0272 (0.0006) & 0.0744 (0.0026) & 0.0191 (0.0002) \\
SRS & - & 0.8113 (0.0289) & 0.9796 (0.0034) & 0.0204 (0.0034) & 0.1887 (0.0289) & 0.0167 (0.0016) \\
\bottomrule
\end{tabular}
\end{table}

\begin{figure}[h!]
\centering
\begin{subfigure}{0.48\textwidth}
\includegraphics[width=\linewidth]{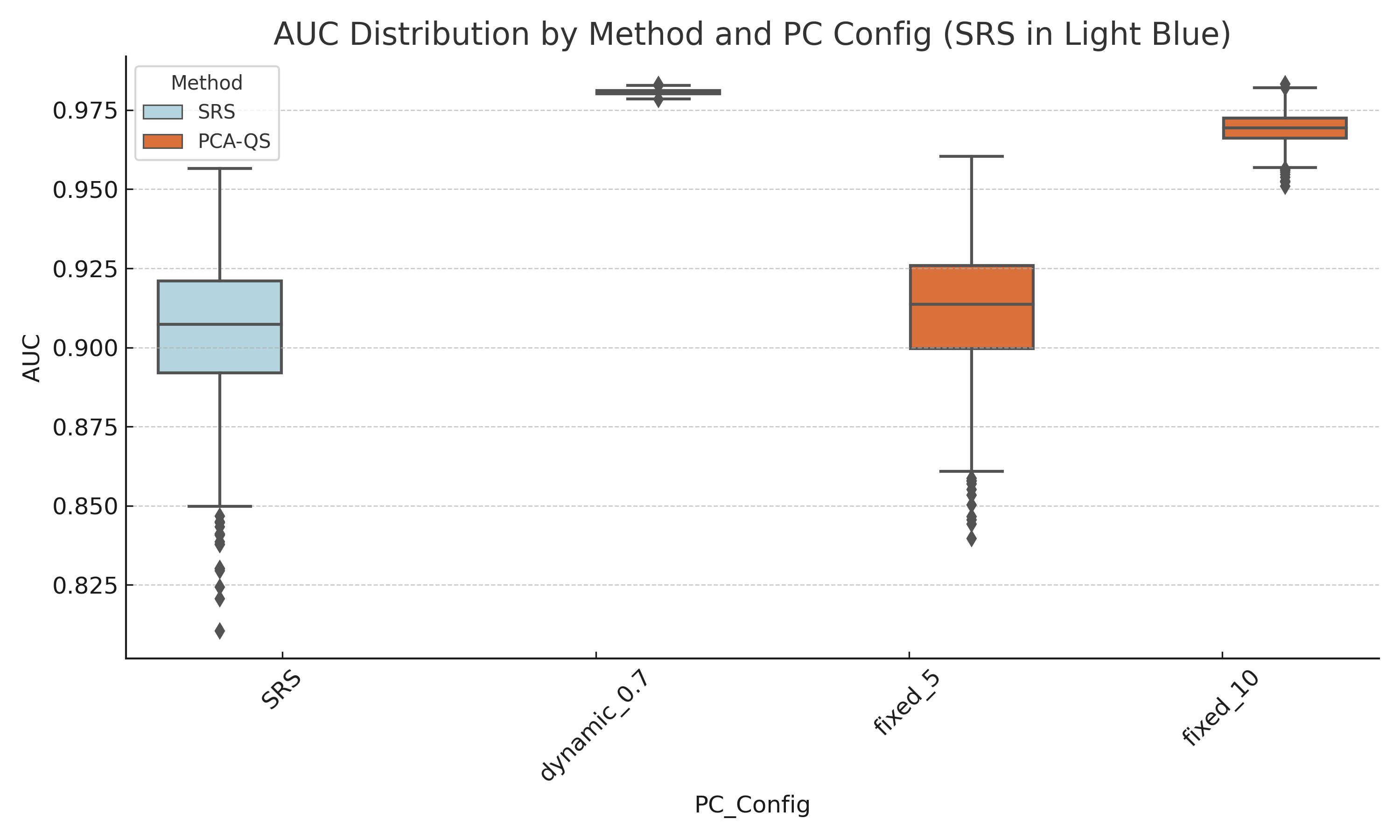}
\caption{AUC}
\end{subfigure}
\hfill
\begin{subfigure}{0.48\textwidth}
\includegraphics[width=\linewidth]{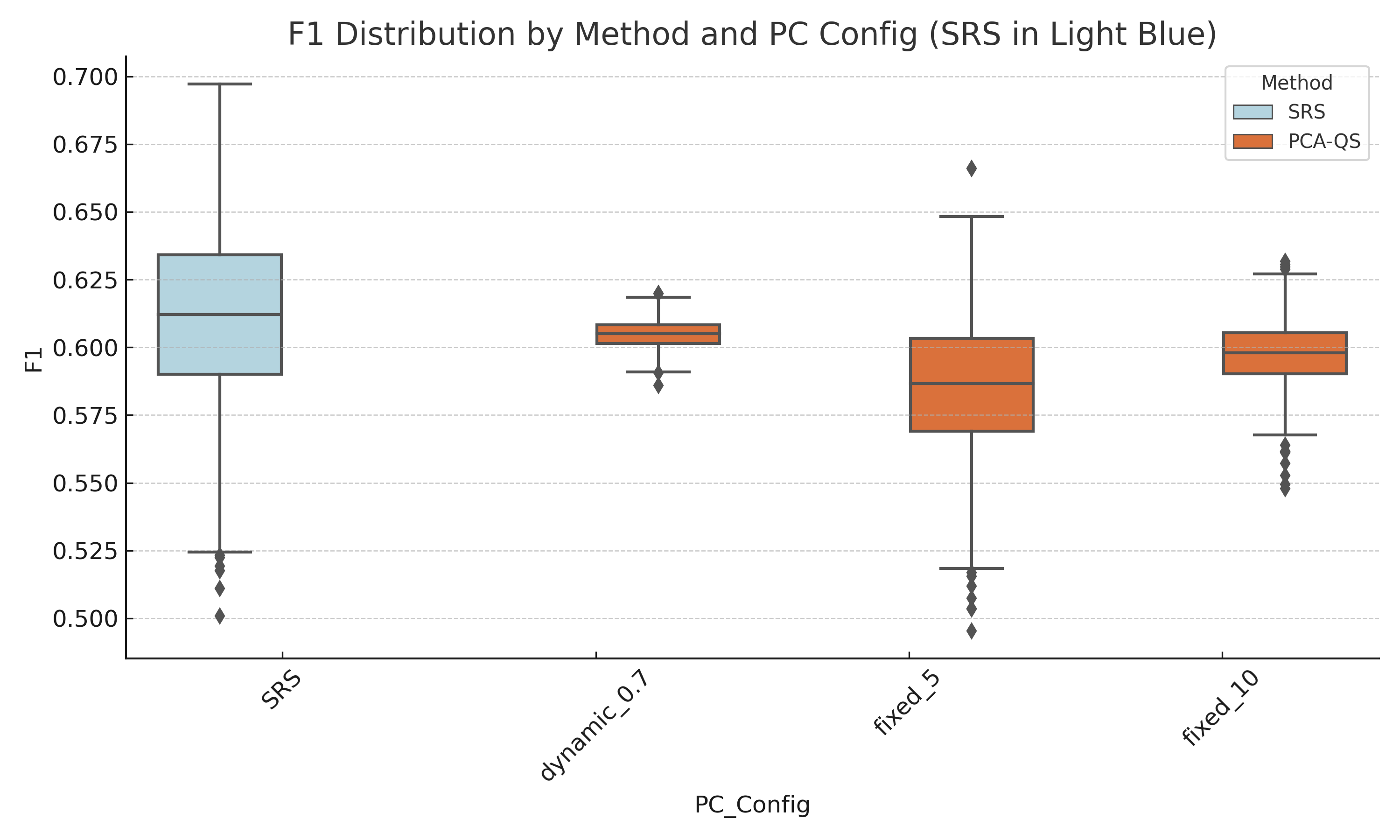}
\caption{F1 Score}
\end{subfigure}
\begin{subfigure}{0.48\textwidth}
\includegraphics[width=\linewidth]{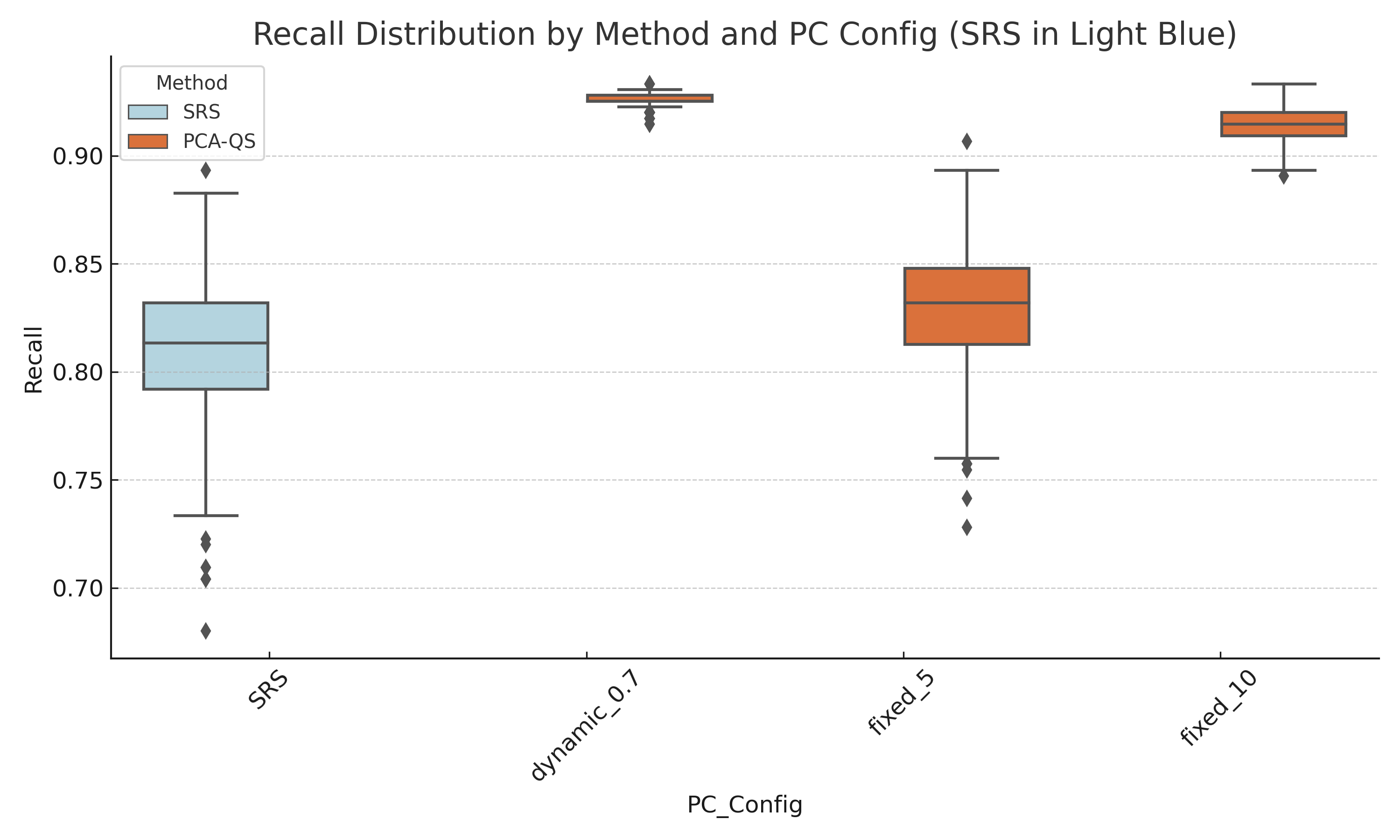}
\caption{Recall}
\end{subfigure}
\hfill
\begin{subfigure}{0.48\textwidth}
\includegraphics[width=\linewidth]{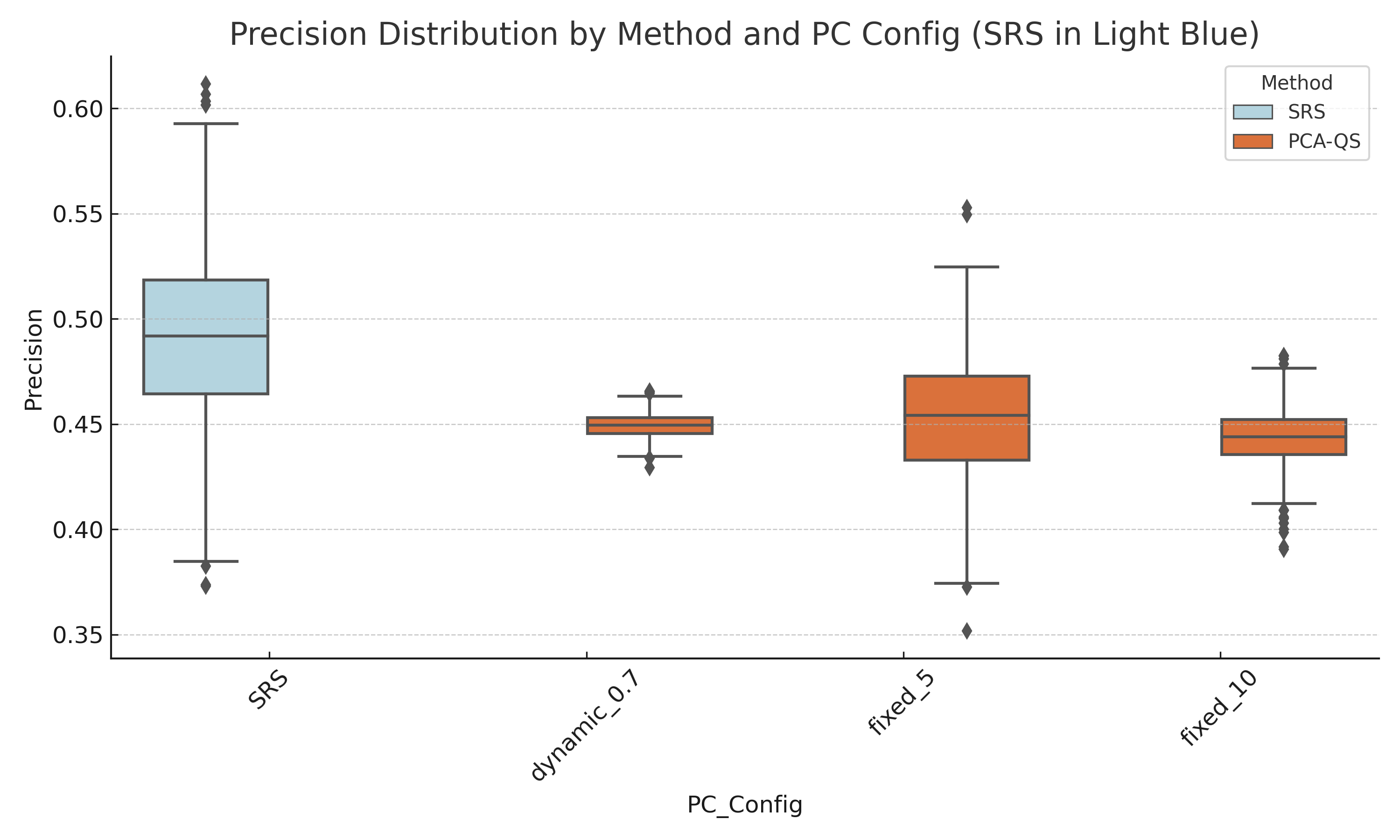}
\caption{Precision}
\end{subfigure}
\caption{Performance variation across 1000 runs for AUC, F1, Recall, and Precision on Credit Default.}
\label{fig:CreditCard_Metrics}
\end{figure}

\paragraph{APS Failure at Scania Trucks Dataset}

The APS Failure dataset was already described in the matrix comparison section; here we revisit it to evaluate classification performance under real-world predictive maintenance. The same three PCA-QS configurations and SRS baseline are compared.

Key observations:
- \textit{AUC}: Dynamic and fixed-10 PCA-QS yield highest AUC with minimal variance, outperforming SRS.
- \textit{F1 and Recall}: PCA-QS maintains stable F1 and robust recall. SRS shows larger fluctuations, which could lead to missed failures.
- \textit{Precision}: SRS sometimes shows slightly higher average precision, but at the expense of lower and unstable recall — risky for fault detection.

\begin{table}[h!]
\centering
\caption{Mean (std) AUC, F1, Recall, Precision over 1000 runs for APS Failure. Dynamic PCA-QS used actual mean PCs = 21.}
\label{tab:performance_precision_aps}
\tiny
\begin{tabular}{llllll}
\toprule
Method & PC Config & AUC & F1 & Recall & Precision \\
\midrule
PCA-QS & Fixed 5 & 0.9122 (0.0205) & 0.5854 (0.0249) & 0.8297 (0.0269) & 0.4530 (0.0285) \\
 & Fixed 10 & 0.9691 (0.0050) & 0.5974 (0.0114) & 0.9143 (0.0069) & 0.4438 (0.0124) \\
 & Dynamic =21 & 0.9807 (0.0008) & 0.6048 (0.0050) & 0.9256 (0.0026) & 0.4492 (0.0055) \\
SRS & - & 0.9055 (0.0221) & 0.6112 (0.0321) & 0.8113 (0.0289) & 0.4918 (0.0401) \\
\bottomrule
\end{tabular}
\end{table}

\begin{table}[h!]
\centering
\caption{Average TP, TN, FP, FN, and Threshold for APS over 1000 runs. Dynamic PCA-QS mean PCs = 21.}
\label{tab:apps_failure}
\tiny
\begin{tabular}{lllllll}
\toprule
Method & PC Config & TP & TN & FP & FN & Threshold \\
\midrule
PCA-QS & Fixed 5 & 311.144 (10.0944) & 15246.521 (45.4812) & 378.479 (45.4812) & 63.856 (10.0944) & 0.0148 (0.0008) \\
 & Fixed 10 & 342.879 (2.6041) & 15194.624 (22.1261) & 430.376 (22.1261) & 32.121 (2.6041) & 0.0167 (0.0005) \\
 & Dynamic =21 & 347.117 (0.9676) & 15199.247 (9.4625) & 425.753 (9.4625) & 27.883 (0.9676) & 0.0191 (0.0002) \\
SRS & - & 304.227 (10.8334) & 15306.375 (53.0452) & 318.625 (53.0452) & 70.773 (10.8334) & 0.0167 (0.0016) \\
\bottomrule
\end{tabular}
\end{table}

\begin{figure}[h!]
\centering
\begin{subfigure}{0.48\textwidth}
\includegraphics[width=\linewidth]{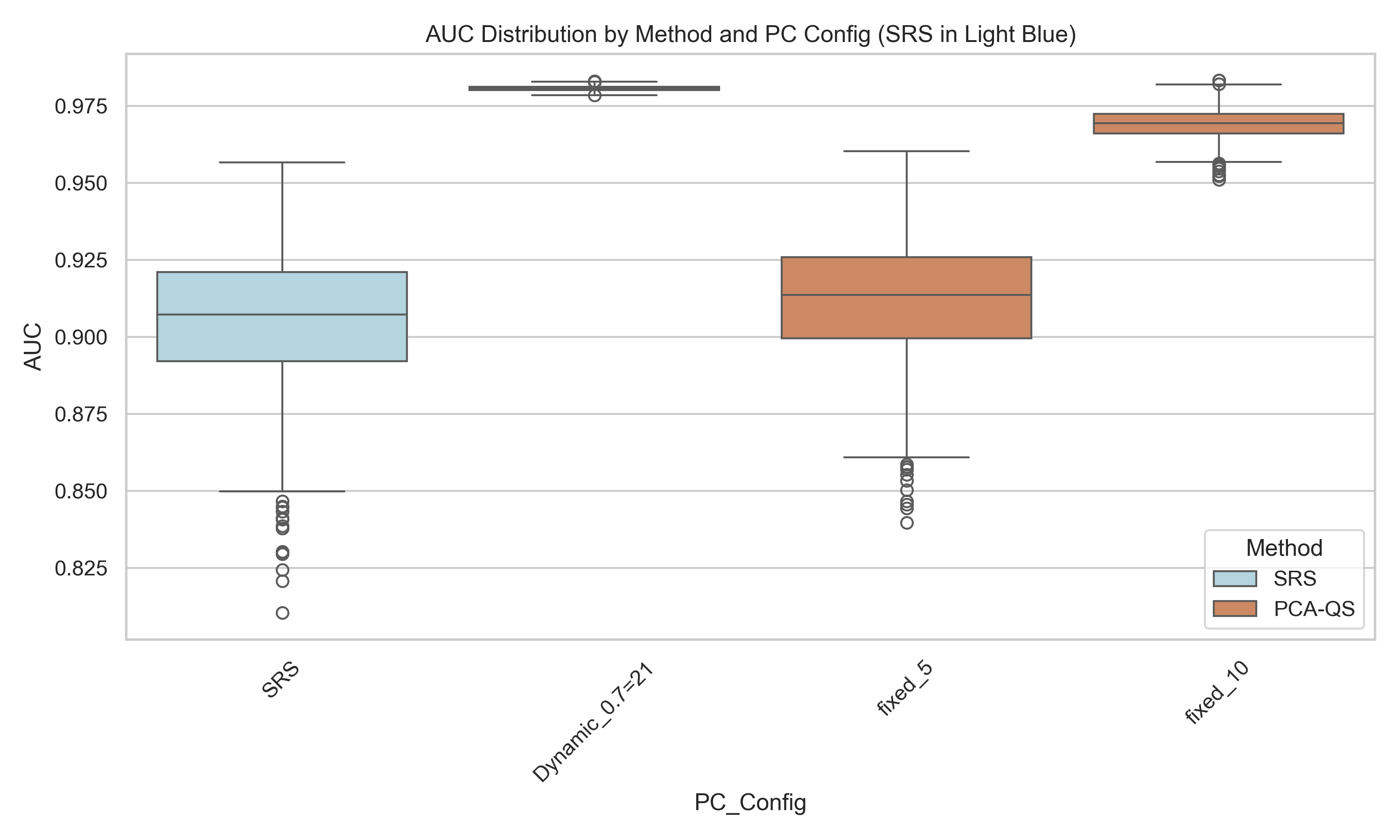}
\caption{AUC}
\end{subfigure}
\hfill
\begin{subfigure}{0.48\textwidth}
\includegraphics[width=\linewidth]{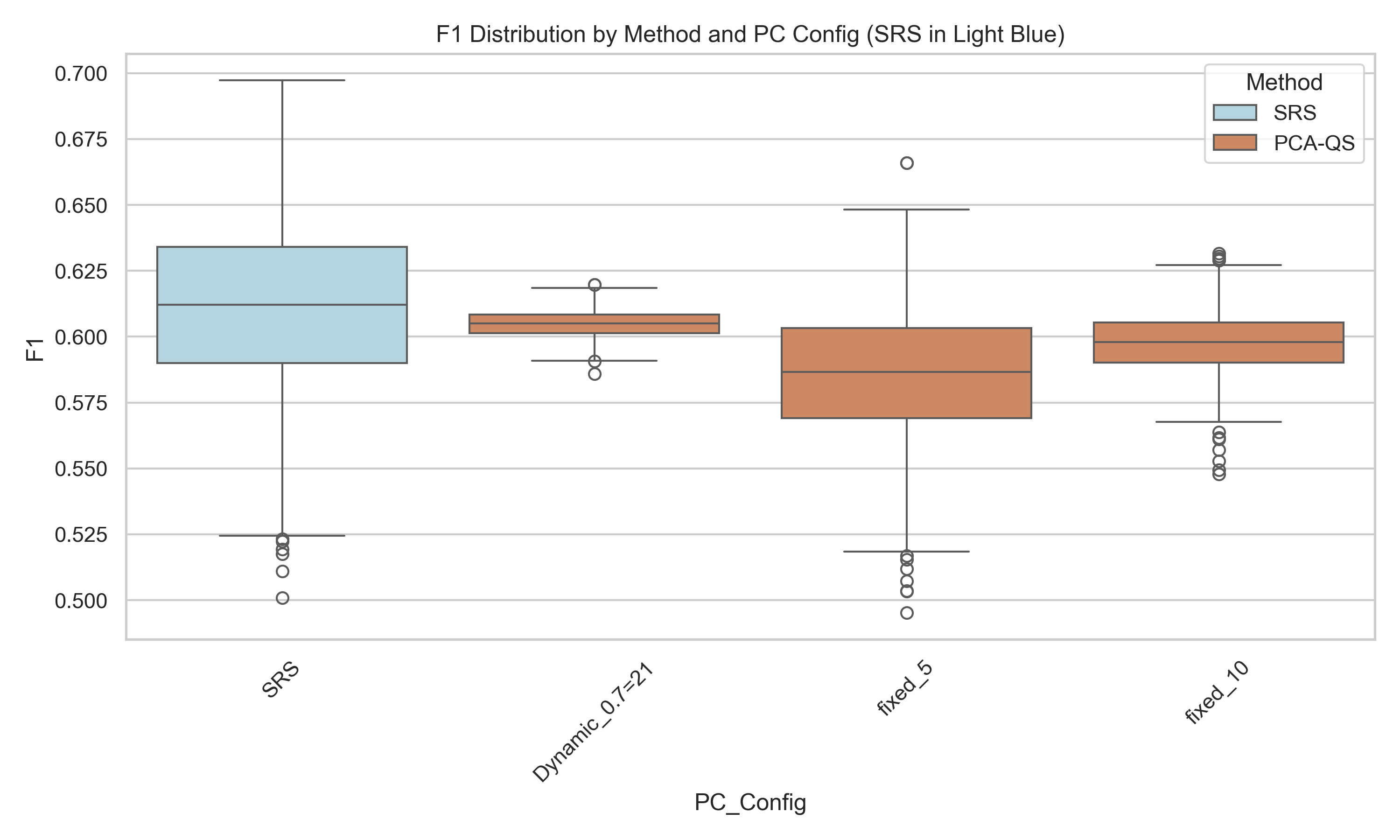}
\caption{F1 Score}
\end{subfigure}
\begin{subfigure}{0.48\textwidth}
\includegraphics[width=\linewidth]{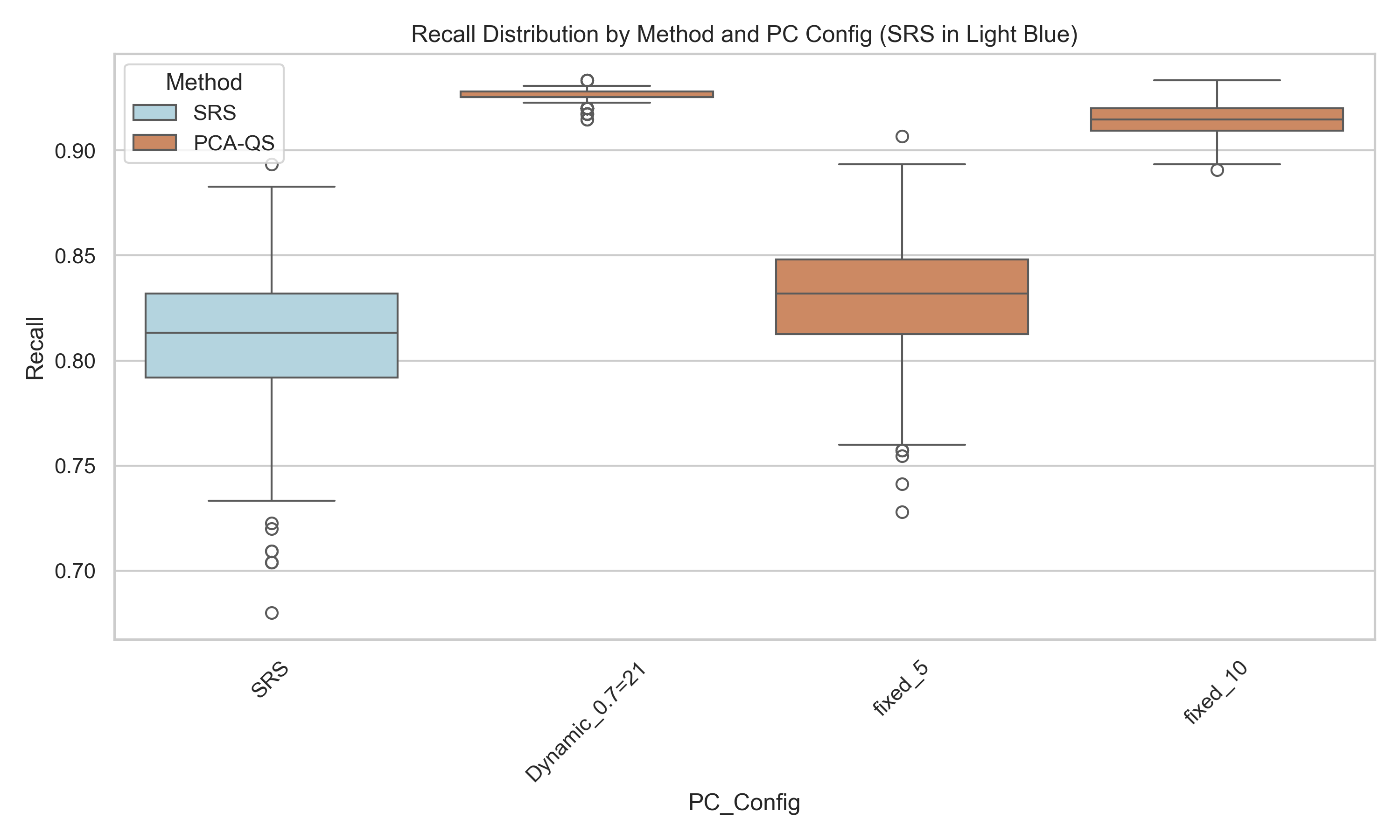}
\caption{Recall}
\end{subfigure}
\hfill
\begin{subfigure}{0.48\textwidth}
\includegraphics[width=\linewidth]{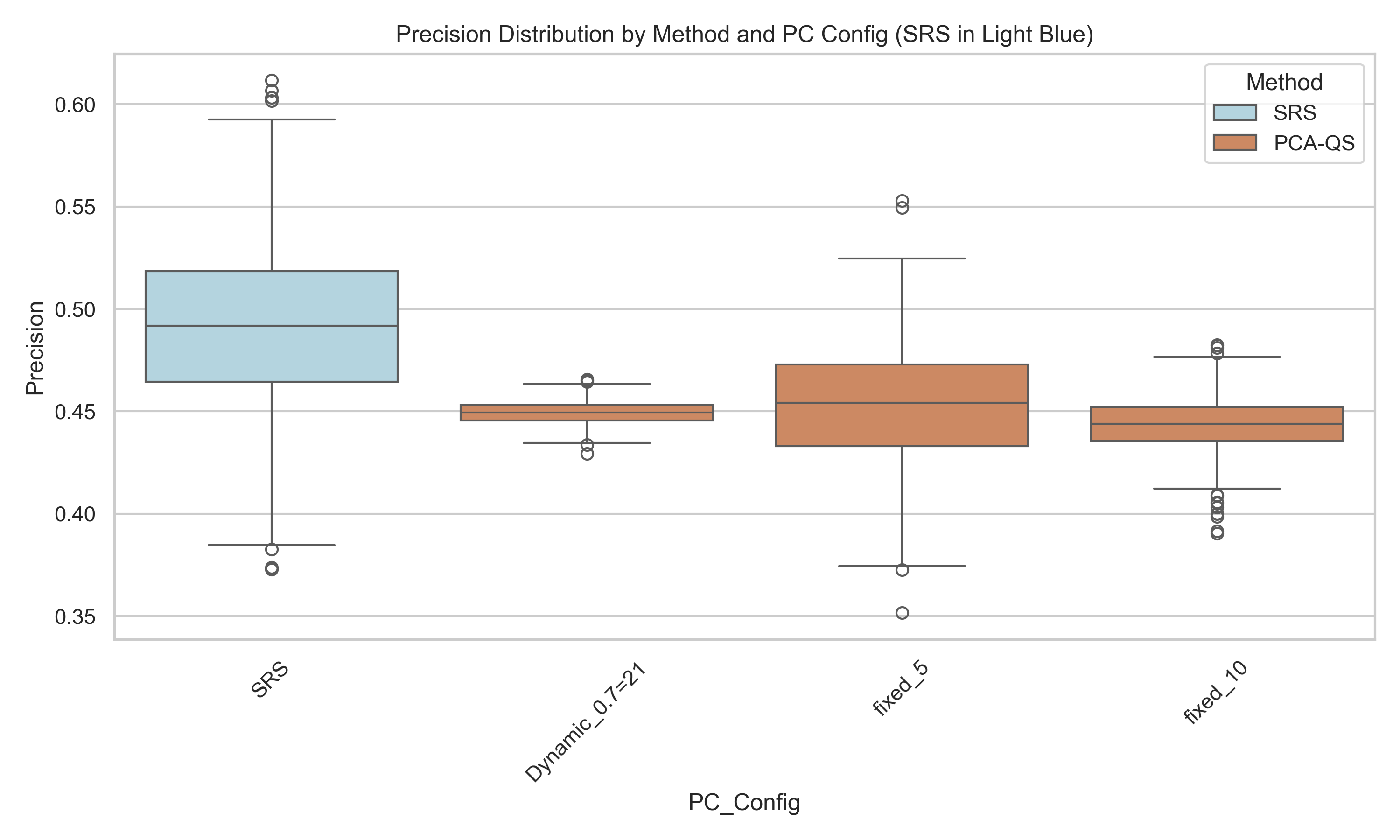}
\caption{Precision}
\end{subfigure}
\caption{Boxplots comparing PCA-QS and SRS under different PC configurations for APS Failure.}
\end{figure}

\paragraph{Summary}

Overall, PCA-QS delivers more reliable and higher classification performance than SRS on both the Credit Default and APS datasets. Dynamic PC configurations consistently maximize AUC and recall with low variance. SRS shows greater variability and less dependable results, reinforcing PCA-QS as the robust choice for practical classification pipelines.

}

\section{Conclusion}
\label{sec:conclusion}

This study presents a comprehensive comparative analysis of Principal Component Analysis-guided Quantile Sampling (PCA-QS) and Simple Random Sampling (SRS) methods, evaluated across varying configurations of dimensionality, sample sizes, and principal component counts. The comparison leverages four statistical similarity metrics — Energy Distance, Maximum Mean Discrepancy (MMD), Mahalanobis Distance, and Kullback–Leibler (KL) Divergence — to quantify the discrepancy between the sampled subsets and the full data distribution. Across all metrics, PCA-QS consistently achieves lower values, indicating superior fidelity in preserving both the statistical and geometric structure of the original dataset. These advantages become more pronounced in higher-dimensional settings and persist across a range of sample sizes and principal component configurations. A composite score aggregating all normalized metrics further confirms the robust performance of PCA-QS. Additionally, clustering analysis delineates specific regimes where SRS performs comparably — typically under low dimensionality and small sample sizes — versus scenarios where PCA-QS offers significant improvements. Overall, the results establish PCA-QS as a robust, scalable, and structure-preserving subsampling method, providing a principled alternative to SRS for large-scale data reduction in modern machine learning workflows.

\appendix
\section{Appendix: Detailed Metric Tables}
\label{app:appendix-tables}

\subsection{Imbalanced Datasets}
\begin{table}[H]
\caption{ UCI Credit Card Dataset}
\label{tab:UCICredit}
\tiny
\centering
\begin{tabular}{cclllll}
\toprule
Deduce & Methods & PC & Energy & Mahalanobis & MMD & KLD \\
\midrule
0.01 & PCA-QS & 2 & 0.0205 (0.0033) & 6.3802 (0.1232) & 0.2033 (0.0593) & 2.0129 (0.3035) \\
 & & 10 & 0.0031 (0.0010) & 6.1531 (0.1387) & 0.0500 (0.0268) & 1.0648 (0.2991) \\
 &  & dynamic & 0.0020 (0.0006) & 6.1261 (0.1404) & 0.0468 (0.0237) & 1.0196 (0.2860) \\
 & SRS &  & 0.0213 (0.0057) & 6.4039 (0.1228) & 0.0996 (0.0508) & 2.1145 (0.4257) \\
0.05 & PCA-QS & 2 & 0.0039 (0.0006) & 6.1850 (0.1417) & 0.0441 (0.0187) & 1.3388 (0.2620) \\
 &  & 10 & 0.0031 (0.0010) & 6.1526 (0.1417) & 0.0507 (0.0264) & 1.0653 (0.3043) \\
 &  & dynamic & 0.0021 (0.0005) & 6.1346 (0.1470) & 0.0464 (0.0245) & 1.0326 (0.2998) \\
 & SRS &  & 0.0050 (0.0013) & 6.2377 (0.1422) & 0.0461 (0.0226) & 1.4237 (0.3627) \\
0.1 & PCA-QS & 2 & 0.0024 (0.0004) & 6.1617 (0.1472) & 0.0460 (0.0231) & 1.1244 (0.2241) \\
 &  & 10 & 0.0031 (0.0011) & 6.1608 (0.1393) & 0.0513 (0.0313) & 1.0690 (0.2961) \\
 &  & dynamic & 0.0021 (0.0006) & 6.1245 (0.1456) & 0.0463 (0.0230) & 1.0369 (0.2963) \\
 & SRS &  & 0.0028 (0.0007) & 6.2069 (0.1424) & 0.0472 (0.0255) & 1.1849 (0.3310) \\
\bottomrule
\end{tabular}
\end{table}

\begin{table}[H]
\caption{MagicGamma Dataset)}
\label{tab:MagicGamma}
\tiny
\centering
\begin{tabular}{cclllll}
\toprule
Deduce & Sampling & PC & Energy & Mahalanobis & MMD & KLD \\
\midrule
0.01 & PCA-QS & 2 & 0.0163 (0.0035) & 4.1176 (0.0813) & 0.0400 (0.0196) & 0.3234 (0.0376) \\
 &  & 10 & 0.0012 (0.0004) & 4.0844 (0.0810) & 0.0191 (0.0133) & 0.0847 (0.0087) \\
 &  & dynamic & 0.0012 (0.0004) & 4.0843 (0.0804) & 0.0193 (0.0130) & 0.0848 (0.0087) \\
 & SRS &  & 0.0211 (0.0079) & 4.1306 (0.0844) & 0.0611 (0.0408) & 0.3123 (0.0498) \\
0.05 & PCA-QS & 2 & 0.0031 (0.0007) & 4.0936 (0.0853) & 0.0137 (0.0077) & 0.1548 (0.0181) \\
 &  & 10 & 0.0012 (0.0005) & 4.0847 (0.0799) & 0.0185 (0.0117) & 0.0852 (0.0087) \\
 &  & dynamic & 0.0012 (0.0004) & 4.0778 (0.0796) & 0.0191 (0.0128) & 0.0848 (0.0087) \\
 & SRS &  & 0.0045 (0.0017) & 4.0931 (0.0846) & 0.0191 (0.0133) & 0.1548 (0.0196) \\
0.1 & PCA-QS & 2 & 0.0018 (0.0004) & 4.0877 (0.0839) & 0.0165 (0.0109) & 0.1172 (0.0125) \\
 &  & 10 & 0.0012 (0.0004) & 4.0800 (0.0823) & 0.0186 (0.0137) & 0.0848 (0.0093) \\
 &  & dynamic & 0.0012 (0.0004) & 4.0845 (0.0837) & 0.0191 (0.0133) & 0.0849 (0.0088) \\
 & SRS &  & 0.0025 (0.0009) & 4.0874 (0.0819) & 0.0190 (0.0126) & 0.1186 (0.0133) \\
\bottomrule
\end{tabular}
\end{table}

\subsection{Time-Series and Signal Data}
\begin{table}[H]
\centering
\caption{EEG Eye State}
\tiny
\begin{tabular}{cclllll}
\toprule
Deduce & Sampling & PC & Energy & Mahalanobis & MMD & KLD \\
\midrule
0.01 & PCA-QS & 2 & 0.0961 (0.0137) & 3.4933 (0.0902) & 5.4727 (0.6681) & 3.4852 (3.4786) \\
 &  & 10 & 0.0007 (0.0003) & 4.0189 (0.1355) & 0.0274 (0.0322) & 3.7867 (3.6362) \\
 &  & dynamic & 0.0005 (0.0002) & 4.0086 (0.1385) & 0.0262 (0.0309) & 3.7882 (3.6250) \\
 & SRS &  & 0.0129 (0.0063) & 4.8454 (0.3042) & 0.1228 (0.4375) & 5.9759 (4.2162) \\
0.05 & PCA-QS & 2 & 0.0035 (0.0007) & 3.6103 (0.1041) & 0.1428 (0.0406) & 3.2883 (3.4485) \\
 &  & 10 & 0.0008 (0.0003) & 4.0224 (0.1372) & 0.0269 (0.0302) & 3.8098 (3.6965) \\
 &  & dynamic & 0.0005 (0.0002) & 4.0137 (0.1379) & 0.0257 (0.0302) & 3.9137 (3.8342) \\
 & SRS &  & 0.0026 (0.0010) & 4.6895 (0.4858) & 0.0305 (0.0396) & 5.6239 (4.1575) \\
0.1 & PCA-QS & 2 & 0.0013 (0.0003) & 3.5717 (0.1048) & 0.0608 (0.0359) & 2.9315 (3.3345) \\
 &  & 10 & 0.0008 (0.0003) & 4.0216 (0.1377) & 0.0267 (0.0308) & 3.8919 (3.6809) \\
 &  & dynamic & 0.0005 (0.0002) & 4.0105 (0.1345) & 0.0266 (0.0309) & 3.8287 (3.6640) \\
 & SRS &  & 0.0014 (0.0005) & 4.4867 (0.5523) & 0.0284 (0.0320) & 5.0408 (3.8959) \\
\bottomrule
\end{tabular}
\end{table}

\begin{table}[H]
\centering
\caption{Epileptic Seizure Dataset}
\label{tab:Epileptic}
\tiny
\begin{tabular}{cclllll}
\toprule
Deduce & Sampling & PC & Energy & Mahalanobis & MMD & KLD \\
\midrule
0.01 & PCA-QS & 2 & 0.1819 (0.0449) & 12.2597 (0.1927) & 2.3995 (0.6149) & 59.9135 (3.6438) \\
 &  & 10 & 0.0032 (0.0007) & 14.7994 (0.5363) & 0.3246 (0.0797) & 19.7332 (1.2963) \\
 &  & dynamic & 0.0032 (0.0006) & 14.8068 (0.5496) & 0.3293 (0.0814) & 19.3603 (1.2623) \\
 & SRS &  & 0.1241 (0.0264) & 12.5550 (0.2385) & 1.7178 (0.5595) & 61.4068 (3.6020) \\
0.05 & PCA-QS & 2 & 0.0257 (0.0038) & 14.8048 (0.3930) & 0.4883 (0.1239) & 45.3470 (2.3583) \\
 &  & 10 & 0.0033 (0.0006) & 14.8170 (0.5080) & 0.3233 (0.0806) & 19.7411 (1.3099) \\
 &  & dynamic & 0.0032 (0.0006) & 14.7984 (0.5327) & 0.3290 (0.0815) & 19.4299 (1.2788) \\
 & SRS &  & 0.0254 (0.0056) & 15.2553 (0.4093) & 0.4696 (0.1197) & 44.9030 (2.4159) \\
0.1 & PCA-QS & 2 & 0.0123 (0.0018) & 15.8887 (0.4347) & 0.3251 (0.0828) & 38.2653 (1.8899) \\
 &  & 10 & 0.0033 (0.0007) & 14.7937 (0.5274) & 0.3220 (0.0799) & 19.6320 (1.2817) \\
 &  & dynamic & 0.0032 (0.0006) & 14.7781 (0.5235) & 0.3294 (0.0807) & 19.4951 (1.3149) \\
 & SRS &  & 0.0128 (0.0025) & 16.0954 (0.4155) & 0.3267 (0.0792) & 38.4042 (2.0699) \\
\bottomrule
\end{tabular}
\end{table}

\subsection{Large-Scale Tabular Data}
\begin{table}[H]
\caption{Higgs Dataset}
\label{tab:higgs}
\tiny
\centering
\begin{tabular}{cclllll}
\toprule
Deduce & Sampling & PC & Energy & Mahalanobis & MMD & KLD \\
\midrule
0.01 & PCA-QS & 2 & 0.0048 (0.0007) & 7.3007 (0.0728) & 0.0531 (0.0160) & 5.2867 (0.0664) \\
 &  & 10 & 0.0028 (0.0005) & 7.2976 (0.0758) & 0.0545 (0.0179) & 4.9860 (0.0600) \\
 &  & dynamic & 0.0028 (0.0005) & 7.2970 (0.0729) & 0.0556 (0.0186) & 4.9819 (0.0586) \\
 & SRS &  & 0.0050 (0.0008) & 7.3001 (0.0693) & 0.0558 (0.0188) & 5.2775 (0.0715) \\
0.05 & PCA-QS & 2 & 0.0028 (0.0004) & 7.2962 (0.0740) & 0.0546 (0.0188) & 4.9816 (0.0613) \\
 &  & 10 & 0.0028 (0.0005) & 7.2972 (0.0729) & 0.0551 (0.0182) & 4.9828 (0.0617) \\
 &  & dynamic & 0.0028 (0.0004) & 7.2948 (0.0738) & 0.0556 (0.0193) & 4.9816 (0.0566) \\
 & SRS &  & 0.0028 (0.0005) & 7.2994 (0.0714) & 0.0543 (0.0177) & 4.9820 (0.0576) \\
0.1 & PCA-QS & 2 & 0.0028 (0.0004) & 7.2954 (0.0741) & 0.0547 (0.0184) & 4.9838 (0.0618) \\
 &  & 10 & 0.0028 (0.0005) & 7.2963 (0.0732) & 0.0547 (0.0175) & 4.9812 (0.0600) \\
 &  & dynamic & 0.0028 (0.0005) & 7.2949 (0.0767) & 0.0546 (0.0180) & 4.9810 (0.0566) \\
 & SRS &  & 0.0028 (0.0005) & 7.2944 (0.0734) & 0.0547 (0.0183) & 4.9874 (0.0607) \\
\bottomrule
\end{tabular}
\end{table}

\begin{table}[H]
\caption{Year Prediction MSD Dataset)}
\label{tab:yearprediction}
\tiny
\centering
\begin{tabular}{ccccccc}
\toprule
Deduce & Sampling & PC & Energy & Mahalanobis & MMD & KLD \\
\midrule
0.01 & PCA-QS & 2 & 0.0046 (0.0005) & 12.5050 (0.2283) & 0.1853 (0.0499) & 23.9607 (0.7412) \\
 &  & 10 & 0.0049 (0.0007) & 12.4285 (0.2361) & 0.1799 (0.0500) & 23.9563 (0.8357) \\
 &  & dynamic & 0.0049 (0.0007) & 12.3990 (0.2425) & 0.1831 (0.0480) & 23.9071 (0.8154) \\
 & SRS &  & 0.0048 (0.0007) & 12.5335 (0.2342) & 0.1837 (0.0487) & 23.8623 (0.7656) \\
0.05 & PCA-QS & 2 & 0.0048 (0.0006) & 12.3958 (0.2368) & 0.1825 (0.0486) & 23.8851 (0.7591) \\
 &  & 10 & 0.0049 (0.0008) & 12.4210 (0.2377) & 0.1796 (0.0457) & 23.9172 (0.7689) \\
 &  & dynamic & 0.0049 (0.0007) & 12.3779 (0.2427) & 0.1799 (0.0467) & 23.8859 (0.7721) \\
 & SRS &  & 0.0048 (0.0007) & 12.4128 (0.2349) & 0.1800 (0.0494) & 23.9038 (0.7856) \\
0.1 & PCA-QS & 2 & 0.0049 (0.0007) & 12.3872 (0.2407) & 0.1813 (0.0477) & 23.8819 (0.7770) \\
 &  & 10 & 0.0049 (0.0008) & 12.4271 (0.2453) & 0.1830 (0.0501) & 23.9855 (0.8135) \\
 &  & dynamic & 0.0049 (0.0007) & 12.3663 (0.2284) & 0.1828 (0.0480) & 23.8659 (0.7869) \\
 & SRS &  & 0.0048 (0.0007) & 12.3911 (0.2383) & 0.1824 (0.0473) & 23.9155 (0.7940) \\
\bottomrule
\end{tabular}
\end{table}

\subsection{Feature Complexity (Text/NLP)}
\begin{table}[H]
\caption{Online News}\label{OnlineNews}
\tiny
\centering
\begin{tabular}{ccccccc}
\toprule
Deduce & Sampling & PC & Energy & Mahalanobis & MMD & KLD \\
\midrule
0.01 & PCA-QS & 2 & 0.0482 (0.0061) & 10.3405 (0.1006) & 0.3796 (0.1371) & 16.9013 (12.4243) \\
 & PCA-QS & 10 & 0.0402 (0.0059) & 10.1516 (0.1707) & 0.2372 (0.0975) & 16.4014 (13.1539) \\
 & PCA-QS & dynamic & 0.0401 (0.0060) & 10.1190 (0.1740) & 0.2361 (0.0963) & 16.7481 (14.5030) \\
 & SRS &  & 0.0452 (0.0068) & 10.3410 (0.1021) & 0.2669 (0.1200) & 16.8300 (13.0912) \\
0.05 & PCA-QS & 2 & 0.0395 (0.0056) & 10.1894 (0.1341) & 0.2436 (0.1008) & 15.4110 (8.6053) \\
 & PCA-QS & 10 & 0.0404 (0.0059) & 10.1471 (0.1630) & 0.2367 (0.0959) & 16.1722 (12.4810) \\
 & PCA-QS & dynamic & 0.0399 (0.0059) & 10.1152 (0.1635) & 0.2402 (0.1008) & 16.0097 (12.0100) \\
 & SRS &  & 0.0402 (0.0057) & 10.2082 (0.1406) & 0.2373 (0.0907) & 16.2688 (13.1679) \\
0.1 & PCA-QS & 2 & 0.0401 (0.0062) & 10.1551 (0.1436) & 0.2438 (0.1014) & 15.9720 (11.4136) \\
 & PCA-QS & 10 & 0.0404 (0.0061) & 10.1394 (0.1584) & 0.2344 (0.0920) & 15.8864 (10.7884) \\
 & PCA-QS & dynamic & 0.0401 (0.0059) & 10.1198 (0.1668) & 0.2464 (0.1083) & 16.5685 (14.2315) \\
 & SRS &  & 0.0399 (0.0059) & 10.1630 (0.1391) & 0.2371 (0.1021) & 16.1921 (12.6247) \\
\bottomrule
\end{tabular}
\end{table}

\end{document}


\maketitle


\section{Quantile-Based Sampling in PCA Space}
\label{app:pcaqs_sampling}

PCA-QS preserves dataset representativity by employing quantile-based stratification in the principal component space. Instead of selecting instances randomly, PCA-QS partitions data into quantile-defined groups along principal components, ensuring that all parts of the distribution are well-represented in the sampled subset.

\subsection{Quantile Stratification for PCA-QS}
Once PCA transformation is performed, data is structured using quantile-based partitioning:
\subsubsection*{Case: Single Principal Component (\(\text{num\_pcs} = 1\))}
For a dataset projected onto the first principal component, we divide it into \( g \) quantile groups:
\begin{equation}
[\text{min}, Q_1), [Q_1, Q_2), \dots, [Q_{g-1}, \text{max}],
\end{equation}
where \( Q_i \) represents the \( i \)-th quantile of the distribution. Sampling is performed proportionally from each quantile group to retain statistical representativeness while reducing dataset size.

\subsubsection*{Case: Multiple Principal Components (\(\text{num\_pcs} > 1\))}
When more than one principal component is retained, quantile stratification extends to a multidimensional space:
\begin{enumerate}
    \item \textbf{Component-wise Quantile Partitioning:} Each principal component \( z_j \) is divided into \( g \) quantile groups.
    \item \textbf{Composite Quantile Groups:} Each sample is assigned a quantile vector based on its quantile memberships across the retained \( k \) components:
    \[
    \{q_{i1}, q_{i2}, \dots, q_{ik}\}.
    \]
    \item \textbf{Structured Sampling:} A fixed proportion of instances is randomly selected from each composite quantile group, maintaining balanced representation.
\end{enumerate}

\subsection{Sampling Strategy and Illustrative Example}
Samples are drawn from each quantile group based on:
\begin{equation}
\text{Sample Size per Group} = \min\left(\lceil \text{retention rate} \times N_g \rceil, N_g\right),
\end{equation}
where \( N_g \) is the number of samples in group \( g \), and the retention rate (e.g., \( \delta = 0.05 \)) determines the proportion of data retained.

\paragraph{Illustrative Example}
Consider a dataset with \( n = 10,000 \) samples and \( p = 50 \) features, where PCA is applied with \( k = 2 \) principal components. With a retention rate \( \delta = 0.05 \):

\begin{enumerate}
    \item The dataset is projected onto a \( 10,000 \times 2 \) matrix \( \mathbf{Z} \).
    \item Each principal component is partitioned into 5 quantile bins, creating a total of \( 5^2 = 25 \) composite quantile groups.
    \item If a group contains \( N_{3,5} = 400 \) samples, the retained subset is:
    \[
    \lceil 0.05 \times 400 \rceil = 20.
    \]
    \item This process is repeated across all quantile groups to form the final dataset.
\end{enumerate}

This structured sampling ensures that statistical properties of the original dataset are preserved while significantly reducing computational complexity.

\section{Intuitive Explanations of Theoretical Results}
\label{app:intuition}

\begin{figure}[h]
    \centering
    \includegraphics[width=0.9\textwidth]{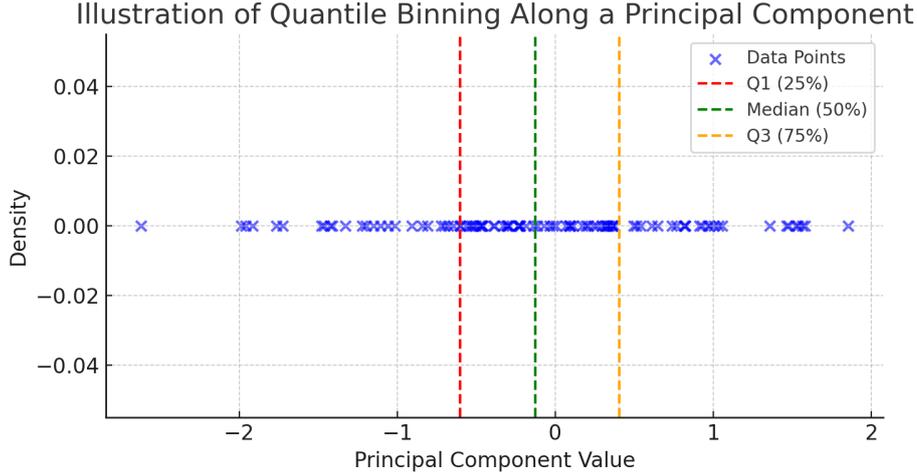}
    \caption{Illustration of quantile binning along a principal component. As the number of samples increases, the estimated quantile boundaries become more stable.}
    \label{fig:quantile_bins}
\end{figure}

\paragraph{Uniform Convergence of Quantiles}
Theorem 1 states that as the number of samples increases, the estimated quantiles of the principal component scores become more stable and accurate. To illustrate, imagine sorting a dataset along its most important directions (principal components) and dividing it into equal-probability bins (quantiles). With a small sample, these estimated quantiles may be imprecise, but as the dataset grows, the proportion of data in each bin converges to its true value. This ensures that PCA-QS preserves the overall structure of the data while reducing its size. Figure~\ref{fig:quantile_bins} illustrates this concept: as the sample size increases, the distribution within each bin becomes more stable, guaranteeing representativeness.

\paragraph{KL Divergence Decay in PCA-QS}
KL divergence measures how much an estimated probability distribution deviates from the true one. The fast decay rate ($O(n^{-1})$) in PCA-QS ensures that as more data points are included in the sampling process, the distribution of the sampled data becomes nearly identical to the original dataset.

\paragraph{Wasserstein Distance in PCA-QS}

The Wasserstein distance measures how much one probability distribution must be transformed to match another, capturing both density differences and spatial relationships. For example, comparing two cities based on their population distributions: KL divergence quantifies differences in density, while Wasserstein distance accounts for how far people would need to move to align the two distributions exactly. In PCA-QS, the Wasserstein distance decay rate, $O(n^{-1/d})$, ensures that the selected samples remain representative while preserving the geometric structure of the original dataset.

\begin{figure}[h]
    \centering
    \includegraphics[width=0.6\textwidth]{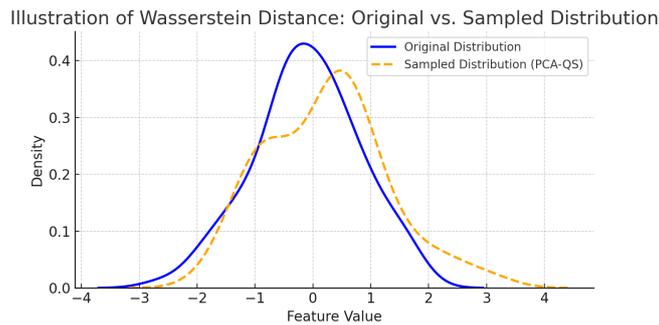}
    \caption{Illustration of Wasserstein distance: The sampled distribution (right) maintains the spatial structure of the original dataset (left).}
    \label{fig:wasserstein}
\end{figure}

However, computing the Wasserstein distance is computationally intensive in high dimensions, which makes it impractical for large-scale numerical comparisons. By projecting data into lower-dimensional principal component subspaces, PCA-QS improves Wasserstein convergence rates while avoiding prohibitive computation, making it an effective method for structure-preserving data reduction.

\subsection{Recommended Settings and Consistency}

The selection of optimal parameters in Principal Component Analysis-based Quantile Sampling (PCA-QS) is inherently application-dependent. There is no universal combination of parameters that is best for all scenarios. Instead, the appropriate parameter choices should align with the specific objectives of the study or application. Based on this consideration, we outline key parameter selection strategies as follows.

For practical use, the choice of PCA-QS settings depends on balancing computational efficiency, statistical representativeness, geometric preservation, and interpretability. For aggressive data reduction with structural retention, use a deduction rate of 0.02 to 0.05, select 6 to 10 principal components, and divide into 10 to 15 quantile groups. For computationally constrained tasks needing a trade-off between speed and accuracy, a deduction rate of 0.02 to 0.05 with 8 to 12 principal components and 10 to 15 quantiles is suggested. When retaining full statistical representativeness is critical, target an explained variance of 85–95\%, use 12 to 15 principal components, 15 to 20 quantile groups, and a deduction rate between 0.05 and 0.10. To preserve local geometric structures such as clusters or outliers, choose 12 to 20 principal components with 20 to 30 quantile groups and a deduction rate around 0.05 to 0.10. If interpretability is the main goal, a lower dimension is better: keep 6 to 10 principal components with an explained variance of 80–85\% and 10 to 15 quantile groups. 
Overall, effective PCA-QS usage depends on adapting these settings to match the goals and constraints of the task, ensuring both robustness and practical feasibility.

\section{Conclusion Remark}
The optimal PCA-QS settings depend on the trade-off between accuracy, computational efficiency, and interpretability. Rather than relying on manually tuned parameters, a well-designed PCA-QS framework should allow users to dynamically prioritize objectives. This ensures flexibility in adapting PCA-QS for diverse analytical tasks while maintaining robust statistical integrity.

Empirical results align with theory: PCA-QS consistently outperforms simple random sampling (SRS) in preserving key data distributions, reducing KL divergence and Wasserstein distance, and performing better as dimensionality grows. Recommended settings for optimal performance are also supported by theory and experiments: use 12–15 principal components, 15–20 quantile groups, a minimum sample size of 1000, and a deduction rate between 0.05 and 0.10.

For high-dimensional data (20+ features), keeping 12–15 PCs and 15–20 quantiles works well with 1000–5000 samples. For moderate dimensions (10–15 features), use about 70–90\% of features as PCs and match quantiles to PCs. For low-dimensional cases ($\leq$5 features), include all PCs and set quantiles equal to the dimension count, with at least 500 samples. In general, aligning the number of quantile bins with the number of principal components ensures balanced stratification and reliable sampling quality.
Overall, this strategy ensures PCA-QS remains theoretically grounded, computationally feasible, and empirically robust across a variety of practical scenarios.

\section*{Python Code for PCA-QS Pipeline}
This section contains the Python code used in the associated paper for implementing the PCA-guided Quantile Sampling (PCA-QS) and Simple Random Sampling (SRS) comparison pipeline. This supplementary material provides the exact script to ensure reproducibility and transparency.

\section*{Python Code}
\label{python:logistic}

\begin{lstlisting}[language=Python]

# ======================== Configuration ========================
config = {
    'random_seed': 123,
    'n_samples': 100000,
    'n_features': 50,
    'n_informative': 30,
    'n_redundant': 15,
    'n_repeated': 10,
    'class_sep': 0.1,
    'flip_y': 0.1,
    'weights': [0.9, 0.1],
    'nonlinear': True,  # Set to False if you want the original GMM
    'test_size': 1000,
    'train_sample_sizes': [1000],
    'n_bins_pca_qs': 10,
    'repeat': 1000,
    'variance_threshold': 0.70,
    'classifier_list': ['logistic', 'random_forest', 'xgboost', 'knn', 'svm'],
    'knn_n_neighbors': 5,
    'csv_output_prefix': 'Correlated_pca_qs_vs_srs',
    'n_jobs': -4
}

# ======================== Imports ==============================
import numpy as np
import pandas as pd
from sklearn.preprocessing import StandardScaler
from sklearn.decomposition import PCA
from sklearn.metrics import roc_auc_score, accuracy_score, confusion_matrix
from sklearn.linear_model import LogisticRegression
from sklearn.ensemble import RandomForestClassifier
from sklearn.svm import SVC
from sklearn.neighbors import KNeighborsClassifier
from joblib import Parallel, delayed
from tqdm import tqdm
import os

try:
    from xgboost import XGBClassifier
    xgb_available = True
except ImportError:
    xgb_available = False

def generate_gmm_data(n_samples, weights, n_features, random_seed=42, nonlinear=False):
    np.random.seed(random_seed)
    centers = [np.zeros(n_features), np.ones(n_features) * 0.5]
    covariances = [np.identity(n_features), np.identity(n_features) * 1.2]
    X_list, y_list = [], []

    for class_idx, (weight, center, cov) in enumerate(zip(weights, centers, covariances)):
        n_class_samples = int(n_samples * weight)
        X_class = np.random.multivariate_normal(mean=center, cov=cov, size=n_class_samples)
        y_class = np.full(n_class_samples, class_idx)
        X_list.append(X_class)
        y_list.append(y_class)

    X = np.vstack(X_list)
    y = np.concatenate(y_list)

    if nonlinear:
        # Add interaction terms
        interaction_terms = np.array([X[:, i] * X[:, j]
                                      for i in range(n_features)
                                      for j in range(i + 1, n_features)]).T

        # Add sine transformation
        sine_terms = np.sin(X)

        # Combine all features
        X = np.hstack([X, interaction_terms, sine_terms])

    return X, y


# ======================== Utility Functions =====================
def compute_fp_fn_rates(cm):
    tn, fp, fn, tp = cm.ravel()
    fpr = fp / (fp + tn) if (fp + tn) > 0 else 0
    fnr = fn / (fn + tp) if (fn + tp) > 0 else 0
    tnr = tn / (tn + fp) if (tn + fp) > 0 else 0
    tpr = tp / (tp + fn) if (tp + fn) > 0 else 0
    return fpr, fnr, tpr, tnr

def evaluate_results(y_true, y_pred, y_prob):
    cm = confusion_matrix(y_true, y_pred)
    acc = accuracy_score(y_true, y_pred)
    auc = roc_auc_score(y_true, y_prob)
    fpr, fnr, tpr, tnr = compute_fp_fn_rates(cm)
    precision = np.sum((y_true == 1) & (y_pred == 1)) / max(np.sum(y_pred == 1), 1)
    recall = tpr
    f1 = 2 * precision * recall / max(precision + recall, 1e-6)
    return acc, auc, fpr, fnr, tpr, tnr, f1

def get_classifier(name, seed):
    if name == 'logistic':
        return LogisticRegression(solver='liblinear', random_state=seed)
    elif name == 'svm':
        return SVC(kernel='rbf', probability=True, random_state=seed)
    elif name == 'random_forest':
        return RandomForestClassifier(random_state=seed, n_jobs=config['n_jobs'])
    elif name == 'xgboost':
        if xgb_available:
            return XGBClassifier(use_label_encoder=False, eval_metric='logloss', random_state=seed, n_jobs=config['n_jobs'])
        else:
            raise ImportError("XGBoost is not installed.")
    elif name == 'knn':
        return KNeighborsClassifier(n_neighbors=config['knn_n_neighbors'], n_jobs=config['n_jobs'])
    else:
        raise ValueError(f"Unknown classifier: {name}")

def pca_qs_sample_indices(X_scaled, train_indices, sample_size, n_bins, variance_threshold):
    # Perform PCA
    X_pca_full = PCA().fit(X_scaled)
    cumulative_variance = np.cumsum(X_pca_full.explained_variance_ratio_)
    n_components = np.searchsorted(cumulative_variance, variance_threshold) + 1
    pca = PCA(n_components=n_components)
    X_pca = pca.fit_transform(X_scaled)

    # Use only training data
    X_pca_train = X_pca[train_indices]

    # Discretize each PC into quantile bins
    quantile_bins = []
    for i in range(n_components):
        quantiles = np.percentile(X_pca_train[:, i], np.linspace(0, 100, n_bins + 1))
        bins = np.digitize(X_pca_train[:, i], quantiles[1:-1], right=True)
        quantile_bins.append(bins)

    quantile_bins = np.stack(quantile_bins, axis=1)
    composite_keys = ['-'.join(map(str, row)) for row in quantile_bins]

    # Map keys to indices
    from collections import defaultdict
    bin_map = defaultdict(list)
    for idx, key in zip(train_indices, composite_keys):
        bin_map[key].append(idx)

    # Sample from each group proportionally
    pca_qs_indices = []
    for group in bin_map.values():
        size = max(1, int(len(group) * sample_size / len(train_indices)))
        sampled = np.random.choice(group, size=min(len(group), size), replace=False)
        pca_qs_indices.extend(sampled)

    return pca_qs_indices, n_components


def run_single_experiment(classifier_name, sample_size, repeat):
    np.random.seed(config['random_seed'] + repeat)

    X, y = generate_gmm_data(
        n_samples=config['n_samples'],
        weights=config['weights'],
        n_features=config['n_features'],
        random_seed=config['random_seed'] + repeat
    )

    scaler = StandardScaler()
    X_scaled = scaler.fit_transform(X)

    test_indices = np.random.choice(len(X_scaled), size=config['test_size'], replace=False)
    train_indices = np.setdiff1d(np.arange(len(X_scaled)), test_indices)
    X_test = X_scaled[test_indices]
    y_test = y[test_indices]

    records = []

    srs_indices = np.random.choice(train_indices, size=sample_size, replace=False)
    X_srs = X_scaled[srs_indices]
    y_srs = y[srs_indices]
    if len(np.unique(y_srs)) >= 2:
        clf_srs = get_classifier(classifier_name, seed=config['random_seed'] + repeat)
        clf_srs.fit(X_srs, y_srs)
        y_prob_srs = clf_srs.predict_proba(X_test)[:, 1]
        threshold_srs = np.mean(y_srs)
        y_pred_srs = (y_prob_srs >= threshold_srs).astype(int)
        acc_srs, auc_srs, fpr_srs, fnr_srs, tpr_srs, tnr_srs, f1_srs = evaluate_results(y_test, y_pred_srs, y_prob_srs)
        records.append({
            'method': 'SRS', 'sample_size': sample_size, 'repeat': repeat,
            'accuracy': acc_srs, 'auc': auc_srs, 'fpr': fpr_srs, 'fnr': fnr_srs,
            'n_pca_components': None,
            'tpr': tpr_srs, 'tnr': tnr_srs, 'f1': f1_srs
        })

    pca_qs_indices, n_components_used = pca_qs_sample_indices(
        X_scaled, train_indices, sample_size,
        config['n_bins_pca_qs'], config['variance_threshold']
    )
    X_qs = X_scaled[pca_qs_indices]
    y_qs = y[pca_qs_indices]
    if len(np.unique(y_qs)) >= 2:
        clf_qs = get_classifier(classifier_name, seed=config['random_seed'] + repeat)
        clf_qs.fit(X_qs, y_qs)
        y_prob_qs = clf_qs.predict_proba(X_test)[:, 1]
        threshold_qs = np.mean(y_qs)
        y_pred_qs = (y_prob_qs >= threshold_qs).astype(int)
        acc_qs, auc_qs, fpr_qs, fnr_qs, tpr_qs, tnr_qs, f1_qs = evaluate_results(y_test, y_pred_qs, y_prob_qs)
        records.append({
            'method': 'PCA-QS', 'sample_size': sample_size, 'repeat': repeat,
            'n_pca_components': n_components_used,
            'accuracy': acc_qs, 'auc': auc_qs, 'fpr': fpr_qs, 'fnr': fnr_qs,
            'tpr': tpr_qs, 'tnr': tnr_qs, 'f1': f1_qs
        })

    return records

# ======================== Main Experiment ======================
def run_experiment_all_classifiers():
    for classifier_name in config['classifier_list']:
        if classifier_name == 'xgboost' and not xgb_available:
            continue

        output_path = f"{config['csv_output_prefix']}_{classifier_name}_results.csv"
        tasks = [
            (classifier_name, sample_size, repeat)
            for sample_size in config['train_sample_sizes']
            for repeat in range(config['repeat'])
        ]

        all_records = Parallel(n_jobs=config['n_jobs'])(
            delayed(run_single_experiment)(clf, size, rep)
            for clf, size, rep in tqdm(tasks, desc=f"Running {classifier_name}")
        )

        flat_records = [rec for sublist in all_records for rec in sublist]
        if flat_records:
            df_results = pd.DataFrame(flat_records)
            df_results.to_csv(output_path, index=False)
            print(f"Saved: {output_path}")

# ======================== Run ======================
if __name__ == '__main__':
    run_experiment_all_classifiers()

\end{lstlisting}